\documentclass[oneside,11pt]{article}

\usepackage[utf8]{inputenc}
\usepackage[T1]{fontenc}

\usepackage{amsthm}
\usepackage{amsmath}
\usepackage{amsfonts}
\usepackage{amssymb}

\usepackage{bbm}

\usepackage{graphicx} 
\usepackage{float}
\usepackage{booktabs}
\usepackage{multirow}
\usepackage{rotating}
\usepackage{array}
\usepackage{xcolor}
\usepackage{subcaption}
\usepackage[ruled]{algorithm2e}

\newcolumntype{R}[1]{>{\raggedleft\let\newline\\\arraybackslash\hspace{0pt}}m{#1}}
\newcolumntype{L}[1]{>{\raggedright\let\newline\\\arraybackslash\hspace{0pt}}m{#1}}

\usepackage{breakcites}

\usepackage{nowidow}

\usepackage{enumitem}

\usepackage[top=1.5cm,bottom=2cm,right=2.5cm,left=2.5cm]{geometry}
\usepackage{titling}

\usepackage{natbib}

\usepackage{ifplatform}

\usepackage{textcomp}

\ifwindows
\fi

\allowdisplaybreaks

\newtheorem{definition}{Definition}

\newtheorem{remark}{Remark}
\newtheorem{proposition}{Proposition}

\newtheorem{algo}{Algorithm}

\renewenvironment{proof}[1]{\textit{Proof#1.}}{\qed\\} 

\usepackage[pdftex,final,bookmarksnumbered,bookmarksopen=false,breaklinks,colorlinks]{hyperref}
\hypersetup{
	final,
	bookmarksnumbered=true,
	bookmarksopen=true,
	bookmarksopenlevel=0,
	unicode=false,
	pdftoolbar=true,
	pdfmenubar=true,
	pdffitwindow=false,
	pdftitle={Toroidal PCA via density ridges},
	pdfdisplaydoctitle=true,
	pdftoolbar=true,
	pdfmenubar=true,
	pdflang={English},
	pdfauthor={Eduardo Garcia-Portugues, Arturo Prieto-Tirado},
	pdfsubject={arXiv paper},
	pdfcreator={Eduardo Garcia-Portugues},
	pdfproducer={Eduardo Garcia-Portugues},
	pdfkeywords={Circular data}{Density ridges}{Dimension reduction} {Torus},
	pdfnewwindow=true,
	breaklinks=true,
	hidelinks,       
	linkcolor=black,
	citecolor=black,
	filecolor=magenta,
	urlcolor=cyan
}

\newif\ifmain
\maintrue
\newif\ifsupplement
\supplementtrue

\newif\iffigstabs
\figstabstrue

\begin{document}

\ifmain

\title{Toroidal PCA via density ridges}
\setlength{\droptitle}{-1cm}
\predate{}%
\postdate{}%
\date{}

\author{Eduardo Garc\'ia-Portugu\'es$^{1,3}$ and Arturo Prieto-Tirado$^{2}$}
\footnotetext[1]{Department of Statistics, Universidad Carlos III de Madrid (Spain).}
\footnotetext[2]{SDG Group (Spain).}
\footnotetext[3]{Corresponding author. e-mail: \href{mailto:edgarcia@est-econ.uc3m.es}{edgarcia@est-econ.uc3m.es}.}
\maketitle

\begin{abstract}
	Principal Component Analysis (PCA) is a well-known linear dimension-reduction technique designed for Euclidean data. In a wide spectrum of applied fields, however, it is common to observe multivariate circular data (also known as toroidal data), rendering spurious the use of PCA on it due to the periodicity of its support. This paper introduces Toroidal Ridge PCA (TR-PCA), a novel construction of PCA for bivariate circular data that leverages the concept of density ridges as a flexible first principal component analog. Two reference bivariate circular distributions, the bivariate sine von Mises and the bivariate wrapped Cauchy, are employed as the parametric distributional basis of TR-PCA. Efficient algorithms are presented to compute density ridges for these two distribution models. A complete PCA methodology adapted to toroidal data (including scores, variance decomposition, and resolution of edge cases) is introduced and implemented in the companion R package \texttt{ridgetorus}. The usefulness of TR-PCA is showcased with a novel case study involving the analysis of ocean currents on the coast of Santa Barbara.
\end{abstract}
\begin{flushleft}
\end{flushleft}

\section{Introduction}
\label{sec:introduction}

Toroidal data is formed by tuples of observations that lie on the $p$-dimensional torus $\mathbb{T}^p =[-\pi, \pi)^p$, $p\geq 1$, where $-\pi$ and $\pi$ are identified. This kind of data is present in a variety of applied fields ranging from bioinformatics, when modeling angles in atom bonds \citep[e.g.,][]{Boomsma2008}, to environmental sciences, when modeling sea conditions \citep[e.g.,][]{Jona-Lasinio2012}, among many others \citep[see, e.g.,][]{Ley2018}. The fact that circular variables present different properties than linear variables renders spurious the application of standard statistical tools, with their adaptations comprising the so-called directional statistics \citep{Mardia1999a}; see, e.g., \cite{Pewsey2021} for a recent review. In particular, Principal Component Analysis (PCA) is a widely used (linear) statistical tool whose extension to circular variables has been challenging due to the ill-posedness of linear relations on $\mathbb{T}^p$. On $\mathbb{T}^2$, PCA produces linear subspaces that are either non-periodic or, if extended periodically by ``wrapping'' at the $-\pi \equiv \pi$ boundaries, would almost surely result in useless perfect fits to a sample. Indeed, infinitely-dense line wrappings are produced by irrational slopes for the first sample principal component, which arise with probability one in a sample generated by a continuous random vector.

There have been several attempts to extend PCA to toroidal data, with all of them facing some relevant limitations or foundational issues; see Section 8.1 in \cite{Pewsey2021} for an overview. These approaches can be divided into two general branches. The first one consists of ``geodesic-based'' methods, like \cite{Fletcher2004}'s Principal Geodesic Analysis (PGA). PGA defines the first principal geodesic as the one passing through the intrinsic mean that minimizes the sum of squared intrinsic residuals. \cite{Nodehi2015} applied PGA to the torus. Both of these approaches inevitably may lead to the aforementioned infinite wrapping. The other branch of techniques is based on embedding toroidal data in Euclidean space to use classical PCA afterward. For example, \cite{Mu2005} proposed dihedral PCA (dPCA), which maps the angles onto $\mathbb{S}^1 \times \mathbb{S}^1$ via $(\phi, \psi) \mapsto (\cos\phi,\sin\phi,\cos\psi,\sin \psi)$ and then applies regular PCA. Although this method transforms the data in a unique way, the geometry of $\mathbb{S}^1 \times \mathbb{S}^1$ is ignored in the subsequent PCA application, leading to substantial transformation-induced artifacts in the resulting PCA scores. \cite{Riccardi2009} introduced angular PCA (aPCA): a centering of the data with the circular mean followed by an application of PCA that disregards periodicity. Another idea was that of characterizing the covariance matrix, which \cite{Kent2009} did on a wrapped normal model using trigonometric moments to facilitate PCA on it. To reduce the distortion generated by transformation-based approaches, \cite{Sittel2017} proposed a modification on aPCA consisting of a shift of the data so that the ``periodic region'' is located at the lowest-density area (determined by histogram approximation) and most of the data is located in the more ``non-periodic region''. Like aPCA, the periodicity of the resulting principal components and scores is not enforced. \cite{Eltzner2018} introduced Torus PCA (T-PCA) by mapping $\mathbb{T}^d$ onto $\mathbb{S}^d$, a geometrically-benign space that, unlike geodesic PCA, produces non-dense linear subspaces. Principal Nested Spheres (PNS) \citep{Jung2012} is then carried out on $\mathbb{S}^d$. Although a principled approach, the transformation in T-PCA is not invariant under permutations of variables, which may yield significantly different outcomes, and is prone to create artifacts in, e.g., datasets with marginal uniform-like distributions. More recently, \cite{zoubouloglou2021scaled} introduced Scaled Torus PCA (ST-PCA), which uses multidimensional scaling maps between $\mathbb{T}^d$ and $\mathbb{S}^d$ designed to alleviate the aforementioned drawbacks in T-PCA, and then applies PNS. However, these maps are computationally demanding and difficult to invert. Despite many partial advances, the development of a flexible and successful PCA on the torus that is free of significant drawbacks remains remarkably open.

Differently from the previous PCA approaches, the purpose of this article is to advance a flexible density-driven PCA on $\mathbb{T}^2$. A landmark in flexible dimension reduction on $\mathbb{R}^p$ was the principal curves of \cite{Hastie1989}. The linear subspaces spanned by PCA's principal directions are self-consistent when the data is normally distributed, i.e., the expectation of all the points projecting onto a point $\mathbf{x} \in \mathbb{R}^p$ of the subspace is also $\mathbf{x}$. This fact motivates the definition of principal curves as those that are self-consistent, although their existence is not guaranteed. Another alternative definition of principal curves was introduced by \cite{delicado2001, delicado2003} in terms of the total variance, which has the advantage of a guaranteed existence. The total variance of a random variable is minimal when its hyperplane is orthogonal to the first principal component. A different approach to flexible PCA was given by \cite{Ozertem2011LocallyDP}, who defined the concept of density ridges, sets of points that characterize the main features of the density, for locally defined principal curves and surfaces. Notably, \cite{Ozertem2011LocallyDP} introduced an estimation approach for density ridges based on the mean shift algorithm. Although the first notions of density ridges were already introduced by \cite{hall1992ridge}, it has been in the last decade that the topic has attracted more attention. To review just a few contributions, \cite{Genovese2014} showed Hausdorff-consistency of the mean-shift estimated ridge sets, \cite{Chen2015} gave a Gaussian-process representation of the asymptotic distribution of the Hausdorff distances between estimated and theoretical ridges, and \cite{Qiao2016} established the pointwise asymptotic normality of the estimated ridges and the asymptotic distribution of its uniform deviation from theoretical ridges.

This article aims to generalize PCA to bivariate toroidal data by making use of density ridges. The contributions towards this goal are manifold. First, we provide efficient algorithms for computing periodic density ridges for two standard toroidal distributions, the Bivariate Sine von Mises (BSvM) \citep{Singh2002} and the Bivariate Wrapped Cauchy (BWC) \citep{Kato2015a}. These two algorithms substantially reduce the computational burden of iterative methods, based on integral curves, by (1) targeting the implicit equations that define the density ridge, (2) leveraging specific solutions of these implicit equations, and (3) exploiting the ridge symmetries induced by the BSvM and BWC densities. Second, we present Toroidal Ridge PCA (TR-PCA), which uses the connected component of the density ridge that passes through the mode(s) of the density as a flexible first principal component in the parametric setting of BSvM and BWC distributions. TR-PCA provides PCA-like scores through the obtention of signed distances along the ridge, signed projections onto the ridge, and Fréchet means within the ridge. These operations are facilitated by Fourier approximations of the ridge curve, which provide a fast analytical handle. Using likelihood theory, TR-PCA automatically handles edge cases arising in practice, such as diagonal ridges, and decides the best-fitting underlying distribution. Third, we empirically evidence the advantages of TR-PCA over aPCA, arguably the most readily implementable alternative, in a series of illustrative numerical examples. In particular, unlike transformation-based approaches, TR-PCA does not introduce distortions and, unlike other approaches like aPCA, yields scores that inherit the periodicity of the data. Fourth, the usefulness of TR-PCA is demonstrated with a novel application on the study of ocean currents at the coast of Santa Barbara, where the obtained principal ridge explains the currents' behavior successfully. A final contribution is the companion R package \texttt{ridgetorus}, which provides an implementation of TR-PCA and allows the end-to-end replicability of the data application.

The organization of the rest of this article is as follows: Sections \ref{sec:Density_ridges}--\ref{sec:ridgesbwc} study the population case of density ridges, and their sample version appears in Section \ref{sec:trpca}, with the definition of TR-PCA, and in Section \ref{sec:application}. More precisely, Section \ref{sec:Density_ridges} provides the definition and some useful properties of density ridges on $\mathbb{R}^p$, as well as the computation methods thereof. Section \ref{sec:ridgesbsvm} is devoted to the computation and properties of the density ridge for the BSvM distribution on $\mathbb{T}^2$. Section \ref{sec:ridgesbwc} gives an analogous construction for the BWC distribution. Then, Section \ref{sec:trpca} introduces an effective parametrization of the connected component of the BSvM/BWC ridge and explains how to obtain scores from it in the presence of a sample, resulting in the definition of TR-PCA in Algorithm \ref{def:TR-PCA}. Section \ref{sec:application} shows an application of TR-PCA on real bivariate data. Finally, a discussion of the main conclusions, alternatives, and limitations of the present work is given in Section \ref{sec:discussion}.

\section{Density ridges}
\label{sec:Density_ridges}

\subsection{Definition}
\label{sec:defrid}

Density ridges are higher-dimensional extensions of the concept of mode that are informative of the main features of a density. A mode is a local maximum of the density function $f$ that, if $f \in \mathcal{C}^2(\mathbb{R}^p)$, $p\geq2$, can be characterized by a null gradient and negative Hessian eigenvalues, if the degenerate cases are excluded. The eigendecomposition of the Hessian of a density function $f \in \mathcal{C}^2(\mathbb{R}^p)$ evaluated at a point $\mathbf{x} \in \mathbb{R}^p$ is given by $\mathrm{H} f(\mathbf{x})=\mathbf{U}(\mathbf{x}) \boldsymbol{\Lambda}(\mathbf{x}) (\mathbf{U}(\mathbf{x}))^\prime$, where $\mathbf{U}(\mathbf{x})=(\mathbf{u}_{1}(\mathbf{x}), \ldots, \mathbf{u}_{p}(\mathbf{x}))$ is a matrix whose columns are the eigenvectors of $\mathrm{H} f(\mathbf{x})$ and $\boldsymbol{\Lambda}(\mathbf{x})=\mathrm{diag}(\lambda_{1}(\mathbf{x}), \ldots,\allowbreak \lambda_{p}(\mathbf{x}))$, $\lambda_{1}(\mathbf{x}) \geq \ldots \geq \lambda_{p}(\mathbf{x})$, contains their eigenvalues. Denoting $\mathbf{U}_{(p-1)}(\mathbf{x}):=(\mathbf{u}_{2}(\mathbf{x}), \ldots, \mathbf{u}_{p}(\mathbf{x}))$ and defining the projected gradient on $\{\mathbf{u}_2(\mathbf{x}),\ldots,\mathbf{u}_p(\mathbf{x}) \}$ as $\mathrm{D}_{(p-1)} f(\mathbf{x}):=\mathbf{U}_{(p-1)}(\mathbf{x}) (\mathbf{U}_{(p-1)}(\mathbf{x}))^{\prime} \mathrm{D} f(\mathbf{x})$, where $\mathrm{D}f(\mathbf{x})$ stands for the (column vector) gradient, the density ridge is defined by \cite{Genovese2014} as follows.

\begin{definition}[Density ridge; \citealt{Genovese2014}]\label{eq:ridge}
	Let $f\in\mathcal{C}^2(\mathbb{R}^p)$, $p\geq2$. The \emph{density ridge} of $f$ is the set
	\begin{align*}
		\mathcal{R}(f):=\big\{\mathbf{x} \in \mathbb{R}^{p}:\big\|\mathrm{D}_{(p-1)} f(\mathbf{x})\big\|=0,\lambda_{2}(\mathbf{x}),\ldots,\lambda_{p}(\mathbf{x})<0\big\}.
	\end{align*}
\end{definition}

\begin{figure}[htpb!]
	\centering
	\includegraphics[width=0.25\textwidth,clip,trim={1cm 0cm 0.5cm 2cm}]{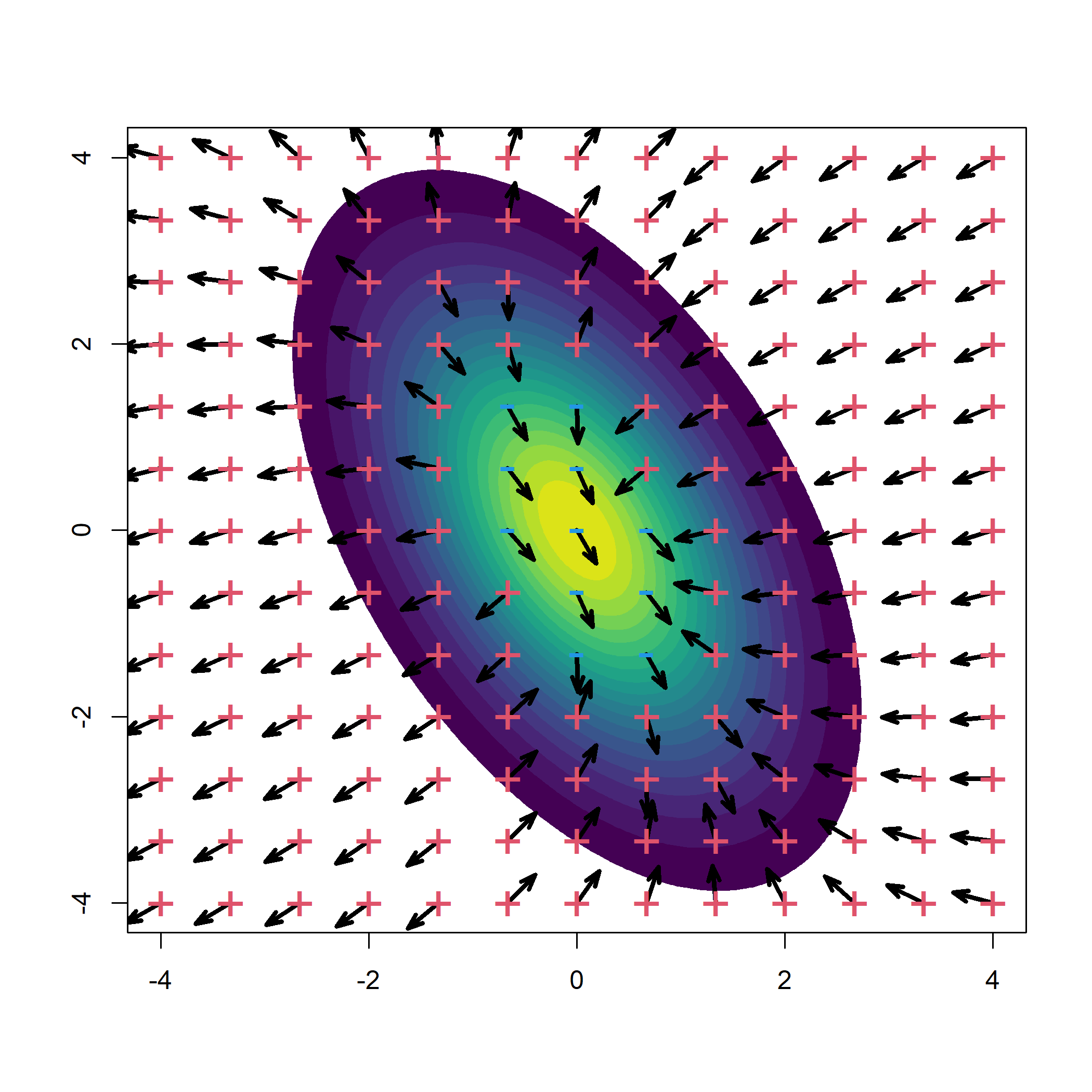}%
	\includegraphics[width=0.25\textwidth,clip,trim={1cm 0cm 0.5cm 2cm}]{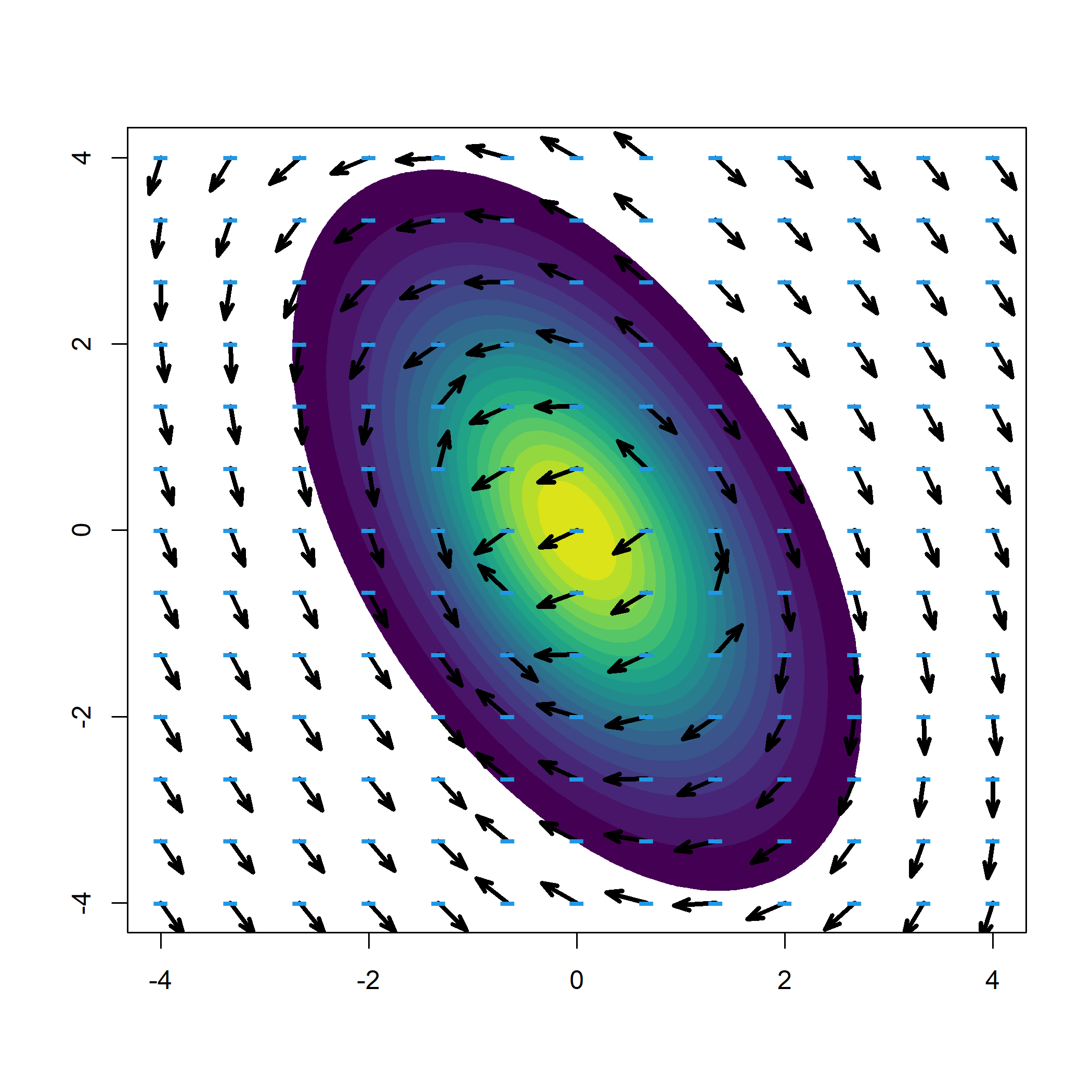}%
	\includegraphics[width=0.25\textwidth,clip,trim={1cm 0cm 0.5cm 2cm}]{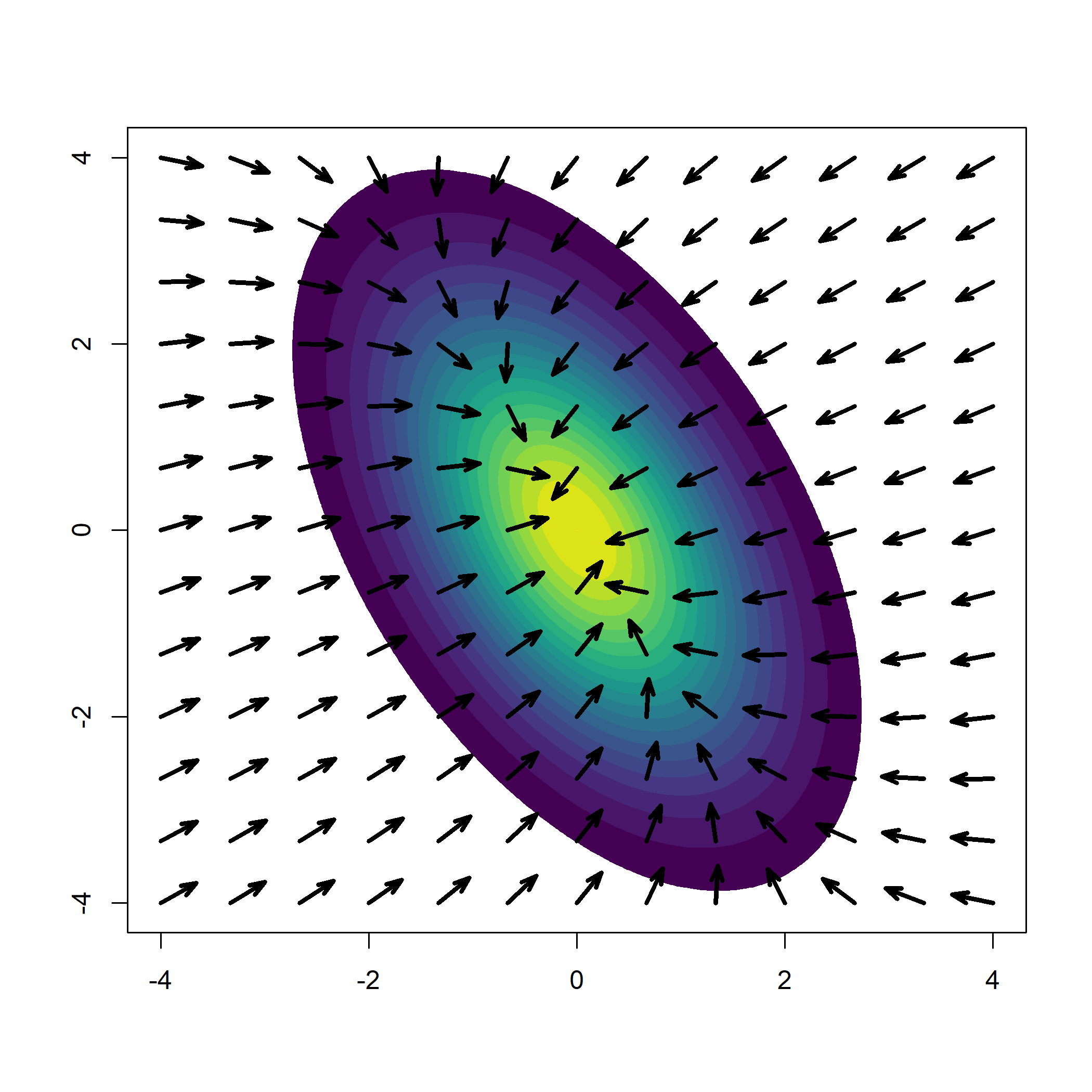}%
	\includegraphics[width=0.25\textwidth,clip,trim={1cm 0cm 0.5cm 2cm}]{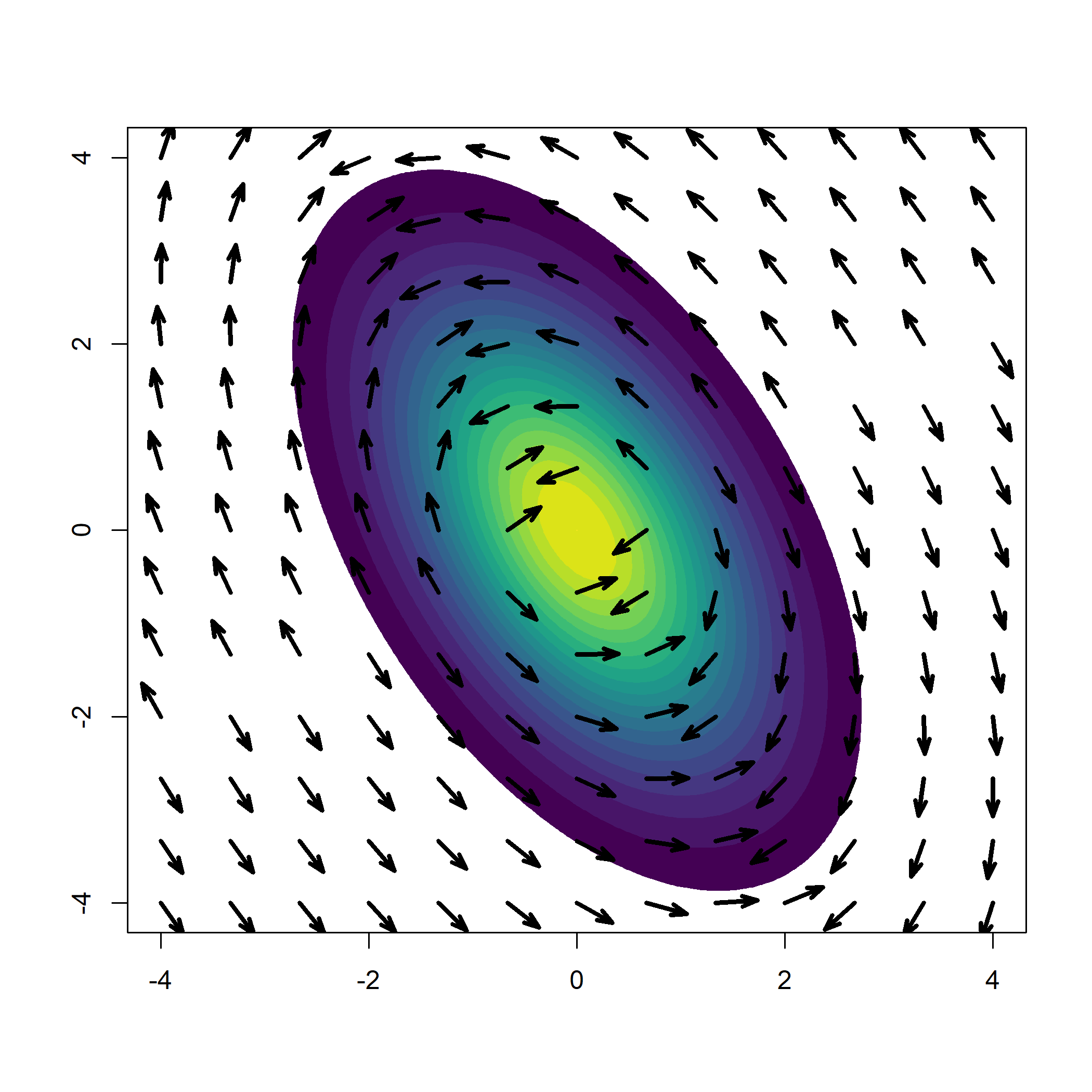}
	\caption{\small From left to right, representation of the eigenvector fields $\mathbf{u}_1,\mathbf{u}_2:\mathbb{R}^2\to\mathbb{R}^2$, and the gradient and projected gradient vector fields $\mathrm{D}f,\mathrm{D}_{1}f:\mathbb{R}^2\to\mathbb{R}^2$ for a Gaussian density $f$ on $\mathbb{R}^2$. Red plus (respectively, blue minus) indicate a positive (negative) eigenvalue $\lambda_j(\mathbf{x})$ associated with $\mathbf{u}_j(\mathbf{x})$, for $j=1,2$, with $\lambda_1(\mathbf{x})\geq \lambda_2(\mathbf{x})$ and $\mathbf{x}\in\mathbb{R}^2$. For visualization purposes, vector fields are unit-norm standardized and low-density regions are shown~in~white.}
	\label{fig:ridges-fields}
\end{figure}

Clearly, two cases imply that $\mathbf{x}\in\mathbb{R}^p$ satisfies $\mathrm{D}_{(p-1)} f(\mathbf{x})=\mathbf{0}$. The first is $\mathrm{D}f(\mathbf{x})=\mathbf{0}$. In that case, if $\lambda_{2}(\mathbf{x}),\ldots,\lambda_{p}(\mathbf{x})<0$, $\mathbf{x}$ is a maximum or a saddle point. The second is that $\mathrm{D}f(\mathbf{x})$ is perpendicular to $\mathbf{U}_{(p-1)}(\mathbf{x})(\mathbf{U}_{(p-1)}(\mathbf{x}))^\prime$, that is, the gradient is parallel to $\mathbf{u}_1(\mathbf{x})$. In this case, the directions of maximum ascent (gradient) and maximum signed curvature ($\mathbf{u}_1(\mathbf{x})$) coincide (see Figure \ref{fig:ridges-fields}), with ``signed'' emphasizing that the maximum is not in absolute value terms since $\lambda_1(\mathbf{x})\in\mathbb{R}$.

\subsection{Some useful properties}
\label{sec:propsrid}

Two useful properties of density ridges are given below. Their proofs are given in Appendix~\ref{Appendix:Proofs}.

\begin{proposition}[Ridge invariance to translations and rotations]\label{prp:1}
	Let $\mathbf{X}$ be a $p$-random vector with density 
	$f\in\mathcal{C}^2(\mathbb{R}^p)$, $p\geq 2$. Let $\mathbf{Y}=\boldsymbol{\mu}+\mathbf{R}\mathbf{X}$ represent a shifting and rotation family of $p$-random vectors spanned by $\mathbf{X}$, where $\boldsymbol{\mu}\in\mathbb{R}^p$ and $\mathbf{R}$ is a $p\times p$ orthogonal matrix. If $f(\cdot;\boldsymbol{\mu},\mathbf{R})$ stands for the density function of $\mathbf{Y}$ and $\mathbf{x}\in\mathbb{R}^p$, then
	\begin{align*}
		\mathbf{x}\in\mathcal{R}(f(\cdot; \boldsymbol{\mu},\mathbf{R}))\iff \mathbf{R}^\prime(\mathbf{x}-\boldsymbol{\mu}) \in \mathcal{R}(f).
	\end{align*}
\end{proposition}

This result yields two convenient handles to manipulate density ridges. First, it reduces the problem of computing density ridges to those that are centered at a certain origin. Second, it shows how to exploit the symmetries of $f$ to reduce the computational costs of evaluating $\mathcal{R}(f)$ (e.g., halve them if $f$ is reflective symmetric in $\mathbb{R}^2$).

\begin{proposition}[Ridges for elliptically-symmetric densities]\label{prp:2}
	Let $f\in\mathcal{C}^2(\mathbb{R}^p)$, $p\geq2$, be an elliptically-symmetric density of the form $f(\mathbf{x})=g((\mathbf{x}-\boldsymbol{\mu})^{\prime}\boldsymbol{\Sigma}^{-1}(\mathbf{x}-\boldsymbol{\mu}))$, with $\mathbf{x},\boldsymbol{\mu}\in\mathbb{R}^p$, $\boldsymbol{\Sigma}$ a $p\times p$ positive definite matrix, and $g\in\mathcal{C}^2(\mathbb{R}^+_0)$ a (strictly) decreasing function. Then, the subspace spanned by the first eigenvector $\mathbf{v}_1$ of $\boldsymbol{\Sigma}$ belongs to the density ridge of $f$:
	\begin{align}
		\{\boldsymbol{\mu}+c \mathbf{v}_1: c\in \mathbb{R}\} \subset \mathcal{R}(f).\label{prp:2:1}
	\end{align}
\end{proposition}

This proposition has several important consequences. First, it proves that $\mathcal{R}(f)$ is intimately related to the first principal component of PCA for elliptically-symmetric densities, since the subspace generated by the latter direction is included in it. A decreasing $g$ implies that $f$ has $\boldsymbol{\mu}$ as a unique mode. This is satisfied by many elliptically-symmetric distributions, with its prime representative being the normal distribution, a class of distributions in which PCA is particularly meaningful. Second, like the first principal component, the density ridge contains the center ($\boldsymbol{\mu}$) in elliptically-symmetric distributions. Third, the proposition shows that the density ridge is more than just the first principal direction subspace. This observation is key for defining the $\boldsymbol{\mu}$-connected density ridge for a mode or mean $\boldsymbol{\mu}$ of $f$ with a view to constructing a density-driven first principal component analog.

\begin{definition}[$\boldsymbol{\mu}$-connected density ridge]\label{def:muridge}
	Let $f\in\mathcal{C}^2(\mathbb{R}^p)$. The \emph{$\boldsymbol{\mu}$-connected density ridge} of $f$, $\mathcal{R}_{\boldsymbol{\mu}}(f)$, is defined as the connected component of $\mathcal{R}(f)$ that contains $\boldsymbol{\mu}\in\mathbb{R}^p$.
\end{definition}

We return to the $\boldsymbol{\mu}$-connected density ridges in Sections \ref{sec:ridgesbsvm} and \ref{sec:ridgesbwc}. Before, we describe in the following subsections two approaches to determining in practice the set of points that conform $\mathcal{R}(f)$ for $f \in \mathcal{C}^2(\mathbb{R}^p)$. Both approaches are exploited in Sections~\ref{sec:ridgesbsvm} and \ref{sec:ridgesbwc}.

\subsection{Integral curve approach}
\label{sec:integralcurve}

\cite{Genovese2014} showed that the vector field of the projected gradient defines a global flow. Hence, the trajectory defined by the projected gradient converges to the points where it is null. This fact allows characterizing the density ridge as an integral curve:
\begin{align*}
	\mathcal{R}(f)=\big\{\mathbf{x} \in \mathbb{R}^{p}: \lim_{t \rightarrow \infty} \phi_{\mathbf{x}_{0}}(t)=\mathbf{x},\ \mathbf{x}_{0} \in \mathbb{R}^{p},\ \lambda_{2}(\mathbf{x}),\ldots,\lambda_{p}(\mathbf{x})<0\big\},
\end{align*}
where $\phi_{\mathbf{x}_{0}}: \mathbb{R} \rightarrow \mathbb{R}^{p}$ is a flow curve in $\mathbb{R}^{p}$ that satisfies $(\mathrm{d}/\mathrm{d} t) \phi_{{\mathbf{x}}_{0}}(t)=\mathrm{D}_{(p-1)} f(\phi_{{\mathbf{x}}_{0}}(t))$, $t>0$, and $ \phi_{{\mathbf{x}}_{0}}(0)=\mathbf{x}_{0}$. This differential equation can be solved numerically using the Euler method until attaining a point where $\mathrm{D}_{(p-1)} f(\mathbf{x})\approx\mathbf{0}$. To move faster in low-density regions and slower in high-density regions, it is useful to consider the normalized projected gradient $\boldsymbol{\eta}_{(p-1)}(\mathbf{x}):=\mathrm{D}_{(p-1)} f(\mathbf{x})/f(\mathbf{x})$:
\begin{align}
	\mathbf{x}_{t+1}=\mathbf{x}_{t}+h \boldsymbol{\eta}_{(p-1)}(\mathbf{x}_{t}), \text{ for } t=0, 1, \ldots.\label{eq:euler}
\end{align}
The above scheme is started from an initial point $\mathbf{x}_0$ and is iterated using a step $h>0$ until a criterion for convergence is met after $N$ iterations. The resulting $\mathbf{x}_N$ (approximately) belongs to $\mathcal{R}(f)$. Then, it is possible to approximately determine $\mathcal{R}(f)$ by running \eqref{eq:euler} from a sufficiently dense grid of initial values $\mathbf{x}_0$.

\subsection{Implicit equation approach}
\label{sec:implicit}

When $p=2$, from Definition \ref{eq:ridge} it can be seen that $\mathcal{R}(f):=\{\mathbf{x} \in \mathbb{R}^{2}:\mathrm{D}_1f(\mathbf{x}) u_{2,1}(\mathbf{x})+\allowbreak\mathrm{D}_2f(\mathbf{x})u_{2,2}(\mathbf{x})=0,\ \lambda_{2}(\mathbf{x})<0\}$, with the subscript indicating the first/second component of the gradient and the first/second component of the \emph{second} eigenvector  $\mathbf{u}_2(\mathbf{x})$ of the Hessian at $\mathbf{x}$. Furthermore, still when $p=2$, the eigenvectors and eigenvalues of the Hessian admit a closed form. For example, \cite{Qiao2016} give:
\begin{align*}
	\mathbf{u}_{2}(u, v, w)&=\!\left(\!\begin{array}{c}
		2 u-2 w+2 v-2 \sqrt{(w-u)^{2}+4 v^{2}} \\
		w-u+4 v-\sqrt{(w-u)^{2}+4 v^{2}}
	\end{array}\right)\!,\\
	\lambda_{2}(u, v, w)&=\frac{u+w-\sqrt{(u-w)^{2}+4 v^{2}}}{2},
\end{align*}
with $u=(\partial^2 /\partial_1 ^2)f$, $v=(\partial^2 /\partial_1 \partial_2)f$, and $w=(\partial^2 /\partial_2 ^2)f$. This means that the implicit equation in terms of the derivatives of $f$ is given by
\begin{align}
	&\mathrm{D}_1\left(2 u-2 w+2 v-2 \sqrt{(w-u)^{2}+4 v^{2}}\right) +\mathrm{D}_2\left(w-u+4 v-\sqrt{(w-u)^{2}+4 v^{2}}\right)=0 \label{eq:implicit}
\end{align}
and that the eigenvalue condition reads as
\begin{align}
	\!\!\!\lambda_{2}(u, v, w)=\frac{u+w-\sqrt{(u-w)^{2}+4 v^{2}}}{2}<0. \label{eq:eigenvalue}
\end{align}
Therefore, it is possible to obtain $\mathcal{R}(f)$ for a bivariate density $f$ by solving Equation \eqref{eq:implicit}, restricted to Equation \eqref{eq:eigenvalue}, along a grid on one of its variables. This approach is much faster than that in Section \ref{sec:integralcurve}, yet it is restricted to $p=2$.

\section{Density ridges for bivariate sine von Mises}
\label{sec:ridgesbsvm}

\subsection{Bivariate sine von Mises}
\label{sec:bsvm}

Let $\Theta_{1}$ and $\Theta_{2}$ be two circular random variables. \cite{Singh2002}'s Bivariate Sine von Mises (BSvM) distribution has density given by
\begin{align}
	f_\mathrm{BSvM}(\theta_{1}, \theta_{2};\mu_1,\mu_2,\kappa_1,\kappa_2,\lambda)=T(\kappa_{1}, \kappa_{2}, \lambda)&\exp \big\{\kappa_{1} \cos (\theta_{1}-\mu_{1})+\kappa_{2} \cos (\theta_{2}-\mu_{2})\nonumber\\
	&\quad\quad+\lambda \sin (\theta_{1}-\mu_{1}) \sin (\theta_{2}-\mu_{2})\big\}\label{eq:bsvm}
\end{align}
for $\theta_{1}, \theta_{2}, \mu_{1}, \mu_{2}\in\mathbb{T}$, $\kappa_{1}, \kappa_{2} \geq 0$, $\lambda\in\mathbb{R}$, and a normalizing constant expressible as $T(\kappa_{1}, \kappa_{2}, \lambda)=4 \pi^{2} \sum_{m=0}^{\infty}\binom{2m}{m}\big(\lambda/2\big)^{2 m} \kappa_{1}^{-m} \mathcal{I}_{m}(\kappa_{1}) \kappa_{2}^{-m} \mathcal{I}_{m}(\kappa_{2})$, where $\mathcal{I}_{m}$ is the modified Bessel function of order $m$. The distribution is pointwise symmetric about $(\mu_1, \mu_2)$, with these two parameters representing the marginal circular means. The parameters $\kappa_1$ and $\kappa_2$ measure the marginal concentrations of the distribution about $\mu_1$ and $\mu_2$, respectively. The parameter $\lambda$ measures dependence: positive/negative values of $\lambda$ correspond to positive/negative correlation between $\Theta_{1}$ and $\Theta_{2}$. If $\lambda=0$, then $\Theta_{1}$ and $\Theta_{2}$ are independent, with each variable having a (univariate) von Mises distribution. Density \eqref{eq:bsvm} can be bimodal. A sufficient unimodality condition is given by $\kappa_1\kappa_2>\lambda^2$ \citep{Mardia2007}. In the bimodal case where $\boldsymbol{\mu}_1,\boldsymbol{\mu}_2\in\mathbb{T}^2$ are the two modes of \eqref{eq:bsvm}, due to the symmetry of the BSvM, it is easy to see that $\mathcal{R}_{\boldsymbol{\mu}_1}(f_\mathrm{BSvM}) = \mathcal{R}_{\boldsymbol{\mu}_2}(f_\mathrm{BSvM})$, thus ensuring that a connected component is well defined.

The BSvM has some properties that make it a candidate toroidal analog to the bivariate normal. Among these are the facts that it is also part of the exponential family and that it has asymptotic bivariate normal distributions under high concentrations \citep{Singh2002}. Furthermore, the marginals of the BSvM are not von Mises, but the conditionals are. Figure \ref{fig:eulerbsvm} shows different forms of the BSvM density. For simplicity, and without loss of generality due to Proposition \ref{prp:1}, we will work with $(\mu_1,\mu_2)=(0,0)$.

No closed expressions for the maximum likelihood estimators of the BSvM parameters exist \citep{Mardia2008}, but these can be computed numerically using moment-based estimators as starting values in the optimization of the log-likelihood. In particular, this enables testing the homogeneity hypothesis $\mathcal{H}_0:\kappa_1 = \kappa_2$ vs. $\mathcal{H}_1:\kappa_1 \neq \kappa_2$ using the Likelihood Ratio Test (LRT) that asymptotically rejects $\mathcal{H}_0$ at the significance level $\alpha$ whenever $-2(\hat{\ell}-\hat{\ell}_0)>\chi^2_{\alpha;1}$, where $\hat{\ell}$ and $\hat{\ell}_0$ are the maximum likelihoods of \eqref{eq:bsvm}, the latter under $\mathcal{H}_0$, and $\chi^2_{\alpha;1}$ is the $\alpha$-upper quantile of the chi-square distribution with one degree of freedom. This homogeneity LRT is convenient for distinguishing in practice the edge case $\kappa_1=\kappa_2$, which has a simple diagonal ridge associated. An analogous LRT can be constructed for testing $\mathcal{H}_0:\lambda = 0$ vs. $\mathcal{H}_1:\lambda \neq 0$, a null hypothesis of independence associated with horizontal/vertical ridges if $\kappa_1$ is smaller/larger than $\kappa_2$.

\subsection{Integral curve approach}
\label{sec:integralcurvebsvm}

The first alternative to compute the $\mathcal{R}(f_\mathrm{BSvM})$ of a $\mathrm{BSvM}(0,0,\kappa_1, \kappa_2, \lambda)$ uses the integral curve approach. With the Euler algorithm, an initial grid of points converges to the density ridge by following the trajectory defined by the projected gradient, as Figure \ref{fig:eulerbsvm} shows. It can be seen that the main part of the density ridge captures the main features of the distribution, shape, and correlation, while also being periodic. The figure also shows that there is a significant number of ``secondary'' ridges.

\subsection{Implicit ridge equation approach and connected ridges}
\label{sec:implicitbsvm}

The implicit equation offers a faster alternative to the computationally-demanding Euler algorithm. In the following, the normalizing constant of the BSvM and a common positive factor will be ignored, as these do not affect the direction of $\boldsymbol{\eta}_{(p-1)}$. It is easy to check that $ \mathrm{D}_1(f_\mathrm{BSvM})\propto-\kappa_1 \sin\theta_1 + \lambda \sin\theta_2 \cos\theta_1$, $ \mathrm{D}_2(f_\mathrm{BSvM})\propto-\kappa_2 \sin\theta_2 + \lambda \sin\theta_1 \cos\theta_2$, $u\propto \mathrm{D}_1^2-\kappa_1 \cos\theta_1 - \lambda \sin\theta_1 \sin\theta_2$, $v\propto \lambda \cos\theta_1 \cos\theta_2+\mathrm{D}_1\mathrm{D}_2$, and $w\propto \mathrm{D}_2^2-\kappa_2 \cos\theta_2 - \lambda \sin\theta_1 \sin\theta_2$. These expressions are readily usable in \eqref{eq:implicit}--\eqref{eq:eigenvalue}. Figure \ref{fig:implicit_eq_bsvm} shows the density ridges obtained with this method, which allow a clear identification of $\mathcal{R}_{\mathbf{0}}(f_\mathrm{BSvM})$.

However, in general, there is no explicit parametrization of $\mathcal{R}_{\boldsymbol{\mu}}(f_\mathrm{BSvM})$ with $\boldsymbol{\mu} \in \mathbb{T}^2$, so a method is needed to filter it from the full $\mathcal{R}(f_\mathrm{BSvM})$ obtained from the implicit equation. The symmetry of $f_{\mathrm{BSvM}}$ about $\boldsymbol{\mu}$ and Proposition \ref{prp:1} imply (see Figure \ref{fig:implicit_eq_bsvm} for graphical insight) that: (1) $\mathcal{R}_{\boldsymbol{\mu}}(f_\mathrm{BSvM}) \subset \{(\theta_1, \theta_2)\in\mathbb{T}^2: \mathrm{sign}(\theta_1/\theta_2) = \mathrm{sign}(\lambda)\}$; and (2) the computation of $\mathcal{R}_{\boldsymbol{\mu}}(f_\mathrm{BSvM})$ reduces to the first or second quadrant, depending on $\mathrm{sign}(\lambda)$. Finally, since $\mathcal{R}_{\boldsymbol{\mu}}(f_\mathrm{BSvM})$ passes through $\boldsymbol{\mu}$ by definition, a connected component can be obtained by iteratively adding the sufficiently close ridge points to the last point added to the set, starting from $\boldsymbol{\mu}$.

\begin{figure}[htpb!]
	\centering
	\includegraphics[width=0.25\textwidth,clip,trim={0cm 0cm 0.75cm 2cm}]{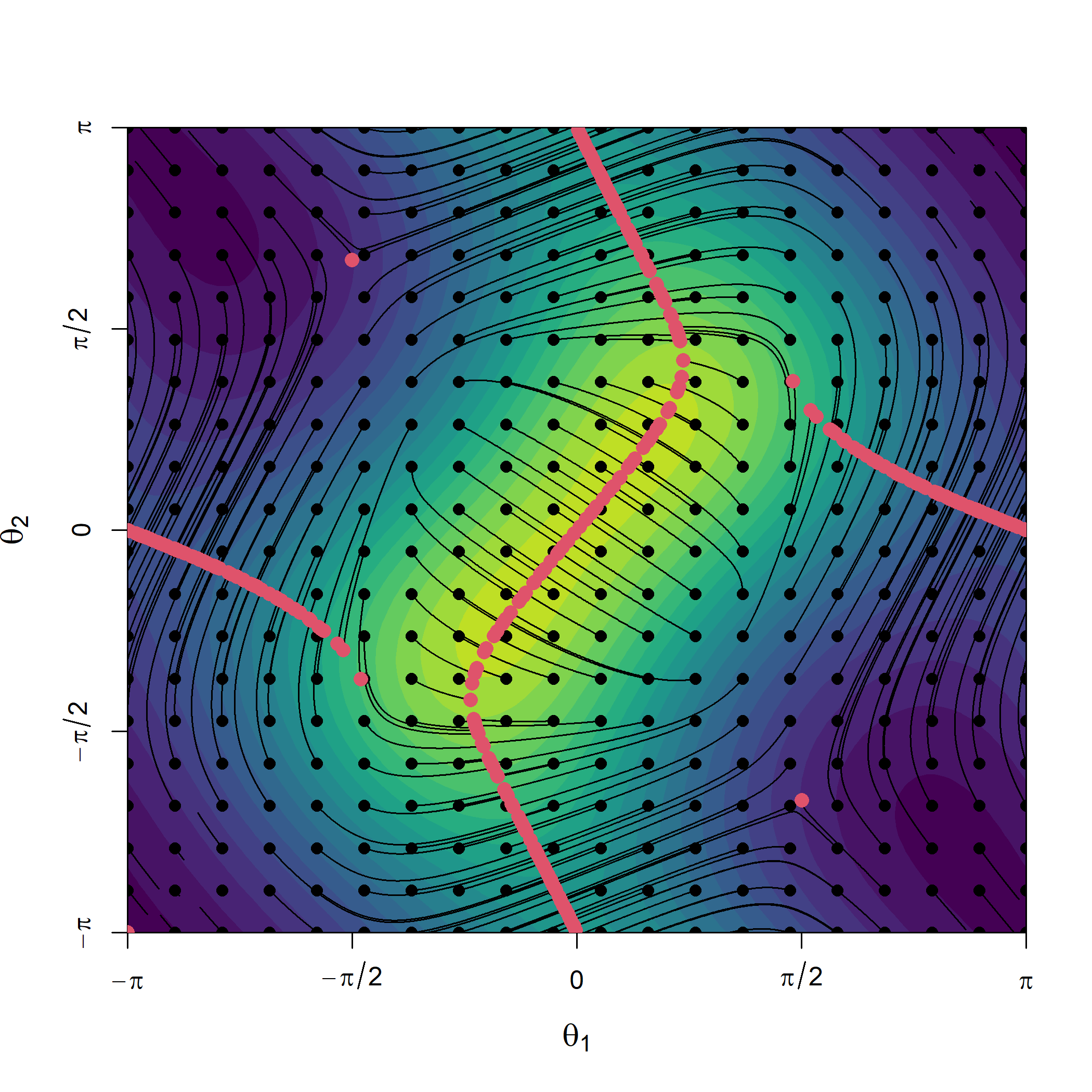}%
	\includegraphics[width=0.25\textwidth,clip,trim={0cm 0cm 0.75cm 2cm}]{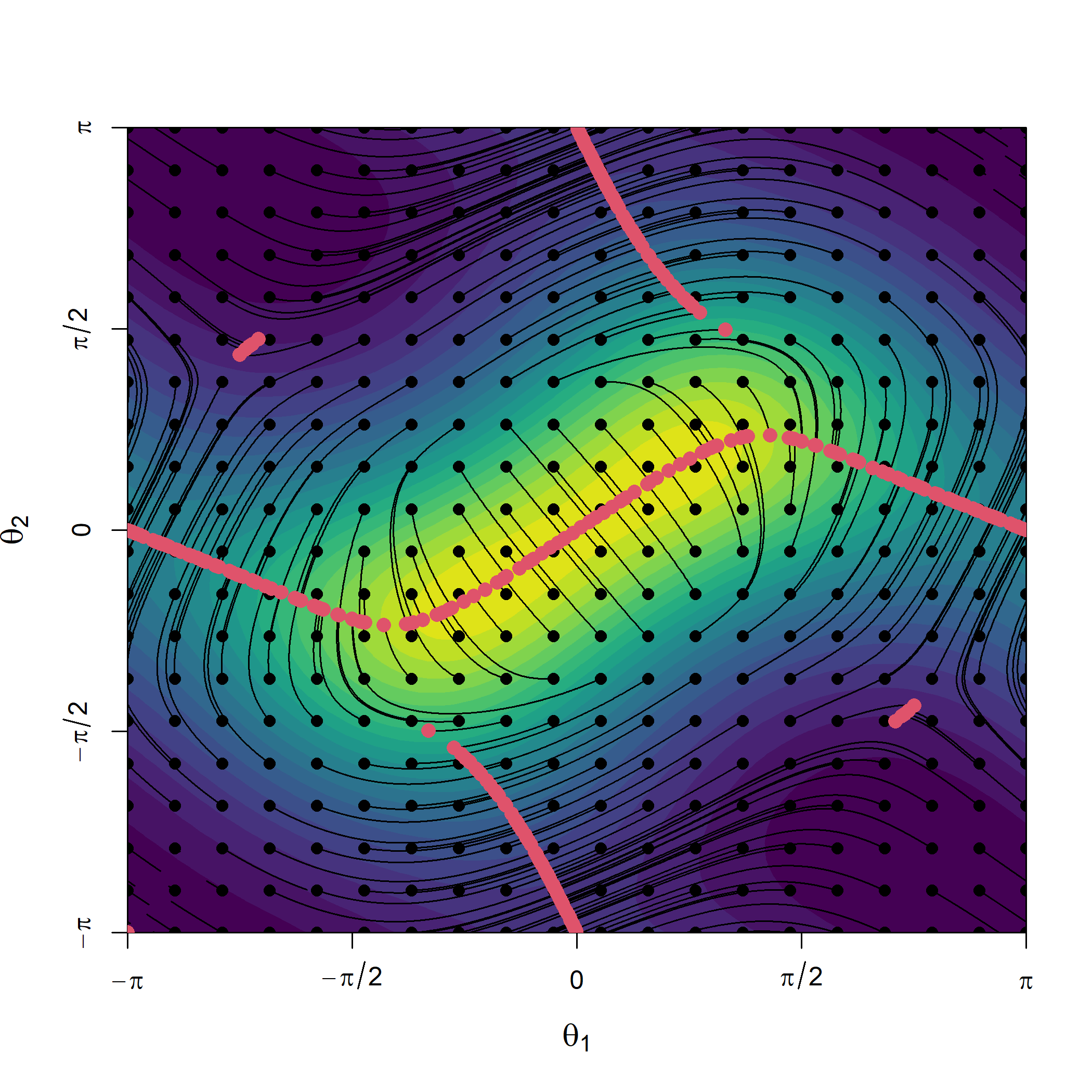}%
	\includegraphics[width=0.25\textwidth,clip,trim={0cm 0cm 0.75cm 2cm}]{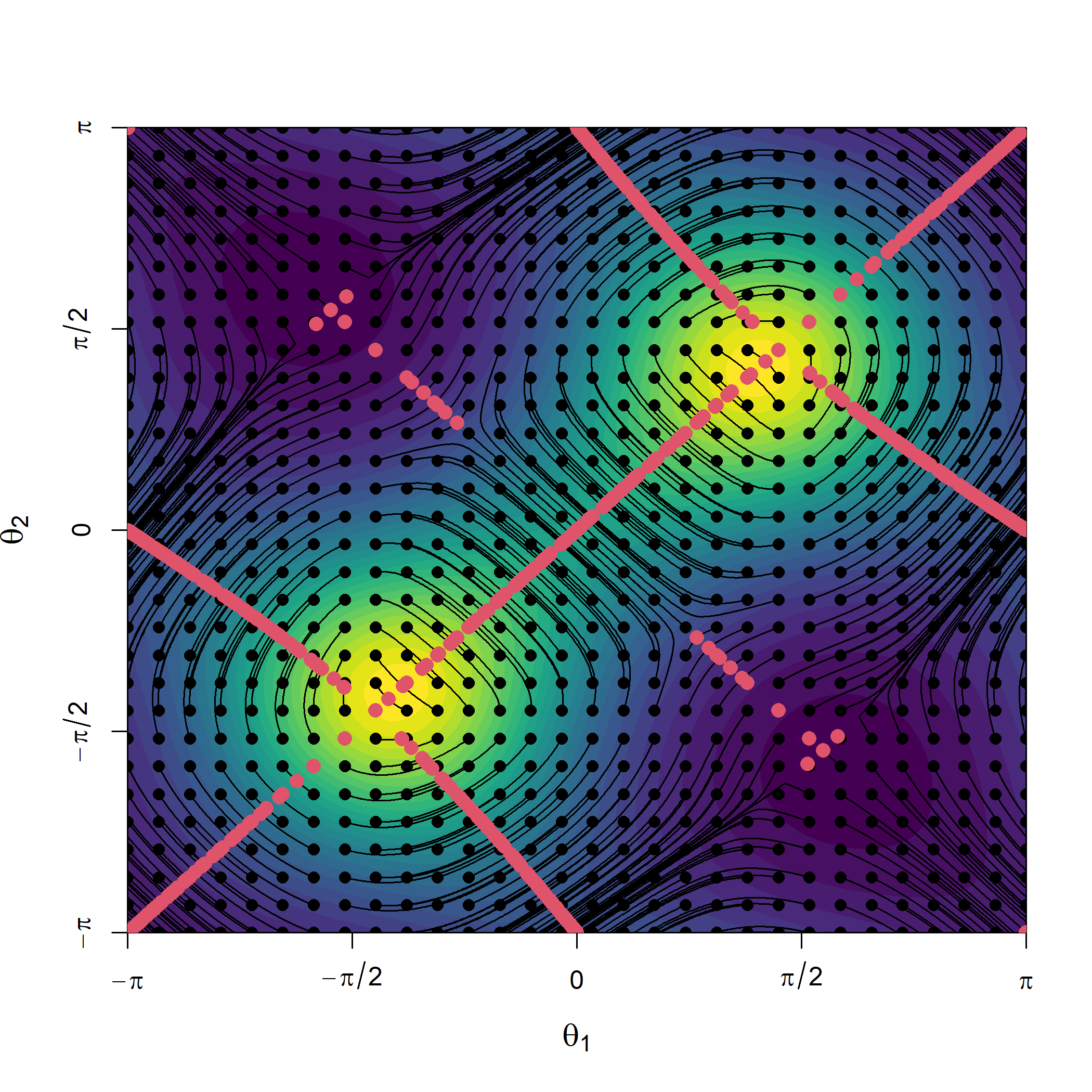}%
	\includegraphics[width=0.25\textwidth,clip,trim={0cm 0cm 0.75cm 2cm}]{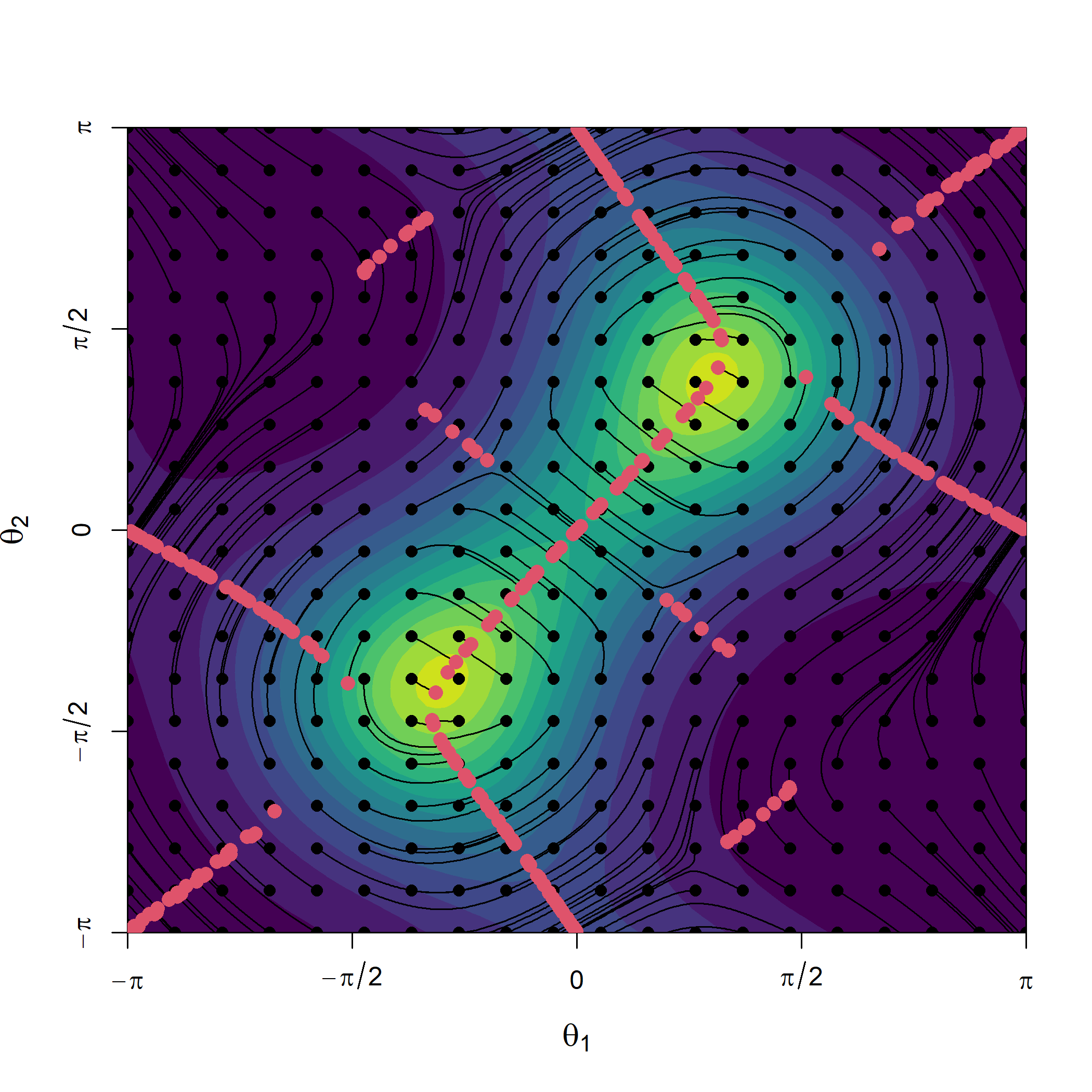}
	\caption{\small Catalog of the Euler-approximated ridges of a zero-centered BSvM with parameters $(\kappa_1, \kappa_2, \lambda) \in\{(0.3, 0.15, 0.25), (0.3, 0.6, 0.5), (0.3, 0.3, 1.0), (1.0, 0.5, 1.5)\}$. The initial set of points, together with their trajectories, are shown in black. The red points represent the final points of the algorithm, which describe $\mathcal{R}(f_\mathrm{BSvM})$. The background shows the density contour of the BSvM. All the contourplots share the same color scale.}
	\label{fig:eulerbsvm}
\end{figure}

\vspace*{-0.75cm}

\begin{figure}[htpb!]
	\centering
	\includegraphics[width=0.25\textwidth,clip,trim={0cm 0cm 0.75cm 2cm}]{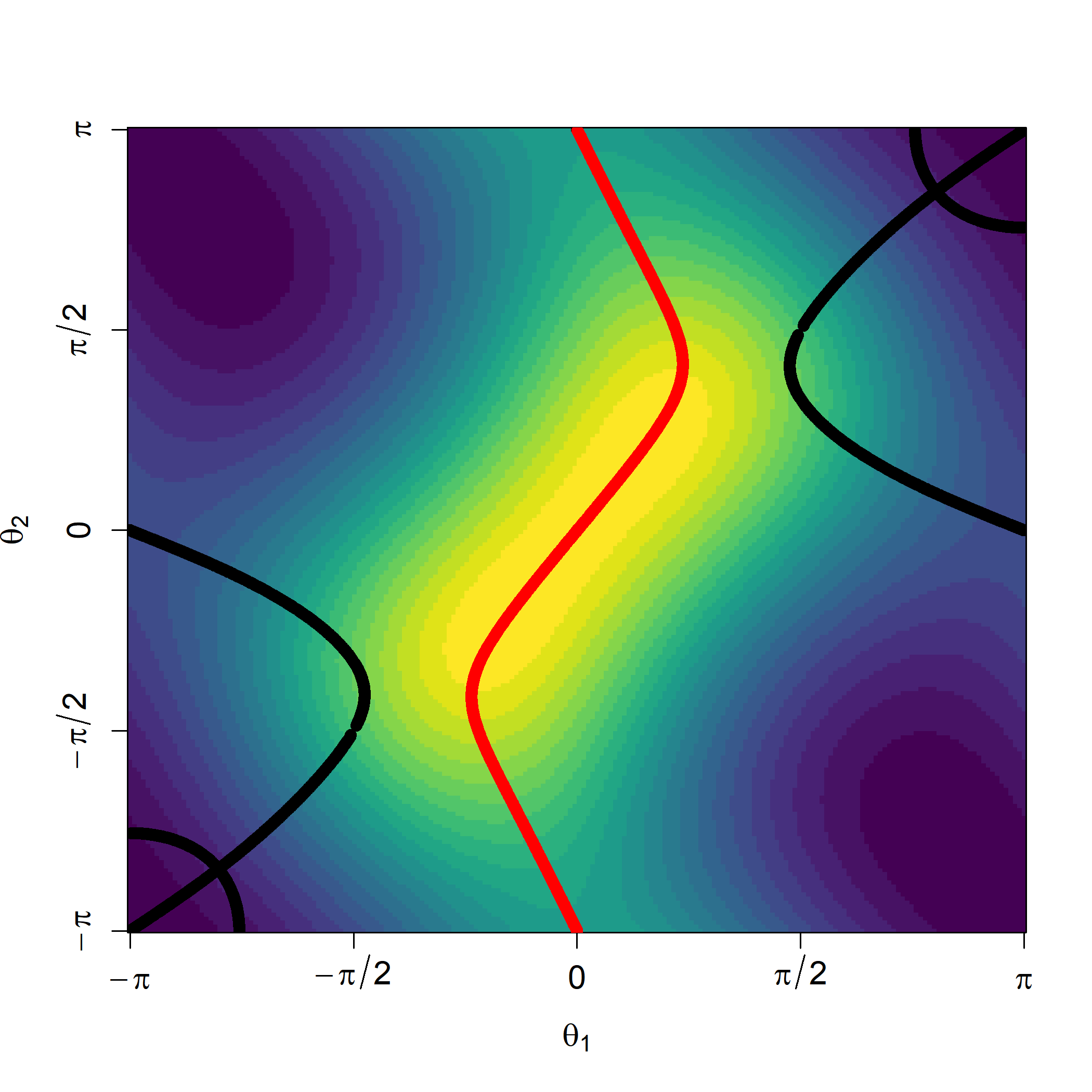}%
	\includegraphics[width=0.25\textwidth,clip,trim={0cm 0cm 0.75cm 2cm}]{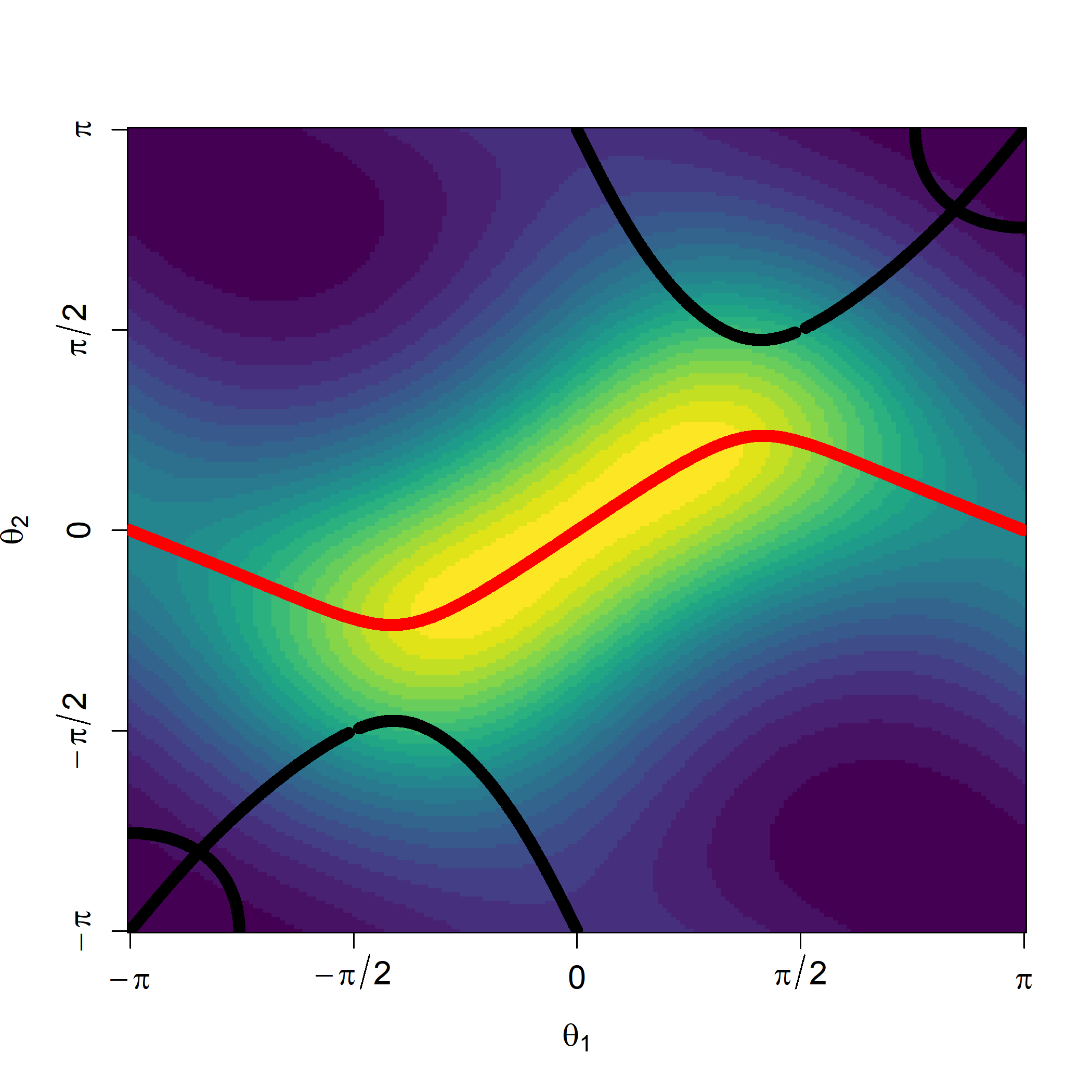}%
	\includegraphics[width=0.25\textwidth,clip,trim={0cm 0cm 0.75cm 2cm}]{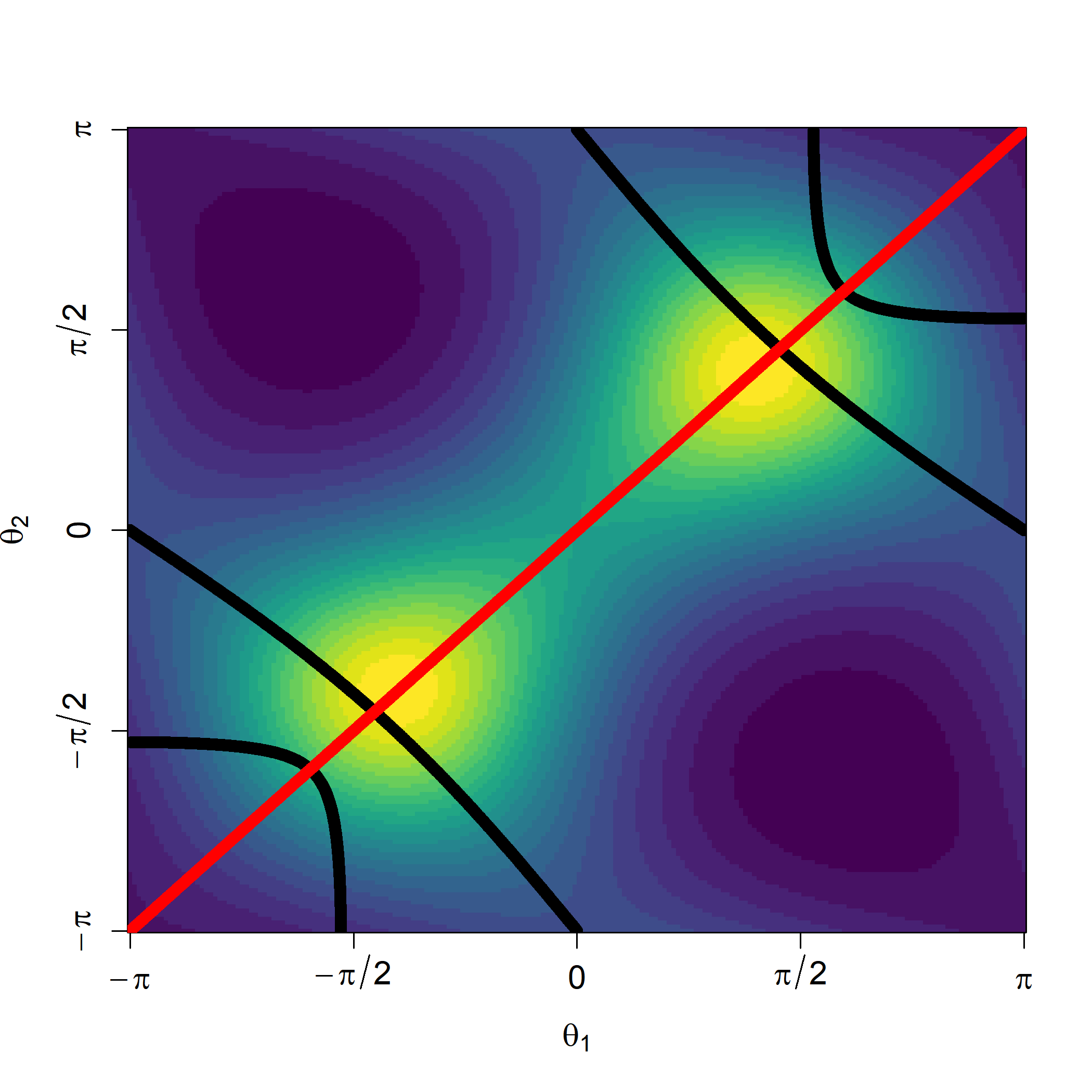}%
	\includegraphics[width=0.25\textwidth,clip,trim={0cm 0cm 0.75cm 2cm}]{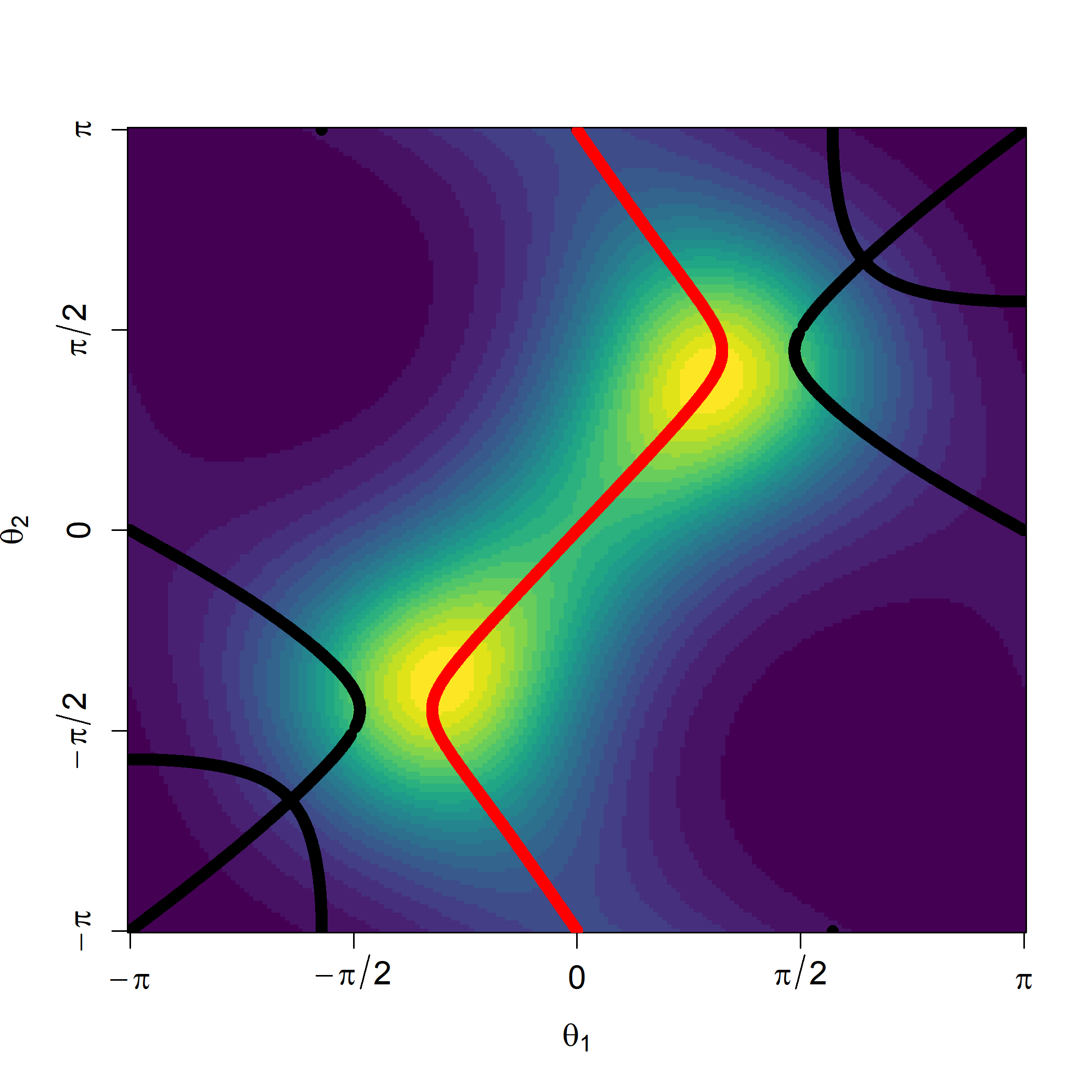}
	\caption{\small Catalog of the implicit-equation-approximated ridges $\mathcal{R}(f_\mathrm{BSvM})$ (black) and $\mathcal{R}_{\mathbf{0}}(f_\mathrm{BSvM})$ (red) of a zero-centered BSvM for the same parameter values as in Figure \ref{fig:eulerbsvm}. The background shows the density contour of the BSvM. All the contourplots share the same color scale.}
	\label{fig:implicit_eq_bsvm}
\end{figure}

The following result presents limit cases for which the density ridge $\mathbf{\mathcal{R}_0}(f_{\mathrm{BSvM}})$ is explicit.

\begin{proposition}\label{prp:limit_cases_vm}
	The $\mathbf{0}$-connected density ridge $\mathbf{\mathcal{R}_0}(f_{\mathrm{BSvM}})$ of a $\mathrm{BSvM}(0,0,\kappa_1, \kappa_2, \lambda)$ admits the following representations:
	\begin{enumerate}[label=(\textit{\roman*}),ref=(\textit{\roman*})]
		
		\item When $\kappa_2>\kappa_1=0$, and $\lambda=0$, $\mathcal{R}_\mathbf{0}(f_\mathrm{BSvM})=\{(\theta_1,\theta_2)\in\mathbb{T}^2: \theta_2=0\}$. \label{prp:limit_cases_vm_2}
		
		\item When $\kappa_1=\kappa_2\geq0$ and $\lambda\in\mathbb{R}$, $\mathcal{R}_\mathbf{0}(f_\mathrm{BSvM})\supset\{(\theta_1,\theta_2)\in\mathbb{T}^2: \cos\theta_1< \vert\lambda\vert/\kappa_1,\ \theta_2=\mathrm{sign}(\lambda)\theta_1 \}$. \label{prp:limit_cases_vm_1}
	\end{enumerate}
\end{proposition}

\begin{remark}\label{rem:1}
	Note that \ref{prp:limit_cases_vm_1} does not characterize $\mathcal{R}_{\mathbf{0}}$, unlike \ref{prp:limit_cases_vm_2}. Although in view of the third leftmost panel in Figure \ref{fig:implicit_eq_bsvm} it seems possible to achieve this characterization with an additional restriction in Definition \ref{def:muridge}, we do not pursue this further due to its limited practical interest: given that $\mathcal{R}_\mathbf{0}(f_\mathrm{BSvM})\not\supset\{(\theta_1,\theta_2)\in\mathbb{T}^2: \theta_2=\mathrm{sign}(\lambda)\theta_1\}$, $\mathcal{R}_\mathbf{0}(f_\mathrm{BSvM})$ will not always wrap periodically at $-\pi \equiv \pi$. To solve this edge-case issue in practice, and in coherence with the other cases, we simply extend $\mathcal{R}_\mathbf{0}(f_\mathrm{BSvM})$ according to the diagonal.
\end{remark}

\section{Density ridges for bivariate wrapped Cauchy}
\label{sec:ridgesbwc}

\subsection{Bivariate wrapped Cauchy}
\label{sec:bwc}

The density of a random vector $(\Theta_{1}, \Theta_{2})$ following the Bivariate Wrapped Cauchy (BWC) distribution proposed by \cite{Kato2015a}~is
\begin{align*}
	f_\mathrm{BWC}(\theta_{1}, \theta_{2};\mu_1,\mu_2,\xi_1,\xi_2,\rho)=&\;c\{c_{0}-c_{1} \cos (\theta_{1}-\mu_{1})-c_{2} \cos (\theta_{2}-\mu_{2})\\
	&\quad-c_{3} \cos (\theta_{1}-\mu_{1}) \cos (\theta_{2}-\mu_{2})-c_{4} \sin (\theta_{1}-\mu_{1}) \sin (\theta_{2}-\mu_{2})\}^{-1},
\end{align*}
for $\theta_{1}, \theta_{2}, \mu_1, \mu_2\in\mathbb{T}$, $0 \leq \xi_{1}, \xi_{2}<1$, and $-1<\rho<1$. The several constants are given as follows: $c=(1-\rho^{2})(1-\xi_{1}^{2})(1-\xi_{2}^{2}) /(4 \pi^{2})$, $c_{0}=(1+\rho^{2})(1+\xi_{1}^{2})(1+\xi_{2}^{2})-8\vert\rho \vert\xi_{1} \xi_{2}$, $c_{1}= 2(1+\rho^{2}) \xi_{1}(1+\xi_{2}^{2})-4\vert\rho\vert(1+\xi_{1}^{2}) \xi_{2}$, $c_{2}=2(1+\rho^{2})(1+\xi_{1}^{2}) \xi_{2}-4\vert\rho\vert \xi_{1}(1+\xi_{2}^{2})$, $c_{3}=-4(1+\rho^{2}) \xi_{1} \xi_ 2+2\vert\rho\vert(1+\xi_{1}^{2})(1+\xi_{2}^{2})$, and $c_{4}=2 \rho(1-\xi_{1}^{2})(1-\xi_{2}^{2})$. Analogously to the BSvM, $\mu_1$ and $\mu_2$ represent the marginal circular means of the density. The parameters $\xi_{1}$ and $\xi_{2}$ regulate the concentrations of the marginal distributions, that of $\Theta_j$ being circular uniform when $\xi_j=0$ ($j=1,2$). When $\xi_{1}, \xi_{2}>0,$ the density is unimodal and pointwise symmetric about $(\mu_{1}, \mu_{2})$. As $\xi_j \to 1$, the marginal distribution of $\Theta_j$ tends to a point mass at $\mu_j$. The association between $\Theta_{1}$ and $\Theta_{2}$ is controlled by the parameter $\rho$, $\rho=0$ corresponding to independence. Positive/negative values of $\rho$ correspond to positive/negative correlation between $\Theta_{1}$ and $\Theta_{2}$. Figure \ref{fig:bwc_ridges} shows different forms of the BWC density. This distribution is always unimodal \citep{Kato2015a}, thus ensuring that $\mathcal{R}_{\boldsymbol\mu}$ is well-defined.

The BWC is closed under marginalization and conditioning, meaning that the resulting marginals and conditional densities belong to the wrapped Cauchy family. This distinctive closedness property is shared with the bivariate normal distribution, whose marginals and conditionals are also normal. The functional form of marginals is immediate from the BWC's construction as a particular \cite{Wehrly1980} model, but the conditionals are not (despite being wrapped Cauchys).

As in the BSvM case, there are no closed expressions for the maximum likelihood estimators for the BWC parameters \citep{Kato2015}, so numerical routines are needed. Analogously to the BSvM case, the LRTs for $\mathcal{H}_0:\xi_1=\xi_2$ vs. $\mathcal{H}_1:\xi_1\neq\xi_2$ (homogeneity) and $\mathcal{H}_0:\rho=0$ vs. $\mathcal{H}_1:\rho\neq0$ (independence) are also relevant to distinguish diagonal and horizontal/vertical ridges in a practical and principled manner.

The integral curve approach for the BWC is completely analogous to the BSvM case.

\subsection{Implicit ridge equation approach and connected ridges}
\label{sec:implicitbwc}

It is simple to check that the derivatives of the BWC density, excluding a common positive factor, are $ \mathrm{D}_1(f_\mathrm{BWC})\propto -c_1 \sin\theta_1 - c_3 \sin\theta_1 \cos\theta_2 + c_4 \sin\theta_2 \cos\theta_1$, $ \mathrm{D}_2(f_\mathrm{BWC})\propto -c_2 \sin\theta_2 - c_3 \sin\theta_2 \cos\theta_1 + c_4 \sin\theta_1 \cos\theta_2$, $u\propto2 \mathrm{D}_1^2f^* -c_1 \cos\theta_1 - c_3 \cos\theta_1 \cos\theta_2 -c_4 \sin\theta_1\sin \theta_2$, $v\propto2\mathrm{D}_1\mathrm{D}_2f^*+c_3\sin\theta_2\sin\theta_1+c_4\cos\theta_1\cos\theta_2$, and $w\propto2 \mathrm{D}_2^2f^* - c_2 \cos\theta_2 - c_3 \cos\theta_2 \cos\theta_1 -c_4 \sin\theta_2\sin \theta_1$, where $f^*=f/c$. Figure \ref{fig:bwc_ridges} shows the density ridges obtained with the implicit equation approach. This figure illustrates that, differently to the sinusoidal shapes of $\mathcal{R}_{\boldsymbol{0}}(f_\mathrm{BSvM})$, the shapes of $\mathcal{R}_{\boldsymbol{0}}(f_\mathrm{BWC})$ involve ridges that connect $(0,0)$ with $(\pm\pi,\pm\pi)$ for unequal marginal concentrations.

The following proposition shows limit cases for which $\mathcal{R}_{\mathbf{0}}(f_\mathrm{BWC})$ is explicit.

\begin{figure*}[htpb!]
	\centering
	\includegraphics[width=0.25\textwidth,clip,trim={0cm 0cm 0.75cm 2cm}]{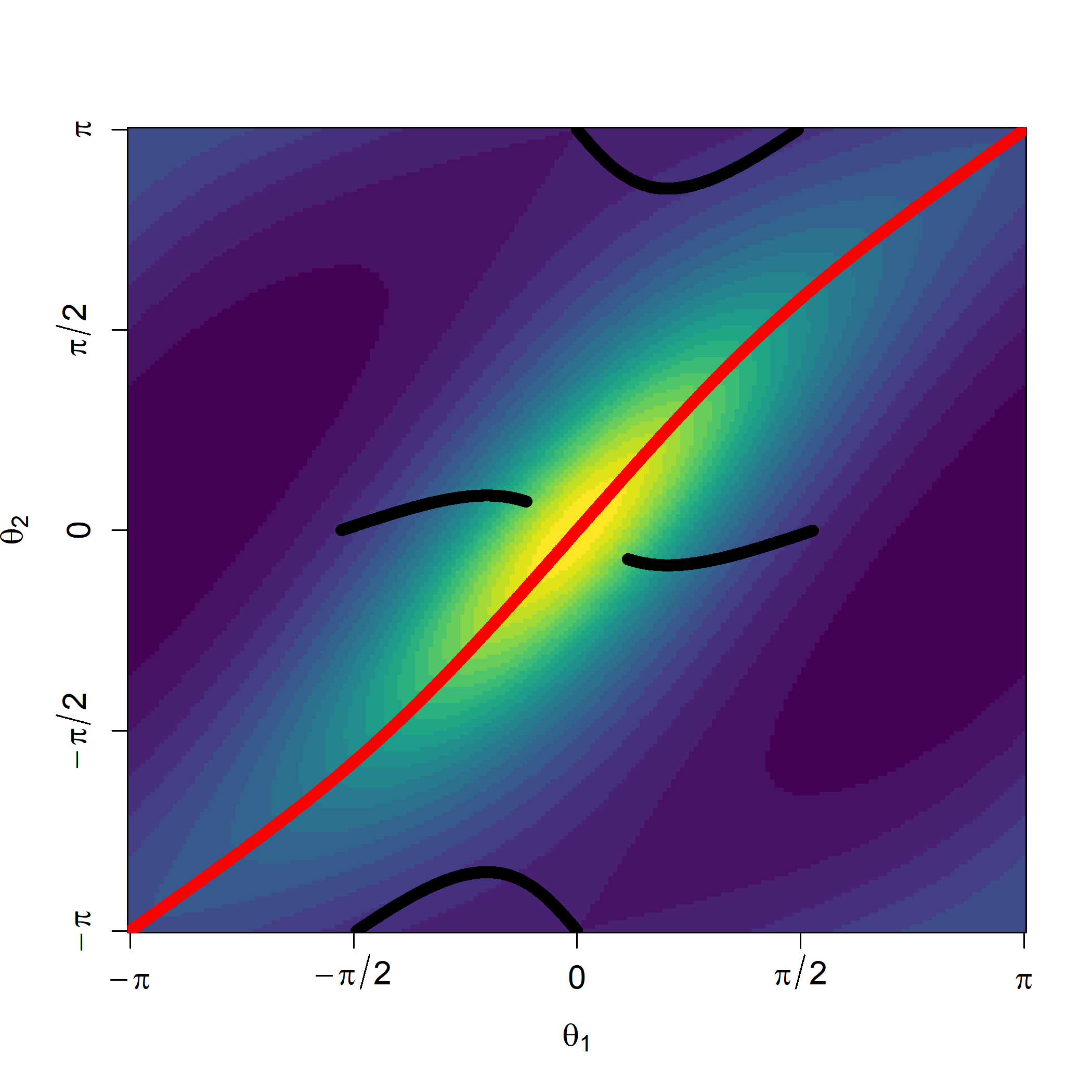}%
	\includegraphics[width=0.25\textwidth,clip,trim={0cm 0cm 0.75cm 2cm}]{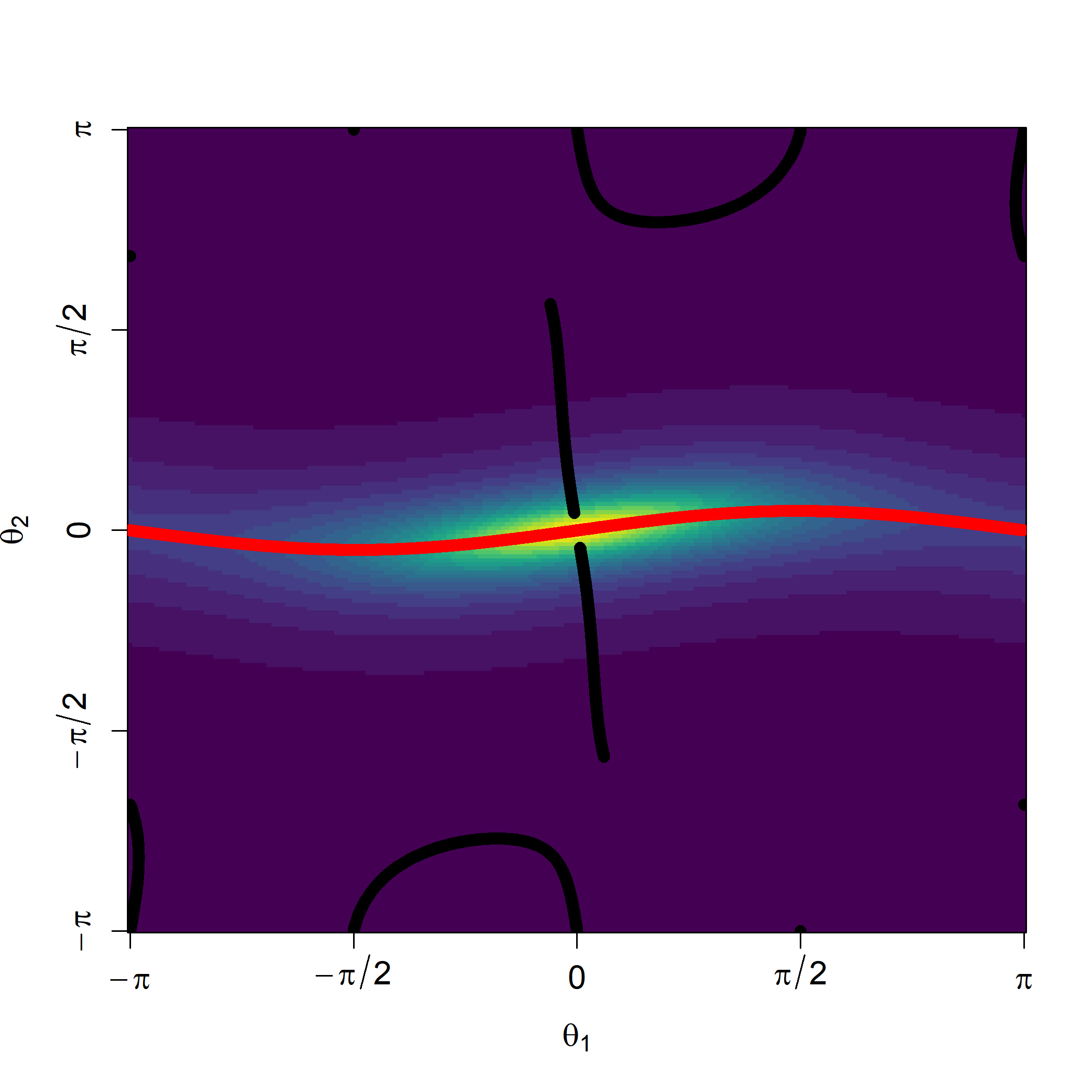}%
	\includegraphics[width=0.25\textwidth,clip,trim={0cm 0cm 0.75cm 2cm}]{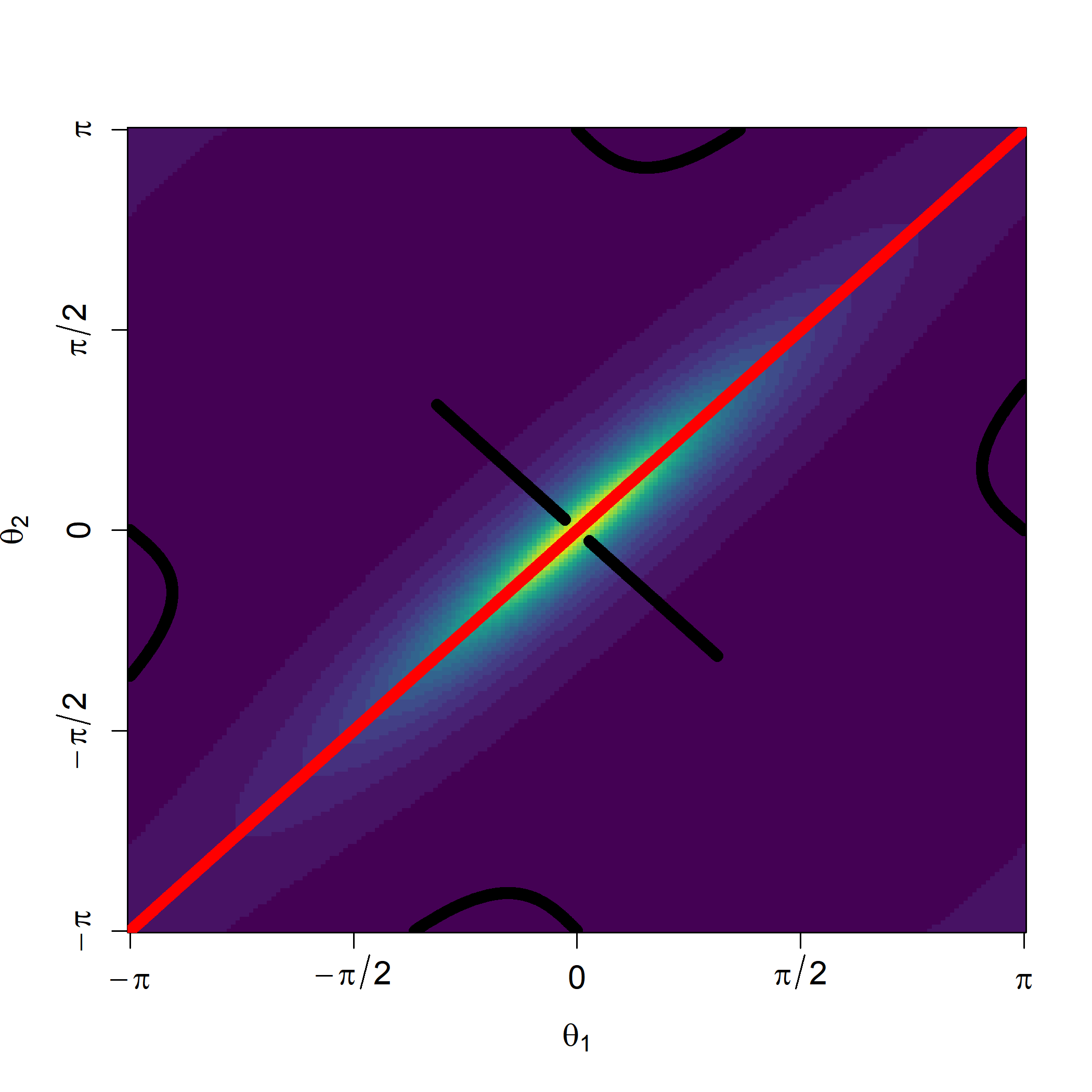}%
	\includegraphics[width=0.25\textwidth,clip,trim={0cm 0cm 0.75cm 2cm}]{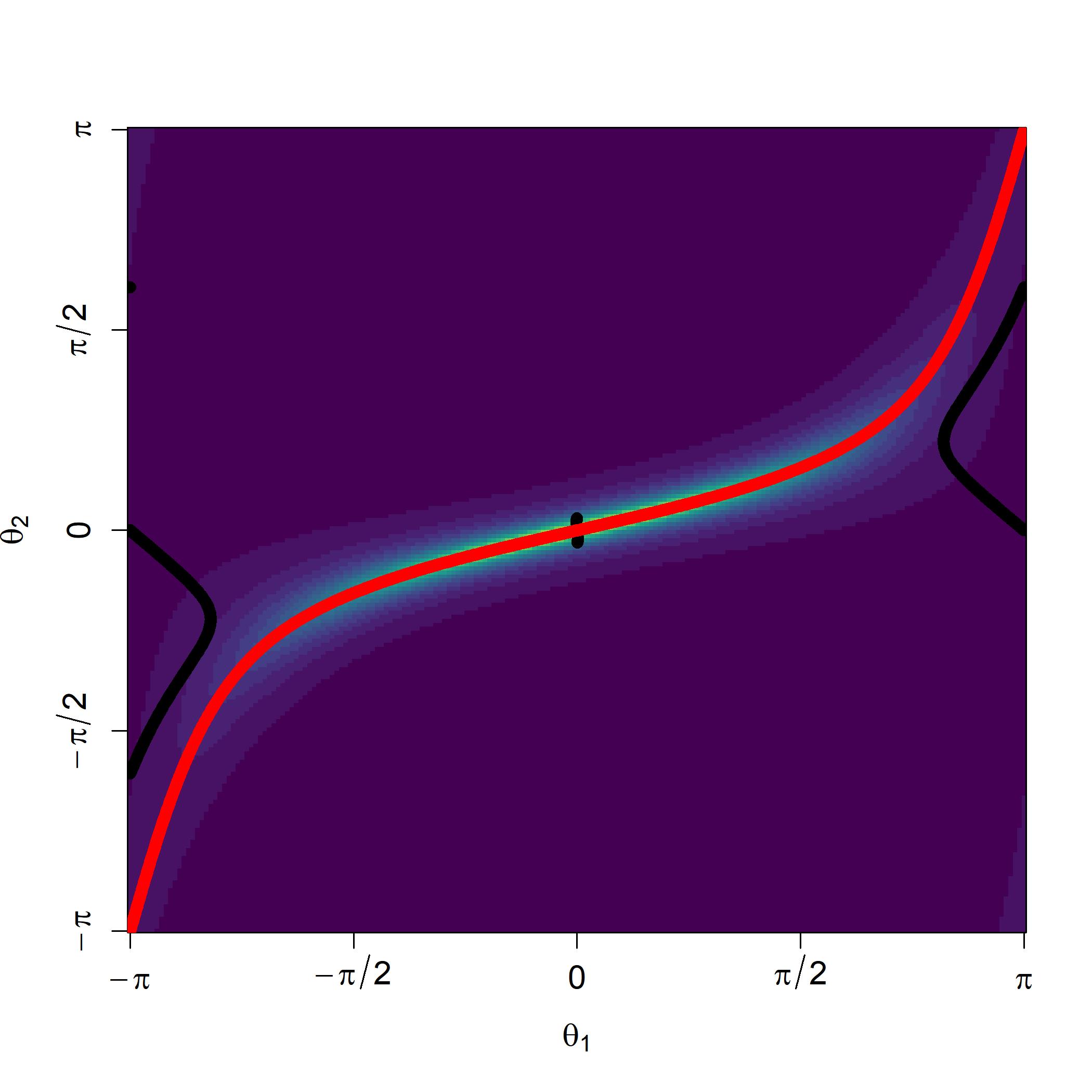}
	\caption{\small Catalog of the implicit-equation-approximated ridges $\mathcal{R}(f_\mathrm{BWC})$ (black) and $\mathcal{R}_{\mathbf{0}}(f_\mathrm{BWC})$ (red) with parameter values $(\xi_1, \xi_2, \rho) \in\{(0.15, 0.075, 0.25),(0.2, 0.7, 0.2),(0.3, 0.3, 0.6),(0.025, 0.6, 0.7) \}$. The background shows the density contour of the BWC. All the contourplots share the same common color scale.}
	\label{fig:bwc_ridges}
\end{figure*}

\begin{proposition}\label{prop:limit_cases_WC}
	The $\mathbf{0}$-connected density ridge $\mathcal{R}_{\mathbf{0}}(f_\mathrm{BWC})$ of a $\mathrm{BWC}(0,0,\xi_1,\xi_2,\rho)$ admits the following explicit representations:
	\begin{enumerate}[label=(\textit{\roman*}),ref=(\textit{\roman*})]
		\item When $\xi_1=\rho=0$ and $\xi_2\in(0,1)$, $\mathcal{R}_{\mathbf{0}}(f_\mathrm{BWC})=\{(\theta_1,\theta_2)\in\mathbb{T}^2: \theta_2=0\}$.\label{prp:limit_cases_wc_1}
		\item When $0\leq \xi_1=\xi_2<1$ and $\rho\in(-1,1)$, $\mathcal{R}_{\mathbf{0}}(f_\mathrm{BWC})\supset\{(\theta_1,\theta_2)\in\mathbb{T}^2: \theta_2=\mathrm{sign}(\rho)\theta_1\}$.\!\!\!\!\label{prp:limit_cases_wc_2}
	\end{enumerate}
\end{proposition}

\begin{remark}\label{rem:2}
	Analogously to Remark \ref{rem:1}, in practical applications we consider $\mathcal{R}_\mathbf{0}(f_\mathrm{BWC}):=\{(\theta_1,\theta_2)\in\mathbb{T}^2: \theta_2=\mathrm{sign}(\rho)\theta_1\}$ for case \ref{prp:limit_cases_wc_2}.
\end{remark}

\section{Toroidal ridge PCA}
\label{sec:trpca}

\subsection{Ridge parametrization}
\label{sec:ridgepar}

Due to the difficulty in explicitly solving the implicit equation \eqref{eq:implicit}, we have not found any simple parametric form for the curve defined by $\mathcal{R}_{\boldsymbol{\mu}}(f)$ beyond limit cases. This poses an important hindrance to the tractability of the distance operation along $\mathcal{R}_{\boldsymbol{\mu}}(f)$ and the projection operation that maps a point on $\mathbb{T}^2$ to its closest point on $\mathcal{R}_{\boldsymbol{\mu}}(f)$, both being core mechanisms for the definition of the forthcoming scores. To prevent this issue from draining the performance of TR-PCA, we consider a Fourier-type parametrization of $\mathcal{R}_{\boldsymbol{\mu}}(f)$ given by
\begin{align}
	\!\!\!\!r_{f,\boldsymbol{\mu},j}(\phi)&:=\mathrm{cmod}\big(\mu_l+ \rho_m(\phi-\mu_j)-\rho_m(0)\big),\label{eq:rf}\\
	\rho_m(\theta)&:=\mathrm{atan2}(\mathcal{S}_{m} (\theta), \mathcal{C}_{m}(\theta)),\nonumber
\end{align}
where $\phi,\theta\in\mathbb{T}$, $\mathrm{cmod}(\cdot):=(\cdot+\pi)\mod2\pi-\pi$, $j,l\in\{1,2\}$, $l\neq j$, and $\mathcal{C}_{m}(\theta) := a_{0}/2 + \sum_{k=1}^m a_k \cos(k\theta)$ and $\mathcal{S}_{m}(\theta) := \sum_{k=1}^m b_k \sin(k\theta)$ are truncated cosine/sine Fourier series with $m\geq1$. The coefficients $\{(a_k,b_k)\}_{k=0}^m$ (with $b_0=0$) are $a_k:=\frac{1}{\pi}\int_{\mathbb{T}} \cos(R_{f,\mathbf{0},j}(\theta)) \allowbreak\cos(k\theta)\,\mathrm{d}\theta$ and $b_k:=\frac{1}{\pi}\int_{\mathbb{T}} \sin(R_{f,\mathbf{0},j}(\theta)) \sin(k\theta)\,\mathrm{d}\theta$, where $\{(\phi,R_{f,\boldsymbol{\mu},1}(\phi)):\phi\in\mathbb{T}\}=\mathcal{R}_{\boldsymbol{\mu}}(f)$ or $\{(R_{f,\boldsymbol{\mu},2}(\phi),\phi):\allowbreak\phi\in\mathbb{T}\}=\mathcal{R}_{\boldsymbol{\mu}}(f)$, depending on which is the index coordinate $j$ (see the next paragraph). The consideration of only the cosine/sine parts in \eqref{eq:rf} is justified by the pointwise symmetry of $\mathcal{R}_{\mathbf{0}}(f_{\mathrm{BSvM}})$ and $\mathcal{R}_{\mathbf{0}}(f_{\mathrm{BWC}})$, which renders null Fourier cosine/sine coefficients for sine/cosine components. In practice, $\{(a_k,b_k)\}_{k=0}^m$ is approximated by Gaussian quadrature using a grid of points in $\mathcal{R}_{\boldsymbol{\mu}}(f)$ determined using the implicit equation method. In numerical experiments, the truncation of the Fourier series \eqref{eq:rf} to $m=15$ was found to be a sensible choice for a large number of parameter specifications of the BSvM and BWC densities. With $m=15$, the maximum distance between the implicit-equation computation of $\mathcal{R}_{\boldsymbol{\mu}}(f)$ and the Fourier-parametrized approximation was found to be smaller than $10^{-2}$. Therefore, we consider $m=15$ to Fourier-parametrize $\mathcal{R}_{\boldsymbol{\mu}}(f_\mathrm{BSvM})$ and $\mathcal{R}_{\boldsymbol{\mu}}(f_\mathrm{BWC})$.

To obtain a one-to-one parametrization for the BSvM and BWC densities, in practice $r_{f,\boldsymbol{\mu},j}$ is indexed along the variable with the smallest concentration (e.g., $\theta_2$ and $\theta_1$ for the first and second leftmost panels in Figure 2, respectively), which is straightforward to identify. For the sake of notational simplicity, henceforth we assume without loss of generality that $j=1$.

To define the first scores, it is fundamental to parametrize the curve $r_{f,\boldsymbol{\mu},1}$ through its arc length from $\phi=\mu_1$ to $\phi=\mu_1+t$, $t\in[0,2\pi)$:
$L(t) := \int_{\mu_1}^{\mu_1+t} \sqrt{1 + (r_{\boldsymbol{\mu},f,1}'(\phi))^2}\,\mathrm{d}\phi$. The length of the curve of $r_{f,\boldsymbol{\mu},1}$ is $R:=\lim_{\theta\to(2\pi)^{-}}L(\theta)$. The arc-length parametrized curve in $\mathbb{T}^2$ is thus
\begin{align}
	\!\!\!\!\!s\in[0, R)\mapsto \mathbf{r}_{f,\boldsymbol{\mu}}(s):=\big((\mathrm{id},r_{f,\boldsymbol{\mu},1})'\circ L^{-1}\big)(s),\!\!\label{eq:rf2}
\end{align}
with $\mathrm{id}$ denoting the identity function. Two further tweaks are required on $\mathbf{r}_{f,\boldsymbol{\mu}}$ to: (1) scale the parametrization to $s\in[0,2\pi)$; (2) center the parametrization to $s\in[-\pi,\pi)$. The first step sets the range of the scores to $2\pi$, just as the original data in $\mathbb{T}$, while the second step is crucial for assigning signs. The definition below summarizes the construction.

\begin{definition}[Scaled-centered arc-length ridge parametrization] \label{def:rf3}
	The \emph{scaled-centered arc-length parametrization} of $\mathcal{R}_{\boldsymbol{\mu}}(f)$ based on \eqref{eq:rf}--\eqref{eq:rf2} is
	$\alpha\in\mathbb{T}\mapsto \tilde{\mathbf{r}}_{f,\boldsymbol{\mu}}(\alpha)$ with
	\begin{align}
		\tilde{\mathbf{r}}_{f,\boldsymbol{\mu}}(\alpha):=\mathbf{r}_{f,\boldsymbol{\mu}}\left([(R/(2\pi))\alpha] \mod R\right).\label{eq:rf3}
	\end{align}
\end{definition}

The proposition below collects two immediate properties of $\tilde{\mathbf{r}}_{f,\boldsymbol{\mu}}$ for the BSvM and BWC densities.

\begin{proposition} \label{prop:rf}
	For $f_\mathrm{BSvM}$ and $f_\mathrm{BWC}$, with $\boldsymbol{\mu}$ being its location parameter, the curve in \eqref{eq:rf3} is such that: 
	\begin{enumerate}[label=(\textit{\roman*}),ref=(\textit{\roman*})]
		\item $\tilde{\mathbf{r}}_{f,\boldsymbol{\mu}}(0)=\boldsymbol{\mu}$ and $\tilde{\mathbf{r}}_{f,\boldsymbol{\mu}}(\pm\pi)\in\{\mathrm{cmod}(\mu_1\pm\pi,\mu_2)',\mathrm{cmod}(\mu_1\pm\pi,\mu_2\pm\pi)'\}$; \label{prop:rf:1}
		\item the signed distance between two points $\boldsymbol{\phi}_i:=\tilde{\mathbf{r}}_{f,\boldsymbol{\mu}}(\alpha_i)$, $\alpha_i\in\mathbb{T}$, $i=1,2$, along the curve $\mathbf{r}_{f,\boldsymbol{\mu}}$ is $(R/(2\pi))\,\mathrm{cmod}(\alpha_1-\alpha_2)$. \label{prop:rf:2}
	\end{enumerate}
\end{proposition}

\subsection{Scores}
\label{sec:scores}

Projections on $\mathcal{R}_{\boldsymbol{\mu}}(f)$ are defined via the fast handle $\tilde{\mathbf{r}}_{f,\boldsymbol{\mu}}$. They involve the toroidal distance $d_{\mathbb{T}^p}(\boldsymbol{\theta},\boldsymbol{\phi}):=\sqrt{\sum_{j=1}^p d_{\mathbb{T}}(\theta_j,\phi_j)^2}$, $p\geq1$, where $d_{\mathbb{T}}(\theta_j,\phi_j)\allowbreak:=\min\{\vert\theta_j-\phi_j\vert, 2\pi-\vert\theta_j-\phi_j\vert\}$, for $\theta_j,\phi_j\in\mathbb{T}$.

\begin{definition}[Ridge projections] \label{def:proj}
	For $\boldsymbol{\theta} \in \mathbb{T}^2$, its \emph{projection} to the Fourier-parametrized $\mathcal{R}_{\boldsymbol{\mu}}(f)$, $\boldsymbol{\mu}\in\mathbb{T}^2$, is
	\begin{align*}
		\mathrm{proj}_{f,\boldsymbol{\mu}}(\boldsymbol{\theta})&:=\tilde{\mathbf{r}}_{f,\boldsymbol{\mu}}\left(\alpha_{f,\boldsymbol{\mu}}(\boldsymbol{\theta})\right),
	\end{align*}
	where the \emph{projection argument} is
	\begin{align*}
		\alpha_{f,\boldsymbol{\mu}}(\boldsymbol{\theta})&:= \arg\min_{\alpha\in\mathbb{T}} d_{\mathbb{T}^2}\left(\boldsymbol{\theta},\tilde{\mathbf{r}}_{f,\boldsymbol{\mu}}(\alpha)\right).
	\end{align*}
\end{definition}

The TR-PCA scores for an arbitrary point $\boldsymbol{\theta}\in\mathbb{T}^2$ are defined as an analogy to ordinary PCA. The first score is the signed distance along $\tilde{\mathbf{r}}_{f,\boldsymbol{\mu}}$ and between $\mathrm{proj}_{f,\boldsymbol{\mu}}(\boldsymbol{\theta})$ and $\boldsymbol{\mu}$. The second score is the signed distance between $\boldsymbol{\theta}$ and $\mathrm{proj}_{f,\boldsymbol{\mu}}(\boldsymbol{\theta})$. The sign is set according to the relative position of the tangent and normal vectors, both at the projection point.

\begin{definition}[Scores in TR-PCA]\label{def:scores}
	For $\boldsymbol{\theta} \in \mathbb{T}^2$, its first TR-PCA score, $s_1(\boldsymbol{\theta})$, is 
	\begin{align*}
		s_{1}(\boldsymbol{\theta}):=\alpha_{f,\boldsymbol{\mu}}(\boldsymbol{\theta}).
	\end{align*}
	
	The second TR-PCA score, $s_2(\boldsymbol{\theta})$, is
	\begin{align}
		\vert s_{2}(\boldsymbol{\theta})\vert&:= (\pi/ m_2)\,d_{\mathbb{T}^2}(\boldsymbol{\theta},\mathrm{proj}_{f,\boldsymbol{\mu}}(\boldsymbol{\theta})),\label{eq:score2}\\
		\mathrm{sign}(s_2(\boldsymbol{\theta}))&:=\mathrm{sign}\big(\angle(\boldsymbol{t}(\boldsymbol{\theta}))-\angle(\boldsymbol{n}(\boldsymbol{\theta}))\big),\nonumber
	\end{align}
	where $\boldsymbol{t}(\boldsymbol{\theta}):= \tilde{\textbf{r}}'_{f,\boldsymbol{\mu}}(\alpha_{f,\boldsymbol{\mu}}(\boldsymbol{\theta}))$, $\boldsymbol{n}(\boldsymbol{\theta}):=\mathrm{cmod}(\mathrm{proj}_{f,\boldsymbol{\mu}}(\boldsymbol{\theta})-\boldsymbol{\theta})$, and $\angle(\mathbf{v}):=\mathrm{atan2}(v_2,v_1)$.
\end{definition}

The first factor in \eqref{eq:score2}, where $m_2:=\max_{\boldsymbol{\theta}\in\mathbb{T}^2} \vert s_2(\boldsymbol{\theta})\vert$, is included to homogenize the scales of both scores, so that $(s_{1}(\boldsymbol{\theta}),s_{2}(\boldsymbol{\theta}))'\in\mathbb{T}^2$.

\subsection{Proportion of variance explained}
\label{sec:prop}

Given a sample $\boldsymbol{\Theta}_1,\ldots,\boldsymbol{\Theta}_n$ in $\mathbb{T}^p$, $n,p\geq1$, its \emph{Fréchet mean} (or intrinsic mean) is defined as
\begin{align*}
	\hat{\boldsymbol{\mu}}_{\mathrm{F}}&:=
	\arg\min_{\boldsymbol{\phi}\in\mathbb{T}^p} \sum_{i=1}^n d_{\mathbb{T}^p}(\boldsymbol{\phi},\boldsymbol{\Theta}_i)^2 \in\mathbb{T}^p.
\end{align*}
The \emph{Fréchet variance} of the sample is the minimum of the previous objective function:
\begin{align*}
	\widehat{\mathrm{var}}_{\mathrm{F}}&:=
	\sum_{i=1}^n d_{\mathbb{T}^p}(\hat{\boldsymbol{\mu}}_\mathrm{F},\boldsymbol{\Theta}_i)^2.
\end{align*}

Due to the product structure of $\mathbb{T}^p$, $\hat{\boldsymbol{\mu}}_{\mathrm{F}}=\big(\hat{\mu}_{\mathrm{F}}^{(1)},\ldots,\hat{\mu}_{\mathrm{F}}^{(p)}\big)'$ and $\widehat{\mathrm{var}}_{\mathrm{F}}=\sum_{j=1}^p \widehat{\mathrm{var}}_{\mathrm{F}}^{(j)}$, where the superscript denotes a marginal Fréchet mean/\allowbreak~variance. For $p=2$, this variance decomposition facilitates the definition of the Proportion of Variance Explained (PVE) in TR-PCA as
\begin{align}
	\text{PVE} := \frac{\widehat{\mathrm{svar}}_{\mathrm{F}}^{(1)}}{\widehat{\mathrm{svar}}_{\mathrm{F}}^{(1)}+\widehat{\mathrm{svar}}_{\mathrm{F}}^{(2)}}, \label{eq:pve}
\end{align}
where $\widehat{\mathrm{svar}}_{\mathrm{F}}^{(j)}$ stands for the Fréchet variance of the $j$th scores $\{s_j(\boldsymbol{\Theta}_i)\}_{i=1}^n$, $j=1,2$.

\subsection{Complete TR-PCA procedure}
\label{sec:fulltrpca}

The complete TR-PCA procedure for a given toroidal sample involves all the concepts introduced so far. It is divided into three main stages.

\begin{algo}[TR-PCA]
	\label{def:TR-PCA}
	Given a sample $\{\boldsymbol{\Theta}_i\}_{i=1}^n$ in $\mathbb{T}^2$, \emph{TR-PCA} proceeds as follows:
	\begin{enumerate}[label=(\textit{\roman*}),ref=(\textit{\roman*})]
		\item Modeling.
		\begin{enumerate}
			\item Fit the BSvM and/or BWC models (Sections \ref{sec:bsvm} and \ref{sec:bwc}) with maximum likelihood estimation. If both models are fit, select the one with the smallest Bayesian Information Criterion (BIC).
			\item Inspect edge cases using LRTs (Sections \ref{sec:bsvm} and \ref{sec:bwc}) at $5\%$ significance level.
			\begin{enumerate}[label=(\textit{b.\roman*}),ref=(\textit{b.\roman*})]
				\item Test diagonal vs. non-diagonal ridges with the independence LRT. \label{b1}
				\item Test straight vs. non-straight ridges with the homogeneity LRT. \label{b2}
			\end{enumerate}
			\item If any of the LRTs does not reject, refit the model with maximum likelihood estimation restricted to the decisions of \ref{b1}--\ref{b2}.
			\item Retrieve $\hat{f}$, $\hat{\boldsymbol{\mu}}$ (location parameter), and $\hat{j}$ (index of the lowest concentration).
		\end{enumerate}
		\item Ridge computation.
		\begin{enumerate}
			\item Determine a grid of $\mathcal{R}_{\mathbf{0}}(\hat{f})$ with the implicit equation approach (Sections \ref{sec:implicitbsvm} and \ref{sec:implicitbwc}).
			\item Construct $r_{\hat{f},\hat{\boldsymbol{\theta}},\hat{j}}$ in \eqref{eq:rf} with $\{(\hat{a}_k,\hat{b}_k)\}_{k=0}^m$ computed with Gauss--Legendre quadrature on the previous grid.
			\item Obtain the arc-length parametrized ridge curve $\tilde{\mathbf{r}}_{\hat{f},\hat{\boldsymbol{\theta}}}$ from Definition \ref{def:rf3}.
		\end{enumerate}
		\item Scores and PVE computation.
		\begin{enumerate}
			\item Compute $\{(s_1(\boldsymbol{\Theta}_i), s_2(\boldsymbol{\Theta}_i))'\}_{i=1}^n$ using Definition \ref{def:scores}.
			\item Obtain the PVE using \eqref{eq:pve}.
		\end{enumerate}
	\end{enumerate}
\end{algo}

\subsection{Illustrative examples}
\label{sec:examps}

We now compare the performance of TR-PCA versus an adaptation of PCA to the torus, angular PCA (aPCA) \citep{Riccardi2009}, which is arguably the most readily implementable alternative to PCA in the torus. aPCA centers the data using the circular mean prior to the application of standard PCA. Hence, periodicity is not preserved, which introduces artifacts on the scores when dealing with non-concentrated data. TR-PCA follows the steps defined in Algorithm \ref{def:TR-PCA}. Figure \ref{fig:TR-PCAperformance} shows this comparison for four samples simulated from Bivariate Wrapped Normal (BWN) and BWC distributions. The BWN distribution is that of a bivariate random vector distributed as $\mathcal{N}_2(\boldsymbol{\mu},\boldsymbol{\Sigma})$, for a mean $\boldsymbol{\mu}\in\mathbb{T}^2$ and a covariance matrix $\boldsymbol{\Sigma}$, after each vector component is transformed by applying $\mathrm{cmod}(\cdot)$.

\begin{figure}[h!]
	\centering
	\includegraphics[width=0.25\textwidth,clip,trim={0cm 0cm 0.75cm 2cm}]{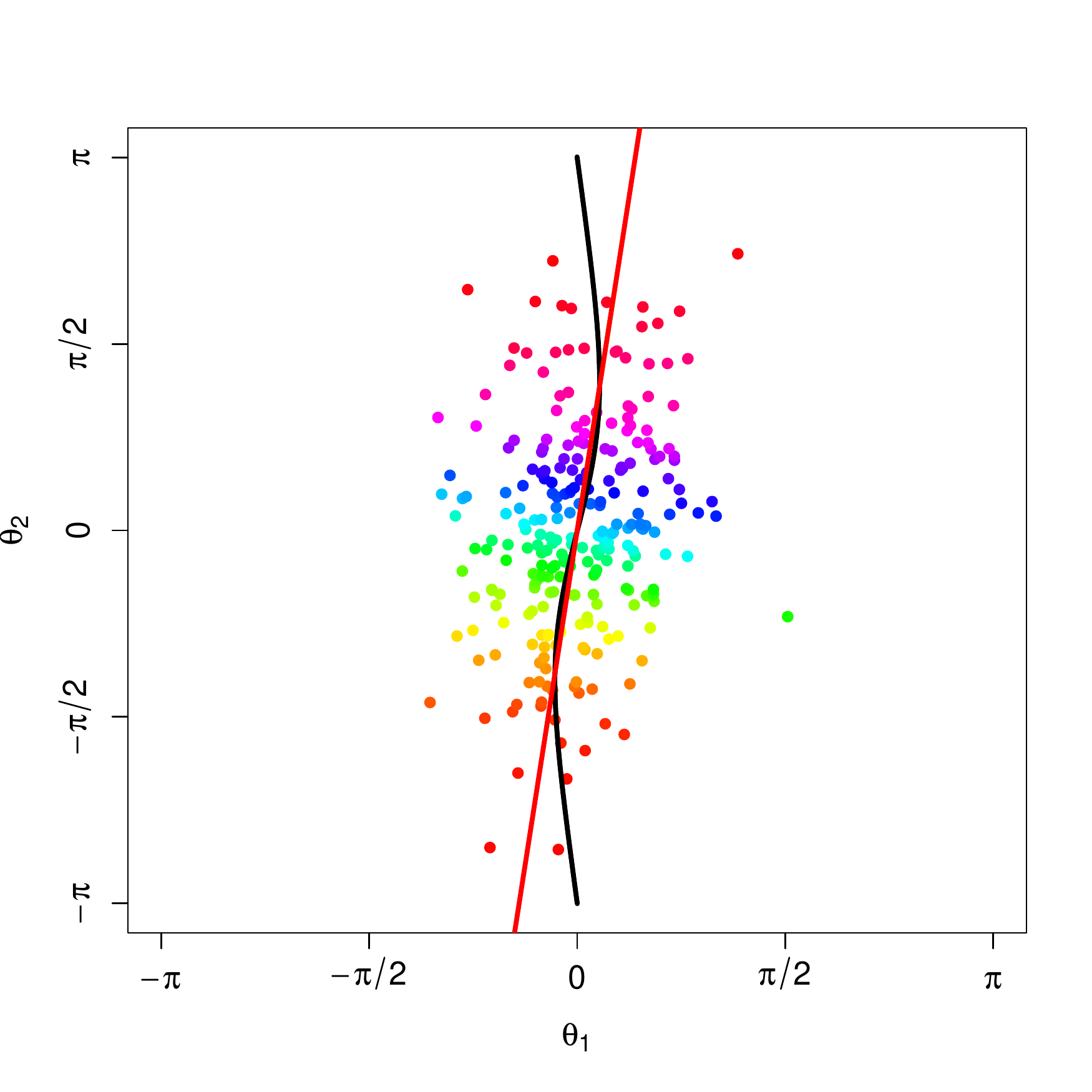}%
	\includegraphics[width=0.25\textwidth,clip,trim={0cm 0cm 0.75cm 2cm}]{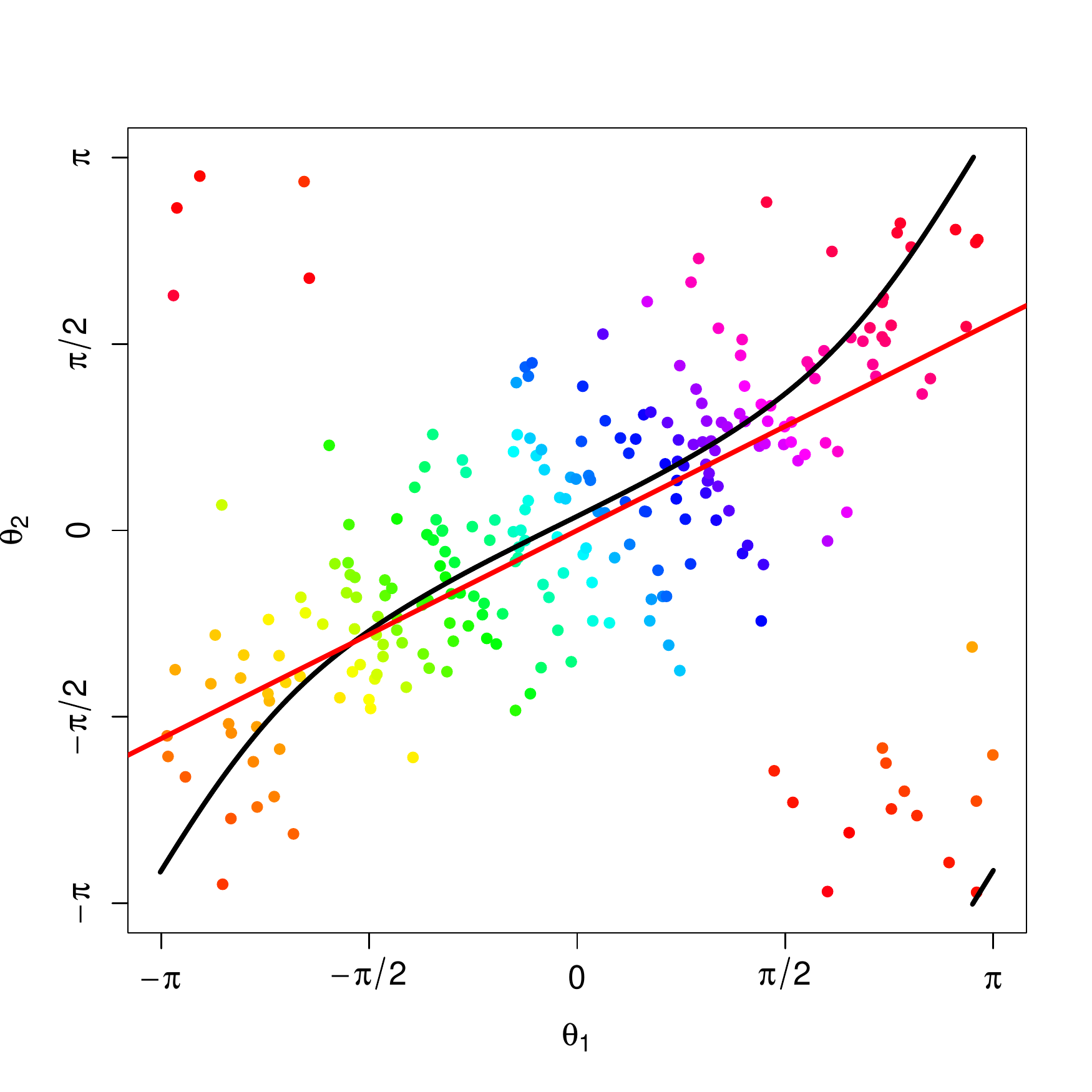}%
	\includegraphics[width=0.25\textwidth,clip,trim={0cm 0cm 0.75cm 2cm}]{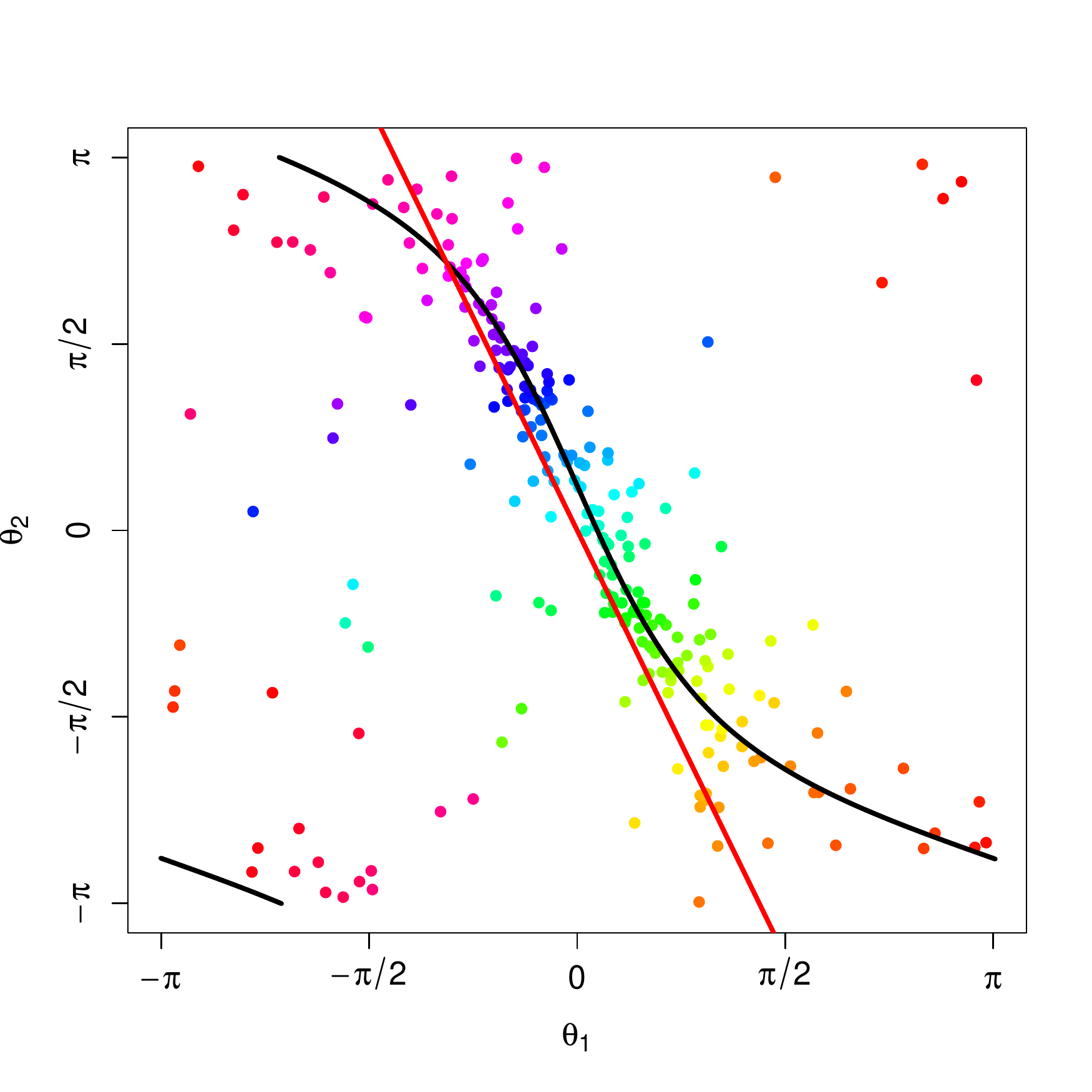}%
	\includegraphics[width=0.25\textwidth,clip,trim={0cm 0cm 0.75cm 2cm}]{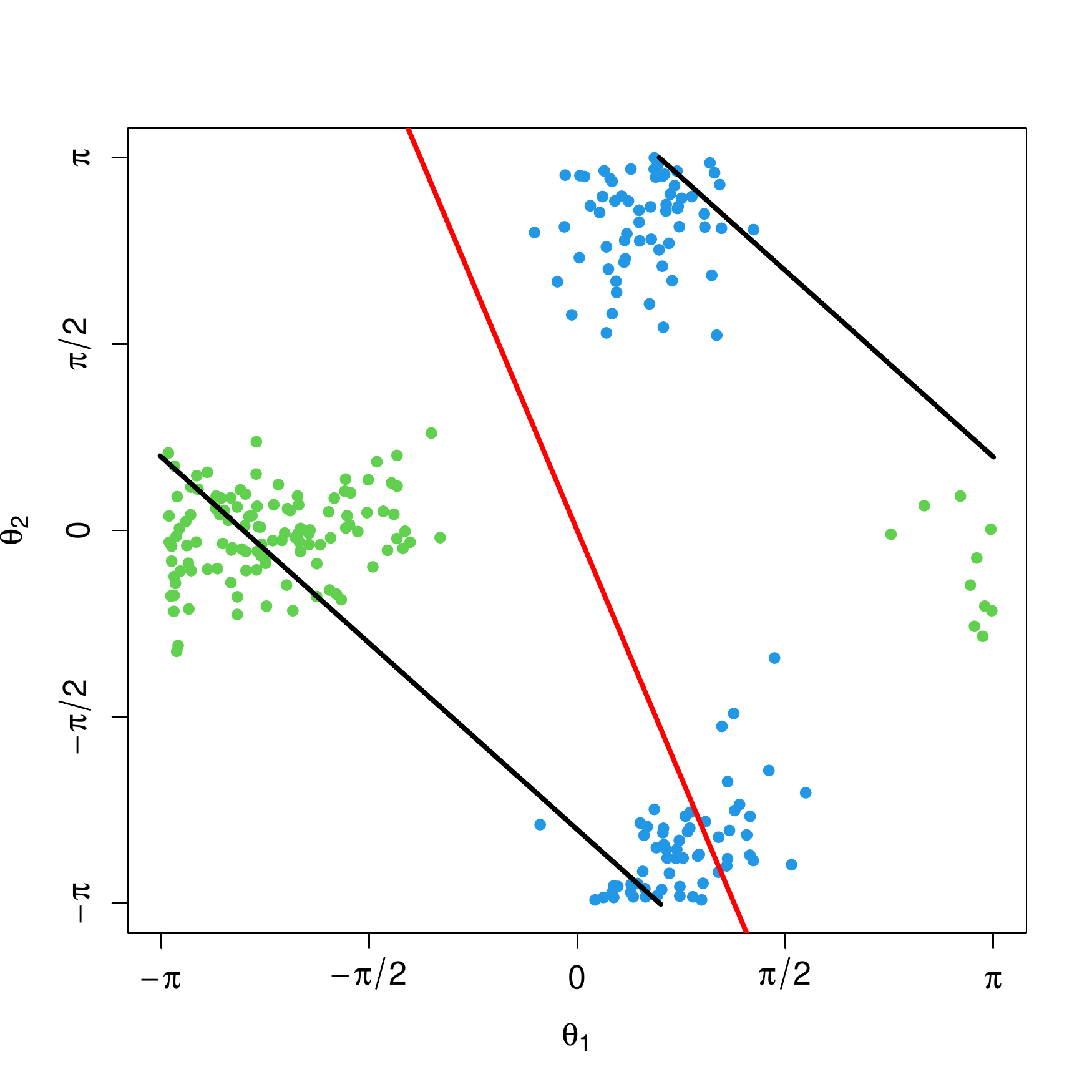}\\%
	\includegraphics[width=0.25\textwidth,clip,trim={0cm 0cm 0.75cm 2cm}]{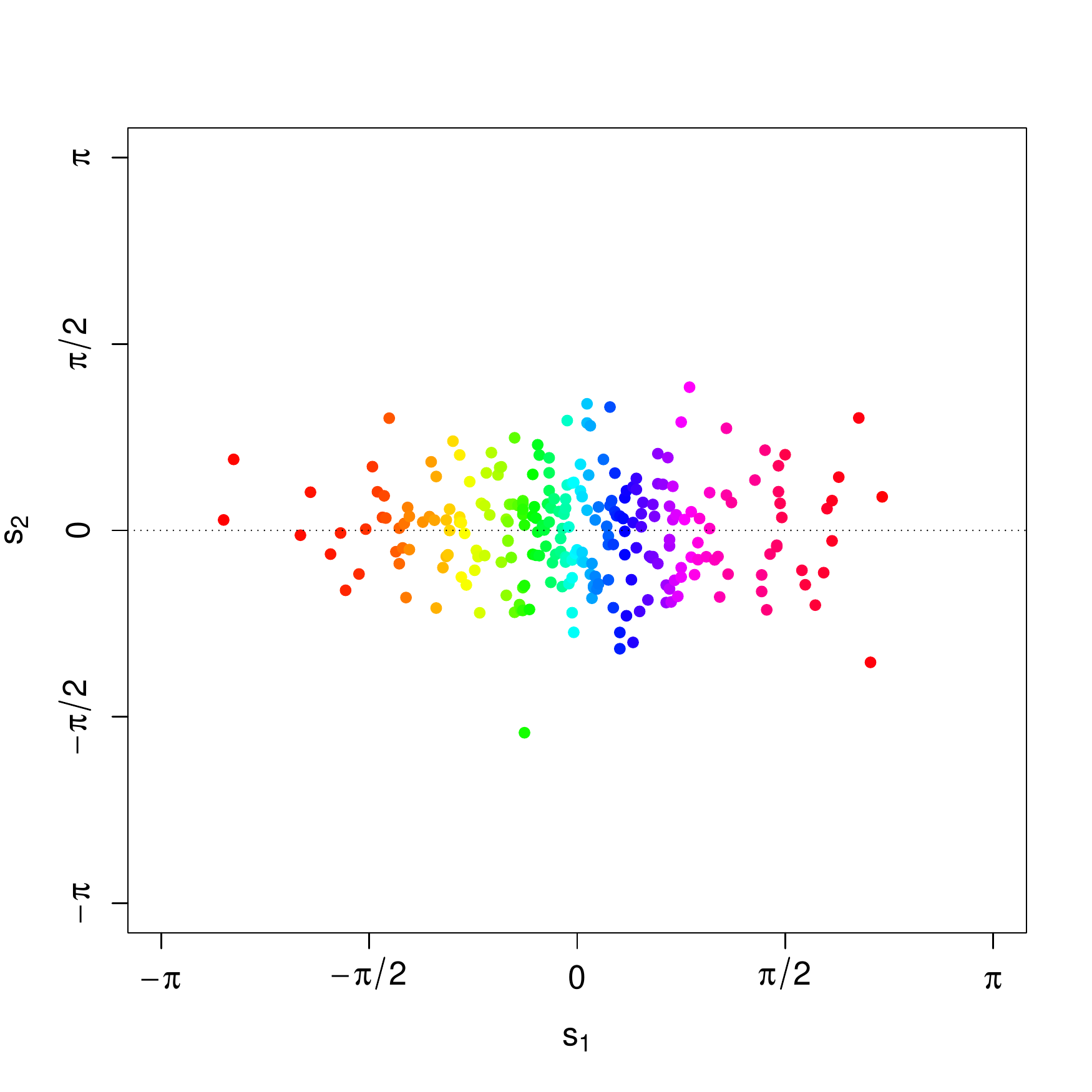}%
	\includegraphics[width=0.25\textwidth,clip,trim={0cm 0cm 0.75cm 2cm}]{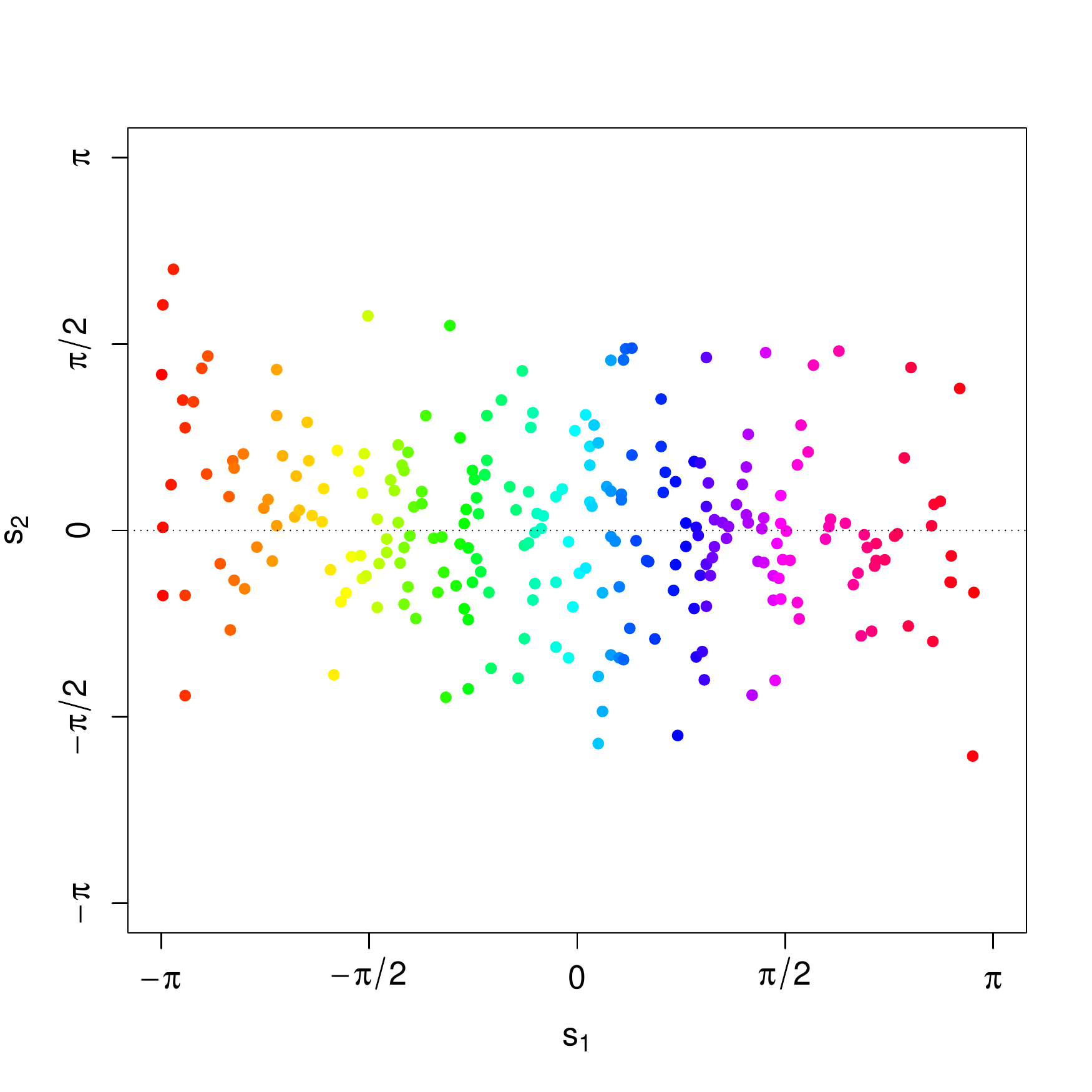}%
	\includegraphics[width=0.25\textwidth,clip,trim={0cm 0cm 0.75cm 2cm}]{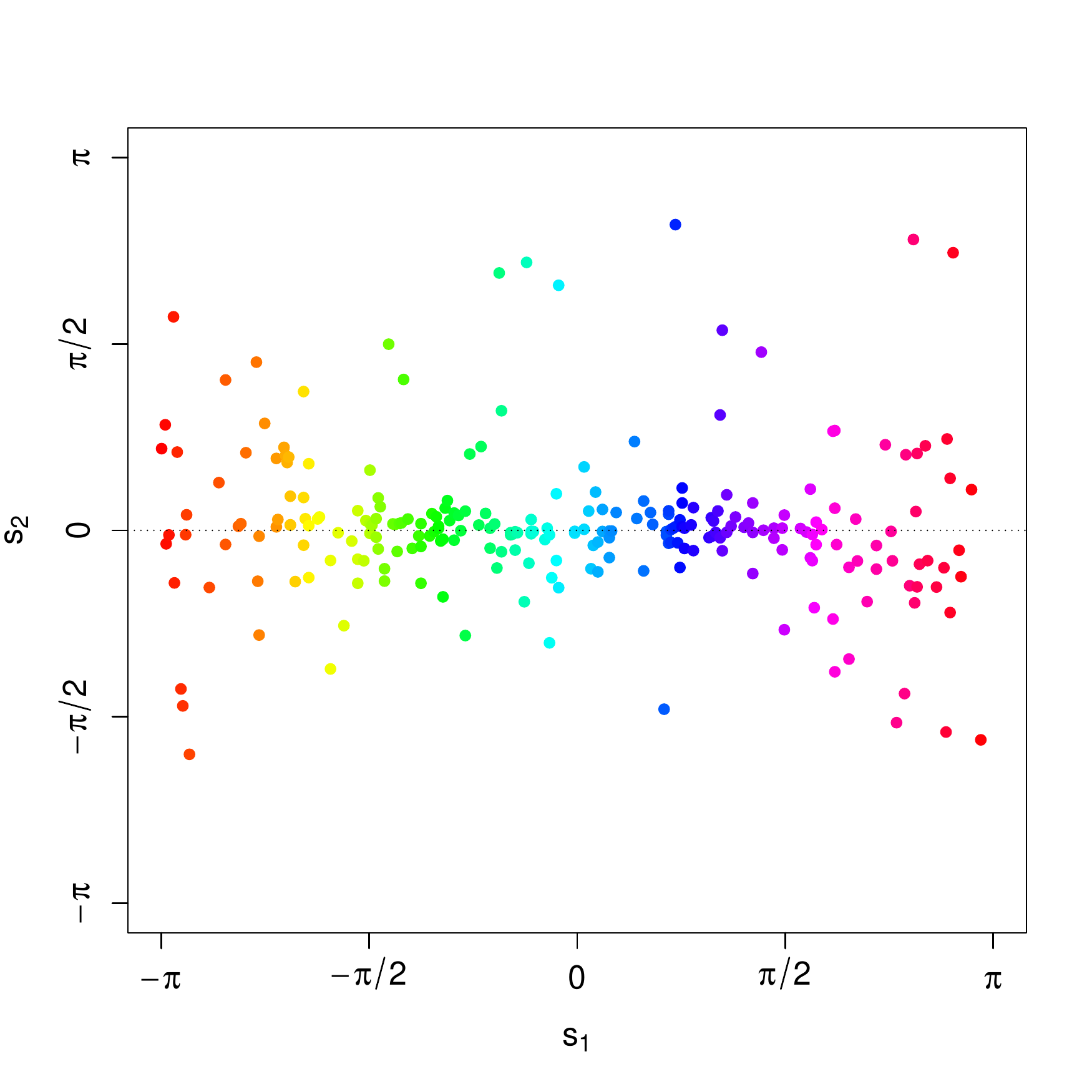}%
	\includegraphics[width=0.25\textwidth,clip,trim={0cm 0cm 0.75cm 2cm}]{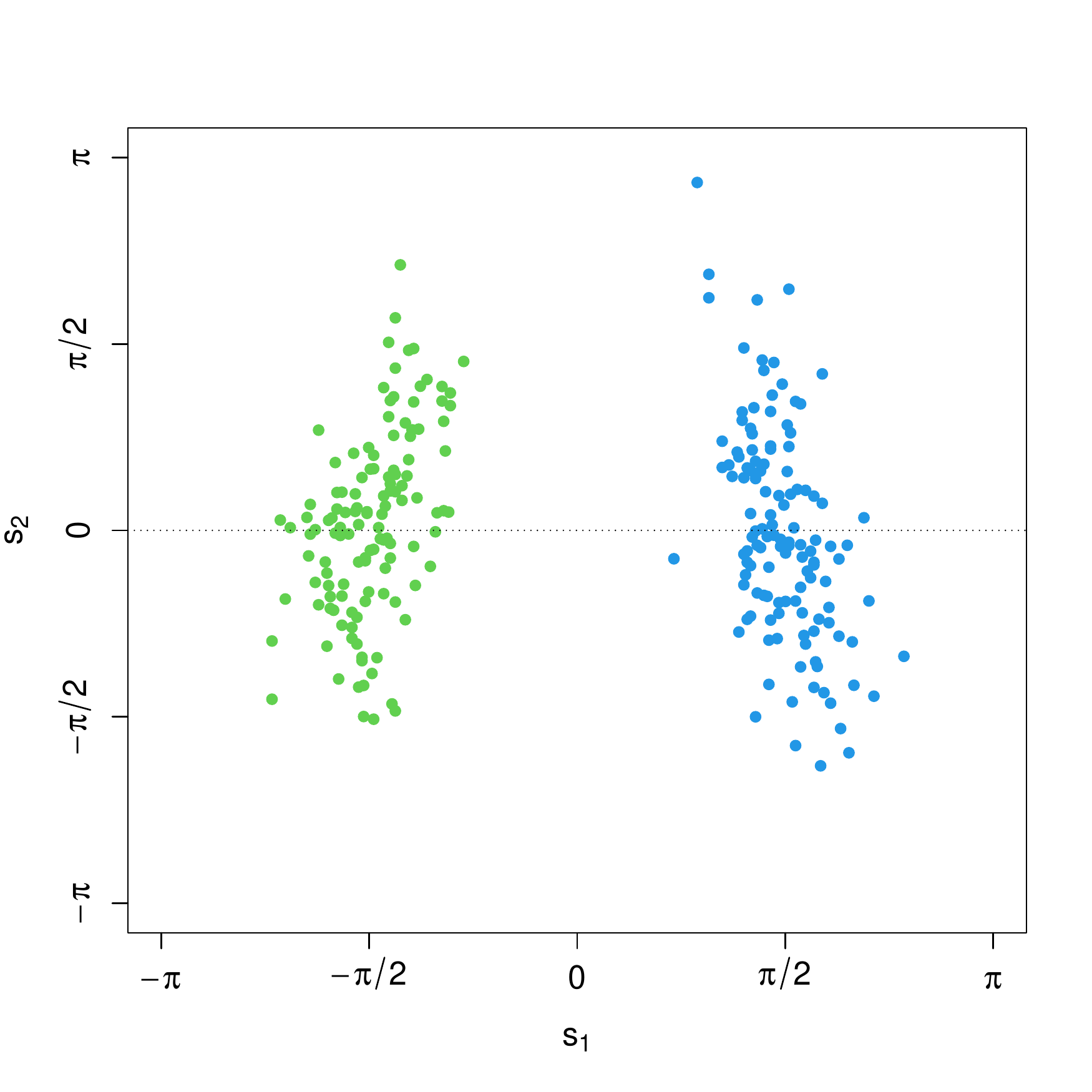}\\%
	\includegraphics[width=0.25\textwidth,clip,trim={0cm 0cm 0.75cm 2cm}]{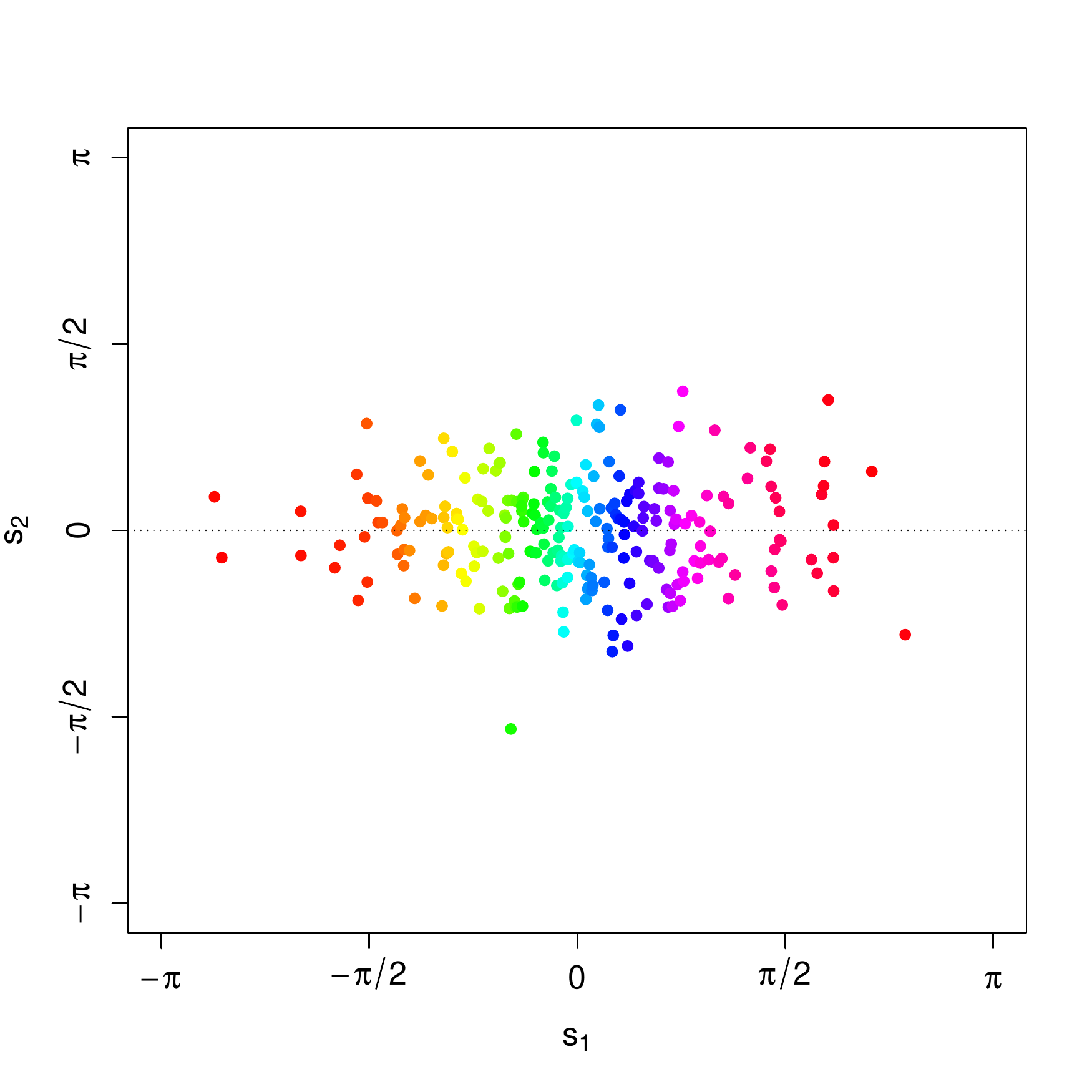}%
	\includegraphics[width=0.25\textwidth,clip,trim={0cm 0cm 0.75cm 2cm}]{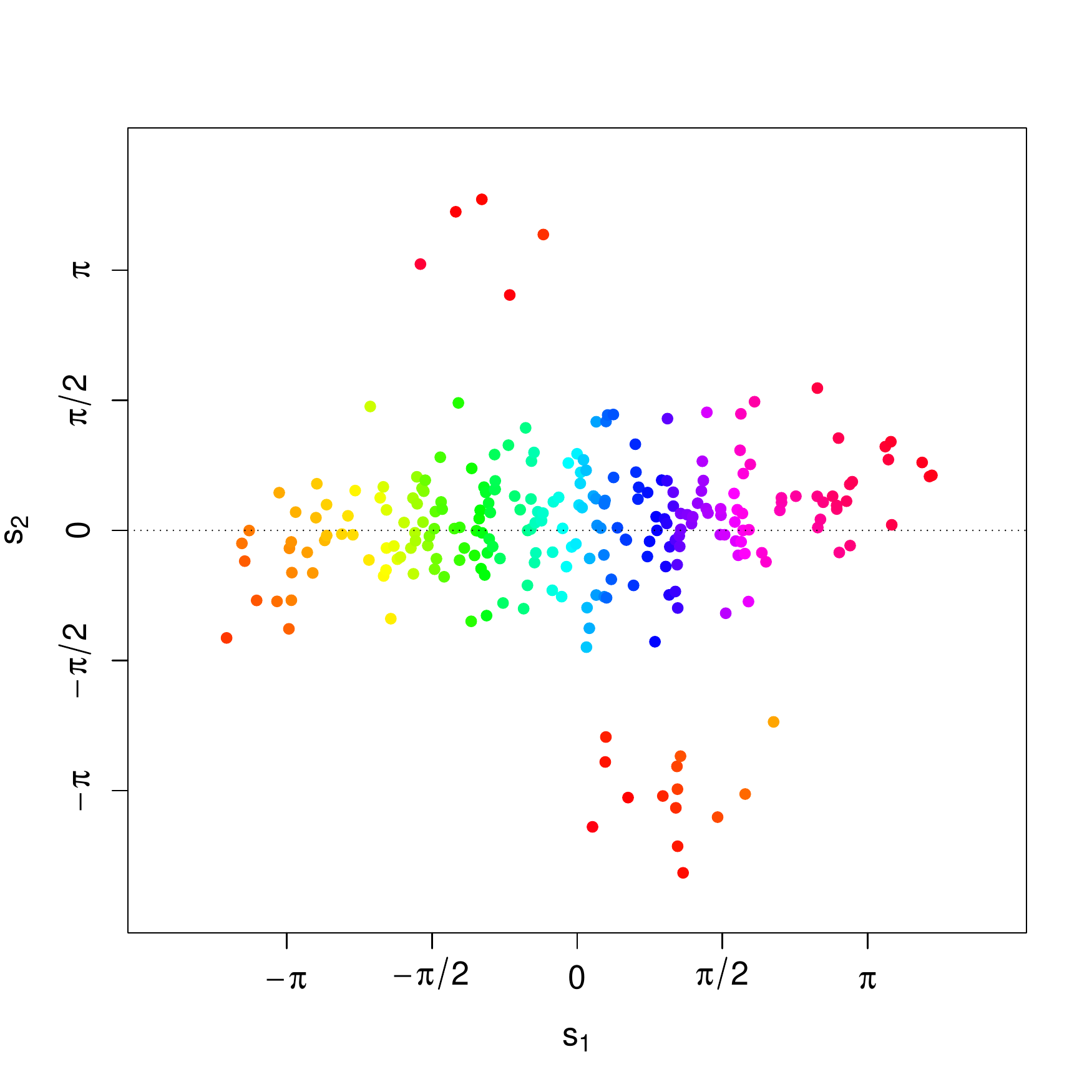}%
	\includegraphics[width=0.25\textwidth,clip,trim={0cm 0cm 0.75cm 2cm}]{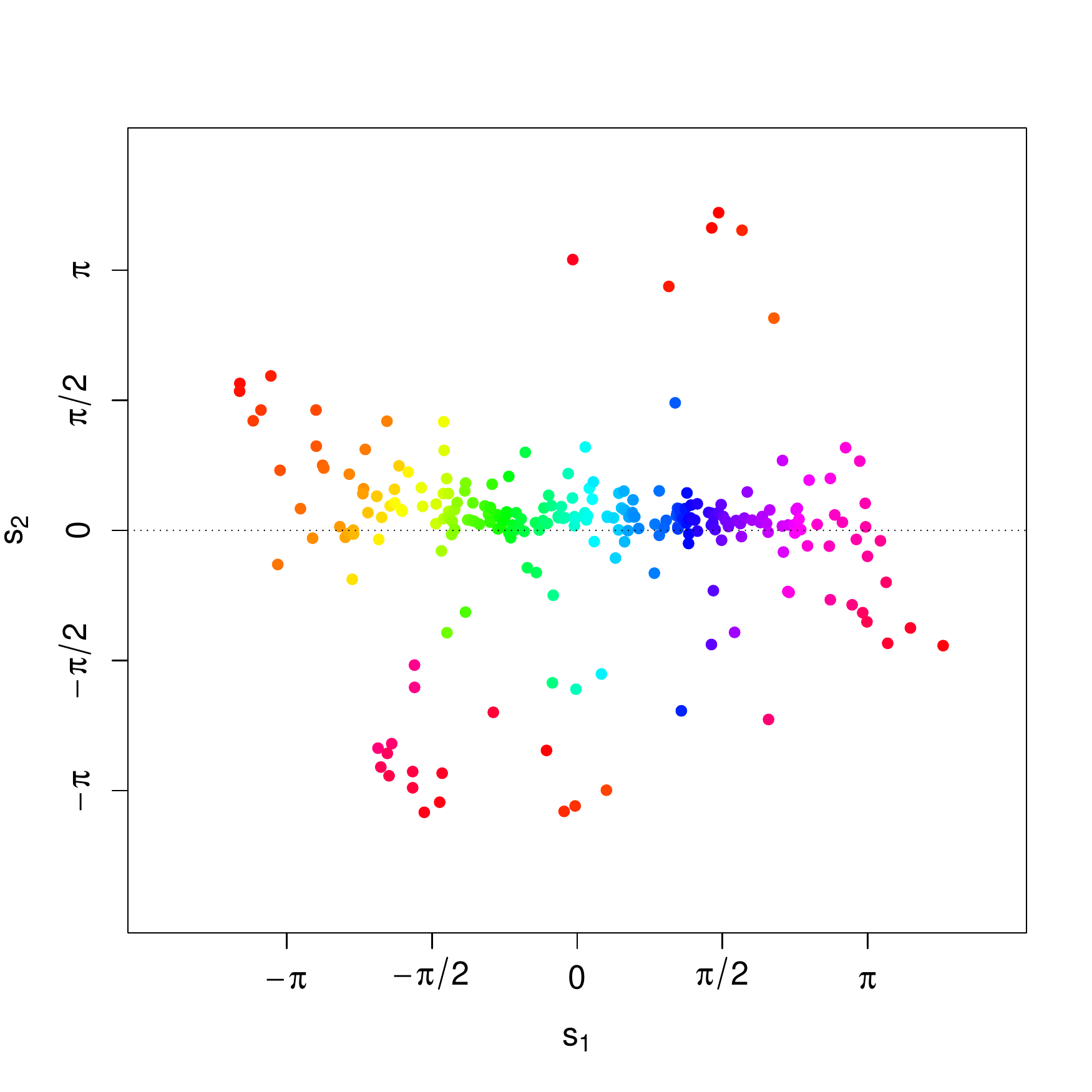}%
	\includegraphics[width=0.25\textwidth,clip,trim={0cm 0cm 0.75cm 2cm}]{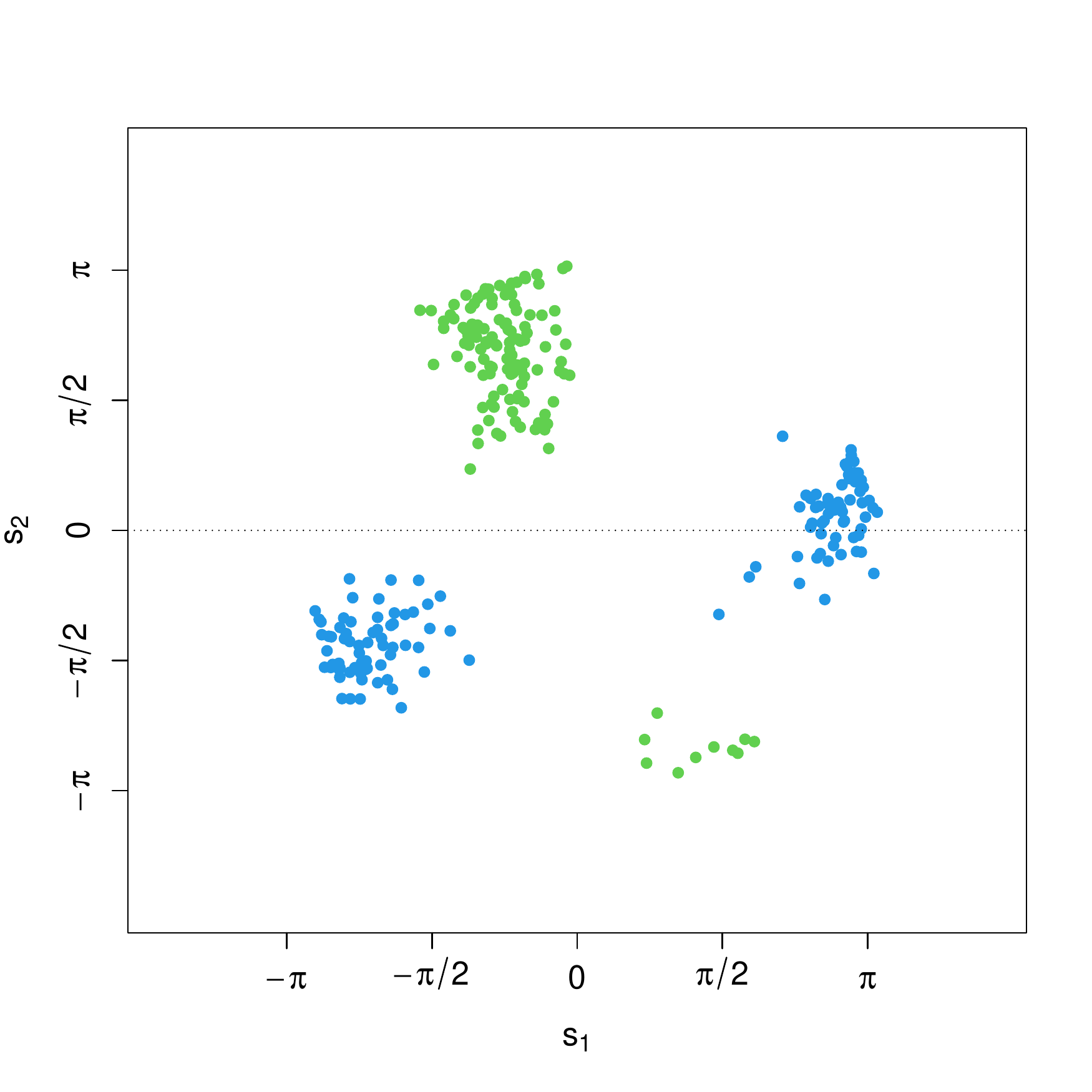}%
	\caption{\small Performance of aPCA (red line) versus TR-PCA (black curve) on four different samples (top row), together with TR-PCA (middle) and aPCA scores (bottom). From left to right: a concentrated BWN with $\boldsymbol{\mu} = (-\pi, 0)^{\prime}$ and $\boldsymbol{\Sigma} = (\sigma_1^2, \rho\sigma_1\sigma_2; \rho\sigma_1\sigma_2, \sigma_2^2)$ with $(\sigma_1^2,\sigma_2^2,\rho)=(0.2,0.8,0.35)$; a more spread BWN with $\boldsymbol{\mu} = (\pi/2, 0)^{\prime}$ and $(\sigma_1^2,\sigma_2^2,\rho)=(3,1.5,0.85)$; a BWC with $\boldsymbol{\mu} = (1,2)^{\prime}$ and $(\xi_1, \xi_2, \rho) = (0.5, 0.1, -0.75)$; and an equal mixture of two BWNs with $\boldsymbol{\mu}_1 =(\pi/2, -\pi/2)^{\prime}$ and $(\sigma_{11}^2,\sigma_{12}^2,\rho_1)=(0.4,0.16,0.35)$, and $\boldsymbol{\mu}_2 =(-\pi/2, \pi/2)^{\prime}$ and $(\sigma_{21}^2,\sigma_{22}^2,\rho_2)=(0.16,0.4,0.35)$. To allow a principled comparison between aPCA and TR-PCA, in the first row the sample has been centered by its circular mean. TR-PCA is invariant to this centering, while unperiodic projection in aPCA depends on it.  The rainbow color palette indicates the main mode of variation. The two clusters in the rightmost column are colored separately.}
	\label{fig:TR-PCAperformance}
\end{figure}

The first scenario (leftmost column) shows the almost equivalence between both methods when dealing with highly-concentrated samples, an expected consequence of the torus being locally Euclidean. Here, both the first scores of TR-PCA and aPCA explain $81\%$ of the variance. When the distribution is more dispersed, as in the other three samples, periodicity becomes relevant and aPCA begins introducing artifacts. In the second scenario, TR-PCA ($83\%$) does not introduce any distortion on the scores thanks to honoring the toroidal geometry, while aPCA ($75\%$) creates some spurious clusters with scores that exit $\mathbb{T}^2$ (observe the vertical scale) and induces a slight rotation on the scores. These artifacts are magnified in the third scenario (TR-PCA: $88\%$; aPCA: $74\%$). Finally, the fourth scenario represents a Simpson's paradox case in which TR-PCA ($78\%$) is able to successfully separate the two clusters along the first scores, while aPCA ($38\%$) fails to do so. In all scenarios, TR-PCA yields periodic scores, unlike aPCA. The PVEs of both methods were computed according to \eqref{eq:pve}.

\section{Data application}
\label{sec:application}

This section illustrates the application of TR-PCA to the study of currents in the Santa Barbara Channel (California, USA). Studying the direction of currents is of crucial importance to understand the supply of nutrients in marine habitats \citep{allen2012influence} and the genetic propagation of marine fauna and flora \citep{white2010ocean}, as well as is important for environmental purposes and prevention of contamination \citep{DiGiacomo2004}. In the present context of increased contamination and climate change, the Santa Barbara Channel is an area well known for the confluence of several important ocean currents and vast marine biodiversity. The complexity of local currents \citep{winant2003characteristic} makes the use of standard statistical tools not directly applicable and motivates the search for new methodologies able to explain their variability. In their Figure 10, \cite{winant2003characteristic} explain that there exists a counterclockwise vortex in the Santa Barbara Channel, which is represented by a general westward flux on its northern coast. In addition, studies such as \cite{auad1998wind} show that there exists a net influx of water in the Santa Barbara Channel that exits heading south through the Santa Cruz Channel. These two facts are taken as a guideline to validate the ability of TR-PCA on indexing the main variability of the data in a fully data-driven way.

The data was obtained from the High-Frequency Radar Database (\url{https://hfradar.ndbc.noaa.gov/}), which maps surface currents and wave fields over wide areas. In particular, the currents' direction, $d = \mathrm{atan2}(u,v)$, is hourly available through the measured eastward and northward surface velocities ($u$ and $v$, respectively) of the water body. We smoothed this direction by taking the daily speed-weighted circular mean in a given zone Z, obtaining the circular variable of the study, $\theta_\mathrm{Z}$. Since large-scale currents involve timespans longer than hours, daily averages allow smoothing of the data while still keeping the variability associated with these ocean currents. We used the data from the three-year period 2019--2021, using complete years so that seasonality events were neither over- nor underrepresented. Four locations were selected: zones A and B, located along the northern coast of the Santa Barbara Channel; and zones C and D, corresponding to the north and south of the Santa Cruz Channel, respectively. These areas are shown in Figure \ref{fig:StaBarbaraMap}. The study focuses on the dependency between $\theta_\mathrm{A}$ and $\theta_\mathrm{B}$ to further analyze the top part of the vortex shown in \cite[Figure 10]{winant2003characteristic} and the water flux parallel to the coast, as well as on the dependency between $\theta_\mathrm{C}$ and $\theta_\mathrm{D}$ for the water output through the inter-island strait \citep{auad1998wind}. The third analysis on $(\theta_\mathrm{A}, \theta_\mathrm{C})$ is performed to investigate the dependence between both regions.

\begin{figure}[!ht]
	\centering
	\includegraphics[width = 0.55\textwidth]{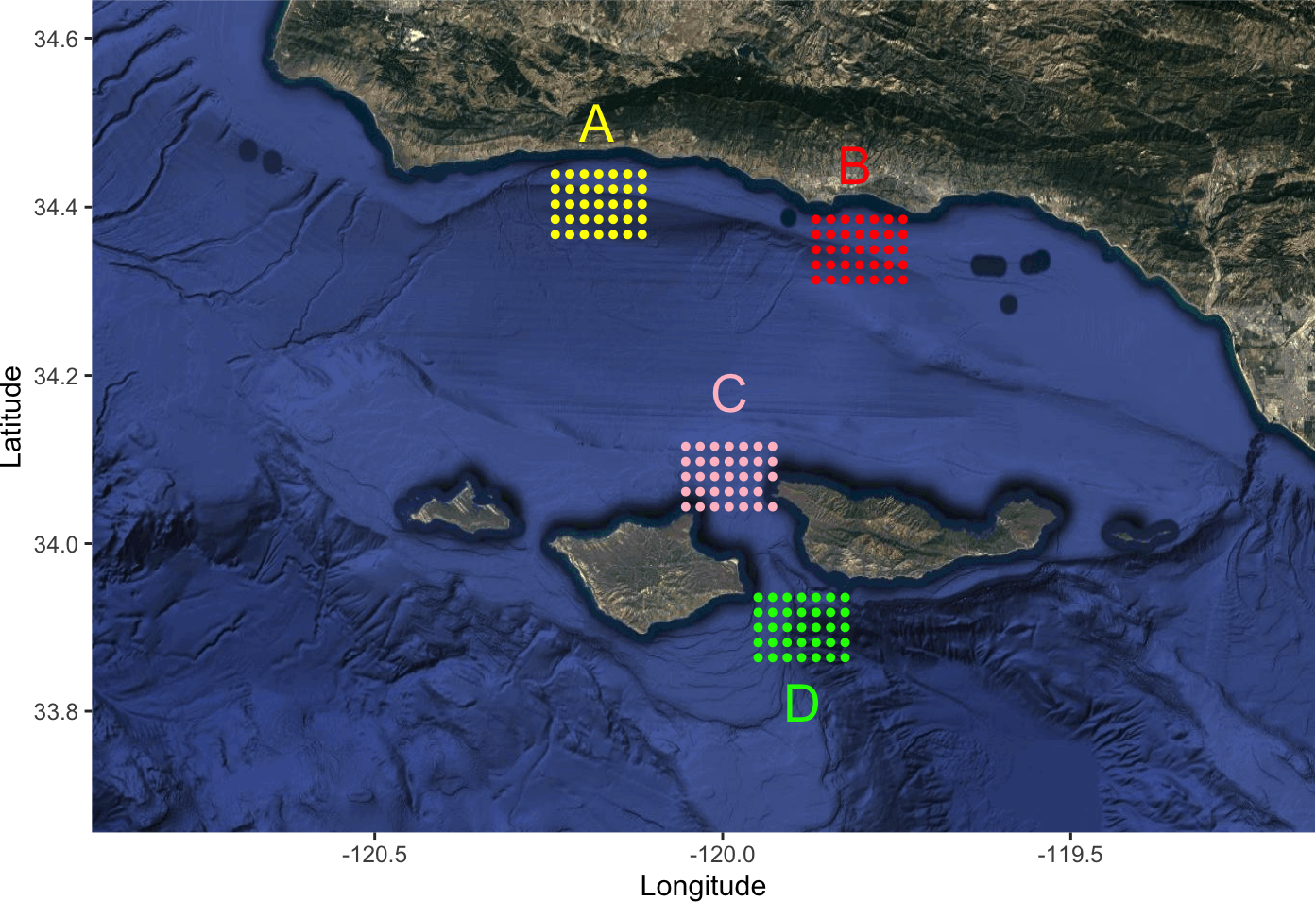}
	\caption{\small Satellite map of Santa Barbara coast with the main zones of the analysis depicted in different colors. The coordinates (in degrees) delimiting the different areas are $\mathrm{A} \equiv (-120.24, -120.12)\times (34.37, 34.44)$, $\mathrm{B} \equiv (-119.86, -119.74) \times (34.31, 34.39)$, $\mathrm{C} \equiv\allowbreak (-120.05, -119.93) \times (34.04, 34.12) $, and $\mathrm{D} \equiv\allowbreak (-119.95, -119.82) \times (33.86, 33.94)$. The connection A--B represents the coast of the Santa Barbara Channel, while C--D is the Santa Cruz Channel.}
	\label{fig:StaBarbaraMap}
\end{figure}

A first analysis of the data's density in the three zones (A--B, A--C, C--D) was performed to check the adequacy of the BSvM and BWC distributions to model the current directions. Toroidal-adapted kernel density estimators for estimating the unknown bivariate densities were compared with the estimated parametric densities of the BSvM and BWC. These estimates are shown in Figure \ref{fig:santabarbarafits}, where it can be seen that the BWC and BSvM give similar shapes, although simplifying some of the asymmetries of the kernel density estimators. To formally assess the existence of dependence in the four scenarios, we performed the $\phi^{(n)}_{\mathrm{dc}}$ test described in \citet[Section 3]{garciaportugues2021}, emphatically rejecting the null hypothesis of independence in all cases.

The BSvM and BWC models were fitted to the samples of $(\theta_\mathrm{A}, \theta_\mathrm{B})$, $(\theta_\mathrm{A}, \theta_\mathrm{C})$, and $(\theta_\mathrm{C}, \theta_\mathrm{D})$ using maximum likelihood. The estimated parameters and log-likelihoods are shown in Table \ref{tab:fits}. In terms of log-likelihoods, it can be seen that the BSvM estimate has better performance in two of the analyses. As expected, the BWC distribution is more concentrated, leading to higher values of the density around the modes.

\begin{table}[h!]
	\centering
	\footnotesize
	\setlength{\tabcolsep}{4pt}
	\begin{tabular}{|c|ccccccc|ccccccc|}
		\toprule
		Zone&$\hat\mu_1$&$\hat\mu_2$&$\hat\kappa_1$&$\hat\kappa_2$&$\hat\lambda$& $\hat{\ell}$&PVE&$\hat\mu_1$&$\hat\mu_2$& $\hat\xi_1$& $\hat\xi_2$& $\hat\rho$&$\hat{\ell}$&PVE \\
		\midrule
		A--B&$-2.63$&$-2.77$ &$2.52$&$0.95$&$\phantom{-}0.50$ &$-3004$&$75\%$&$\mathbf{-2.75}$& $\mathbf{-3.08}$ &$\mathbf{0.67}$ & $\mathbf{0.51}$ &$\mathbf{\phantom{-}0.09}$ &$\mathbf{-2977}$&$\mathbf{75\%}$ \\
		A--C&$\mathbf{-2.66}$ &$\mathbf{\phantom{-}0.40}$ &$\mathbf{2.69}$ & $\mathbf{0.65}$ &$\mathbf{-1.12}$ &$\mathbf{-3042}$&$\mathbf{80\%}$ &$-2.71$ &$\phantom{-}0.41$ &$0.66$ & $0.25$&$-0.24$ &$-3115$&$79\%$ \\
		C--D&$\mathbf{\phantom{-}0.73}$&$\mathbf{-2.11}$&$\mathbf{0.60}$ &$\mathbf{2.21}$ &$\mathbf{-1.91}$ &$\mathbf{-3107}$&$\mathbf{79\%}$&$-0.13$ & $-1.82$ &$0.31$ &$0.58$ &$-0.33$&$-3235$&$74\%$ \\
		\bottomrule
	\end{tabular}
	\caption{\small Estimated parameters, log-likelihood ($\hat{\ell}$), and Proportion of Variance Explained (PVE) by TR-PCA for the different analyses. Left and right blocks give the BSvM and BWC fits, respectively. The bold font indicates the best-performing model in terms of the log-likelihood. \label{tab:fits}}
\end{table}

The estimated ridges $\mathcal{R}_{\hat{\boldsymbol{\mu}}}(\hat{f})$ for the three bivariate analyses are shown in Figure \ref{fig:santabarbarafits}. It can be seen how $\mathcal{R}_{\hat{\boldsymbol{\mu}}}(\hat{f})$ indexes the main variability of the sample and informs on the positioning of high-density regions. On the one hand, the best-fitting models capture the local correlations of the currents' directions about their modes, which are seen to be positive in the A--B case, and negative in the A--C and C--D cases. On the other hand, the curve $\mathcal{R}_{\hat{\boldsymbol{\mu}}}(\hat{f})$ allows synthesizing the behavior of the currents in a sensible way. For example, in the A--B case, the BSvM model not only reproduces the W--W mode (western direction both at zone A and B) found in previous studies \citep{winant2003characteristic}, but also provides additional information in terms of the sinusoidal-like dependence between A--B. A final insight revealed by the estimated ridges is that there is a high negative correlation in C--D. Due to the geographical shape of the strait, water can flow through it in two main directions: SE--SE and NW--NW, as previous studies show \citep{auad1998wind}. As seen in Figure \ref{fig:santabarbarafits}, both main modes are captured fairly well: there is a high concentration toward SE--S and a secondary concentration at NW--W. Furthermore, it can also be seen that the higher-density transition region between the two modes has a negative correlation, a result that cannot be easily obtained by only analyzing the modes.

\begin{figure}[h!]
	\centering
	\includegraphics[width=0.25\textwidth,clip,trim={0cm 0cm 0.75cm 2cm}]{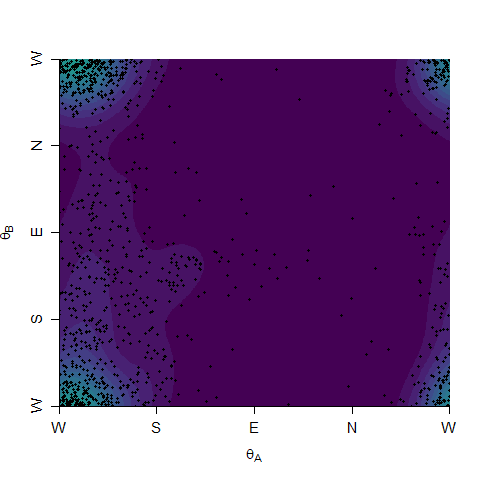}%
	\includegraphics[width=0.25\textwidth,clip,trim={0cm 0cm 0.75cm 2cm}]{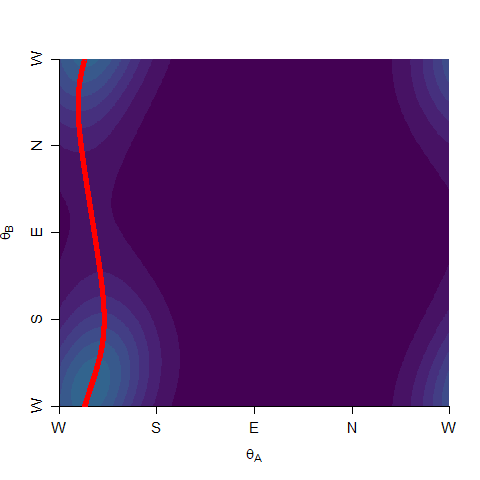}%
	\includegraphics[width=0.25\textwidth,clip,trim={0cm 0cm 0.75cm 2cm}]{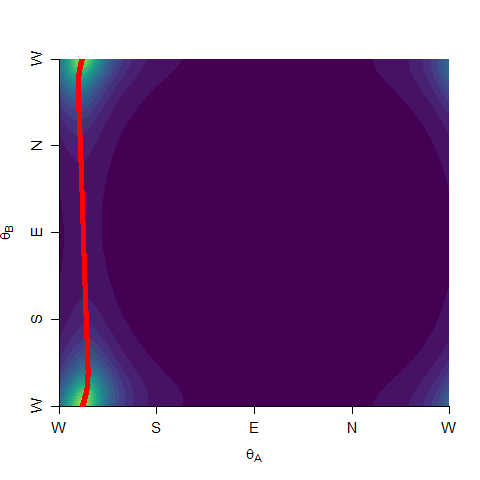}%
	\includegraphics[width=0.25\textwidth,clip,trim={0cm 0cm 0.75cm 2cm}]{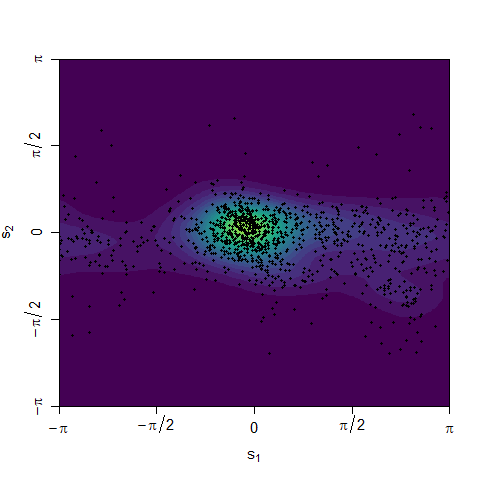}\\%
	\includegraphics[width=0.25\textwidth,clip,trim={0cm 0cm 0.75cm 2cm}]{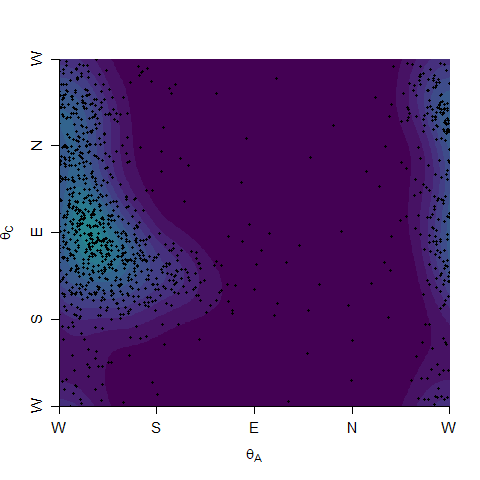}%
	\includegraphics[width=0.25\textwidth,clip,trim={0cm 0cm 0.75cm 2cm}]{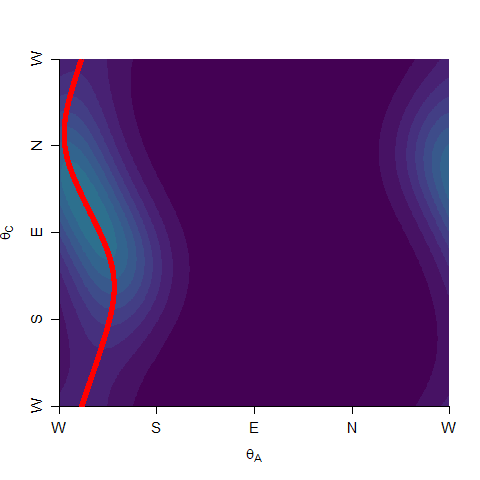}%
	\includegraphics[width=0.25\textwidth,clip,trim={0cm 0cm 0.75cm 2cm}]{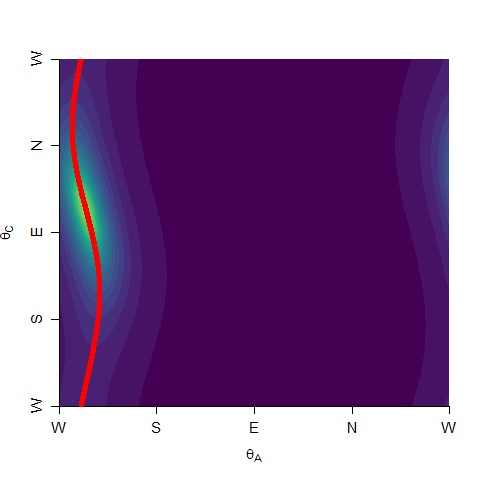}%
	\includegraphics[width=0.25\textwidth,clip,trim={0cm 0cm 0.75cm 2cm}]{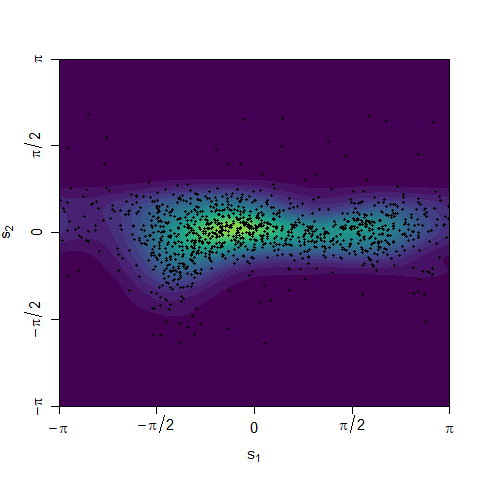}\\%
	\includegraphics[width=0.25\textwidth,clip,trim={0cm 0cm 0.75cm 2cm}]{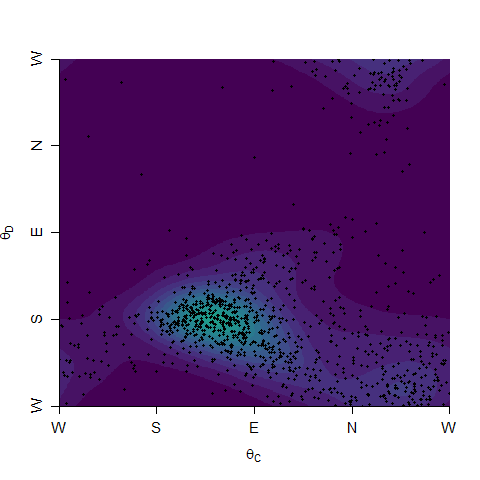}%
	\includegraphics[width=0.25\textwidth,clip,trim={0cm 0cm 0.75cm 2cm}]{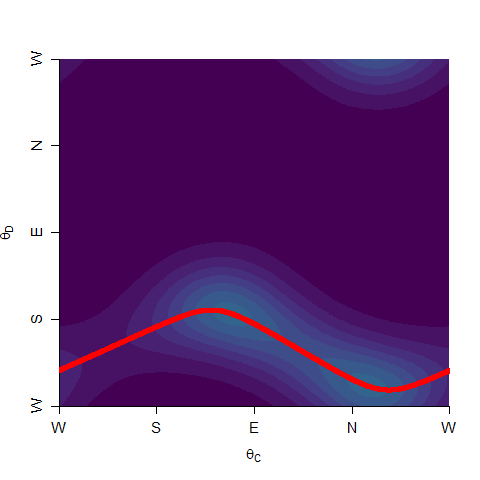}%
	\includegraphics[width=0.25\textwidth,clip,trim={0cm 0cm 0.75cm 2cm}]{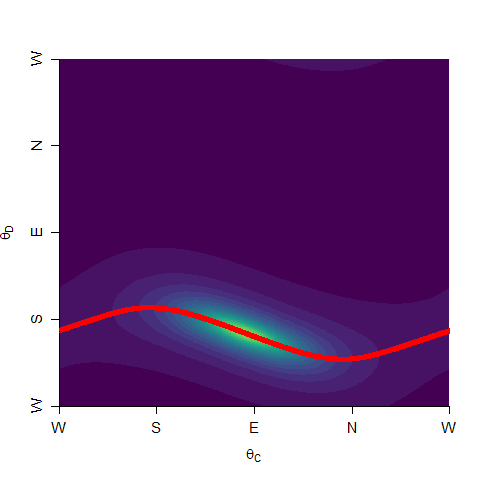}%
	\includegraphics[width=0.25\textwidth,clip,trim={0cm 0cm 0.75cm 2cm}]{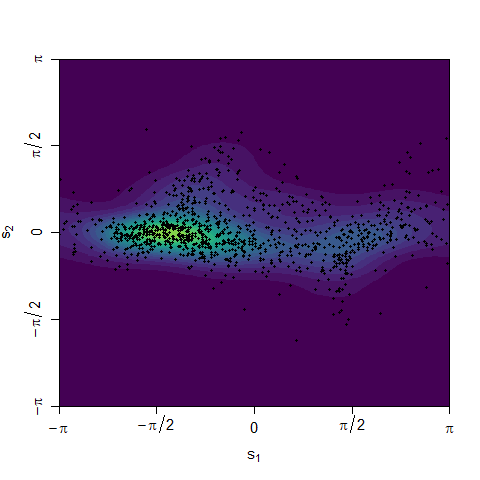}
	\caption{\small From left to right, the columns represent the density contours of kernel density estimates, fitted BSvM densities, fitted BWC densities, and kernel density estimates of the TR-PCA scores of the best-performing model. The second and third columns show the estimated ridge curves $\mathcal{R}_{\hat{\boldsymbol{\mu}}}(\hat{f})$ (red curves) from the sample (black points in the first column) for the four bivariate analyses and two models. From top to bottom, rows represent zones A--B, A--C, and C--D. The contourplots within the same row share a common color scale in the first three columns. The parameters of the fits are shown in Table \ref{tab:fits}.}
	\label{fig:santabarbarafits}
\end{figure}

The summarizing capability of $\mathcal{R}_{\hat{\boldsymbol{\mu}}}(\hat{f})$ can be further exploited by visualizing a march along it. Figure \ref{fig:santabarbaraindexing} shows this march in the case C--D using the BSvM fit shown previously in Figure \ref{fig:santabarbarafits}. Unlike previous studies that were limited to finding the main modes of direction \citep{auad1998wind, winant2003characteristic}, this parametrization also allows an estimation of the variability when the flow has other directions. For example, the net influx of water outside the strait is reproduced in the second leftmost plot in Figure~\ref{fig:santabarbaraindexing}.

\begin{figure}[h!]
	\centering
	\includegraphics[width=0.25\textwidth,clip]{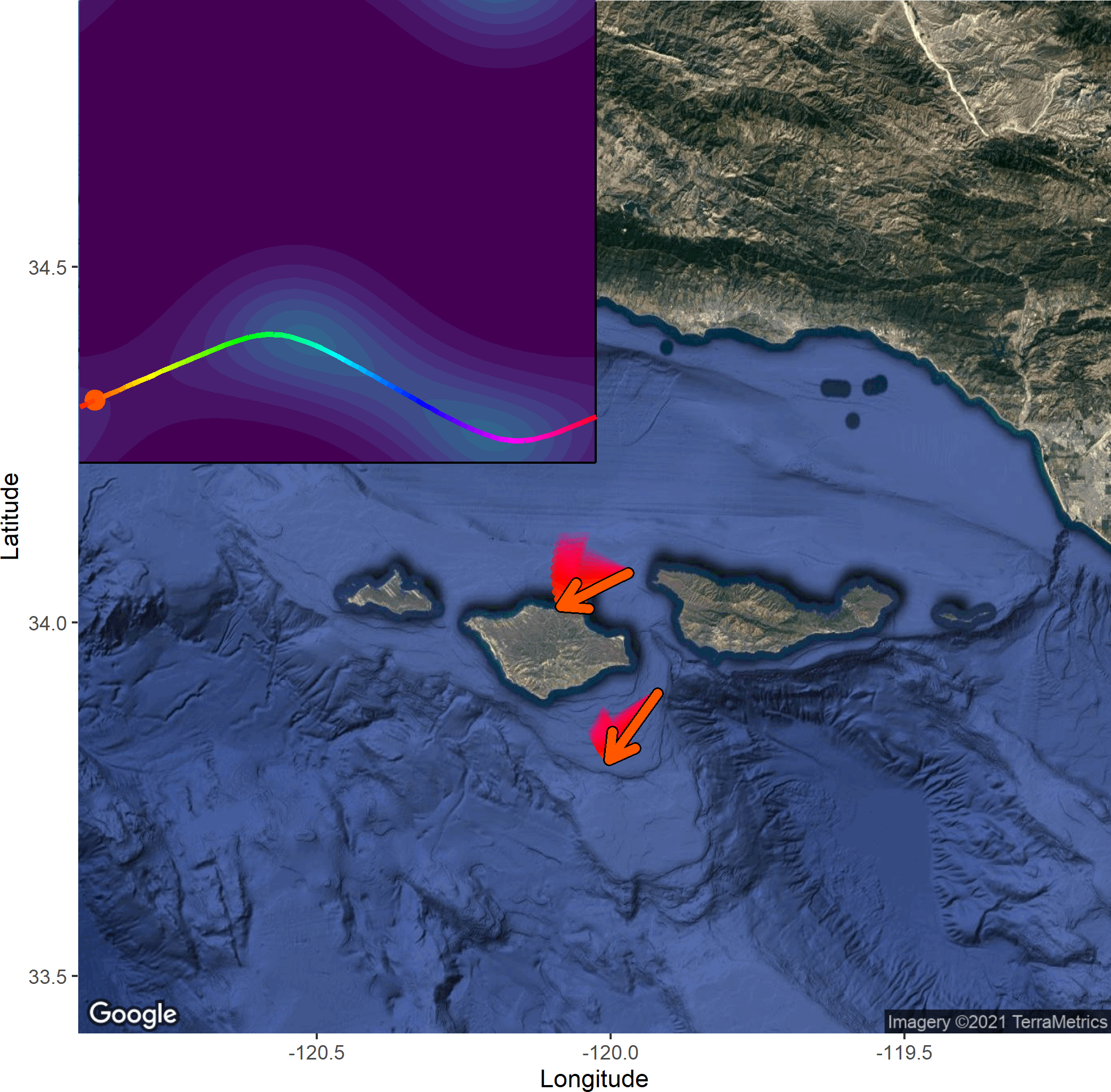}%
	\includegraphics[width=0.25\textwidth,clip]{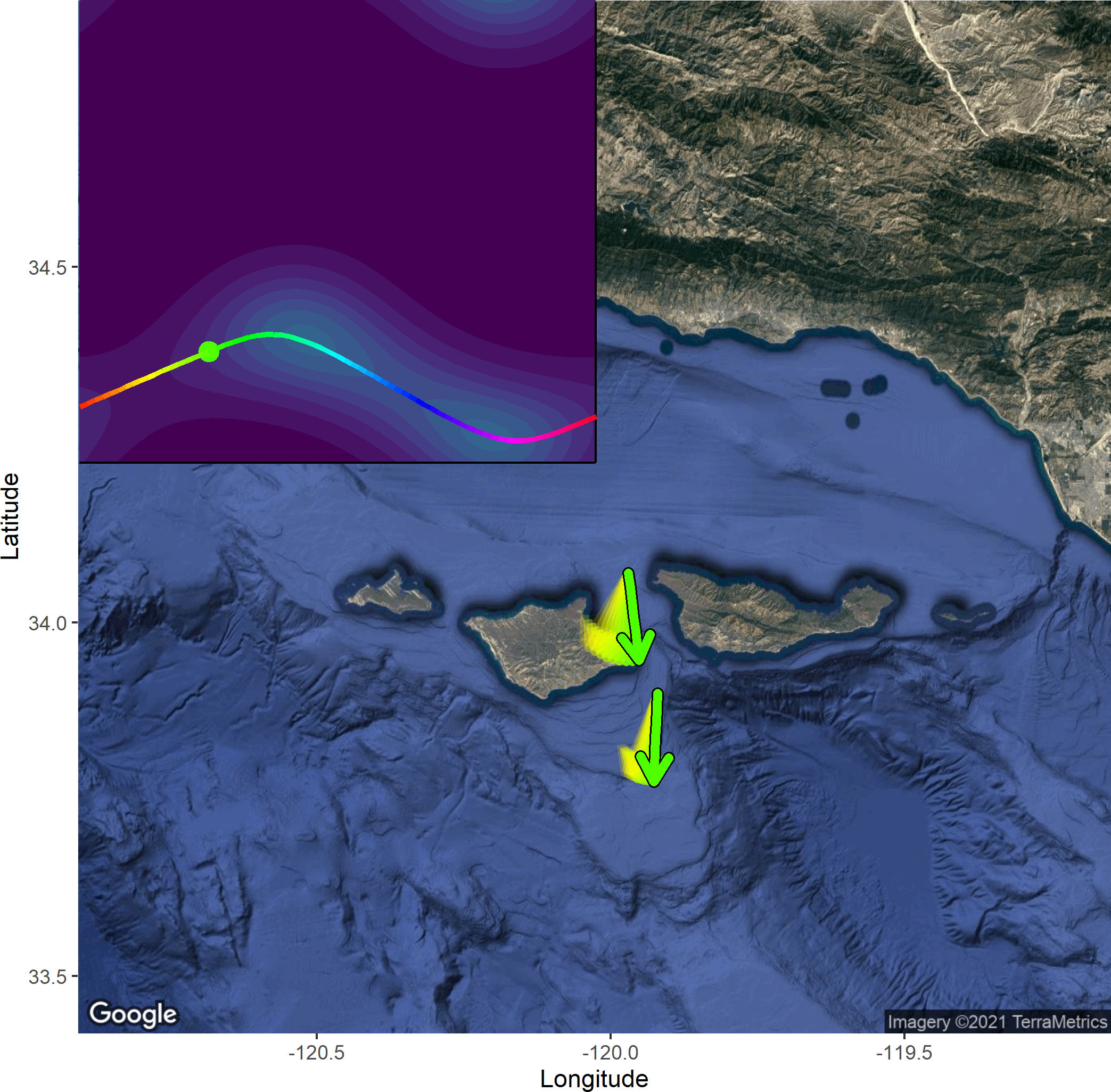}%
	\includegraphics[width=0.25\textwidth,clip]{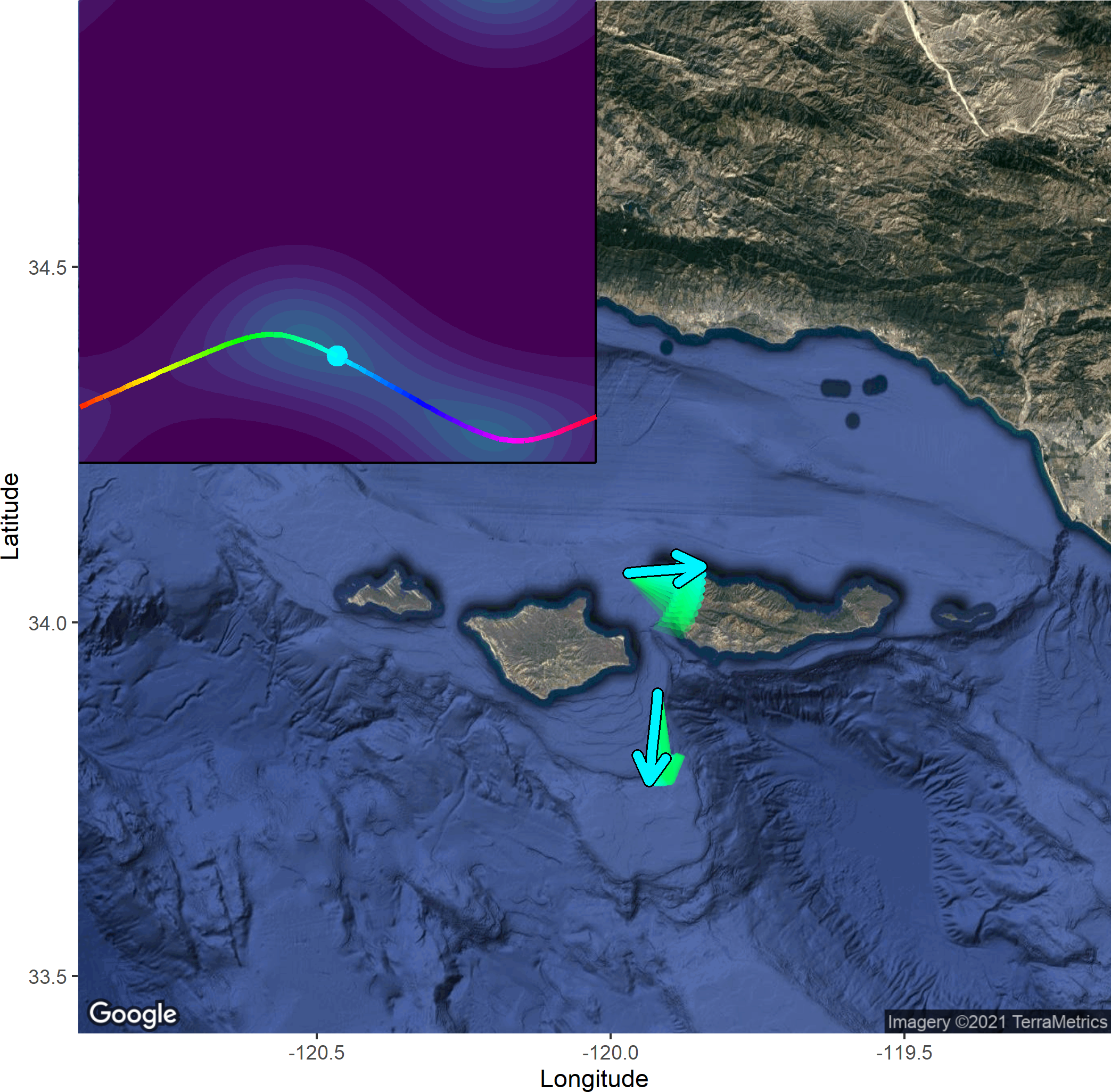}%
	\includegraphics[width=0.25\textwidth,clip]{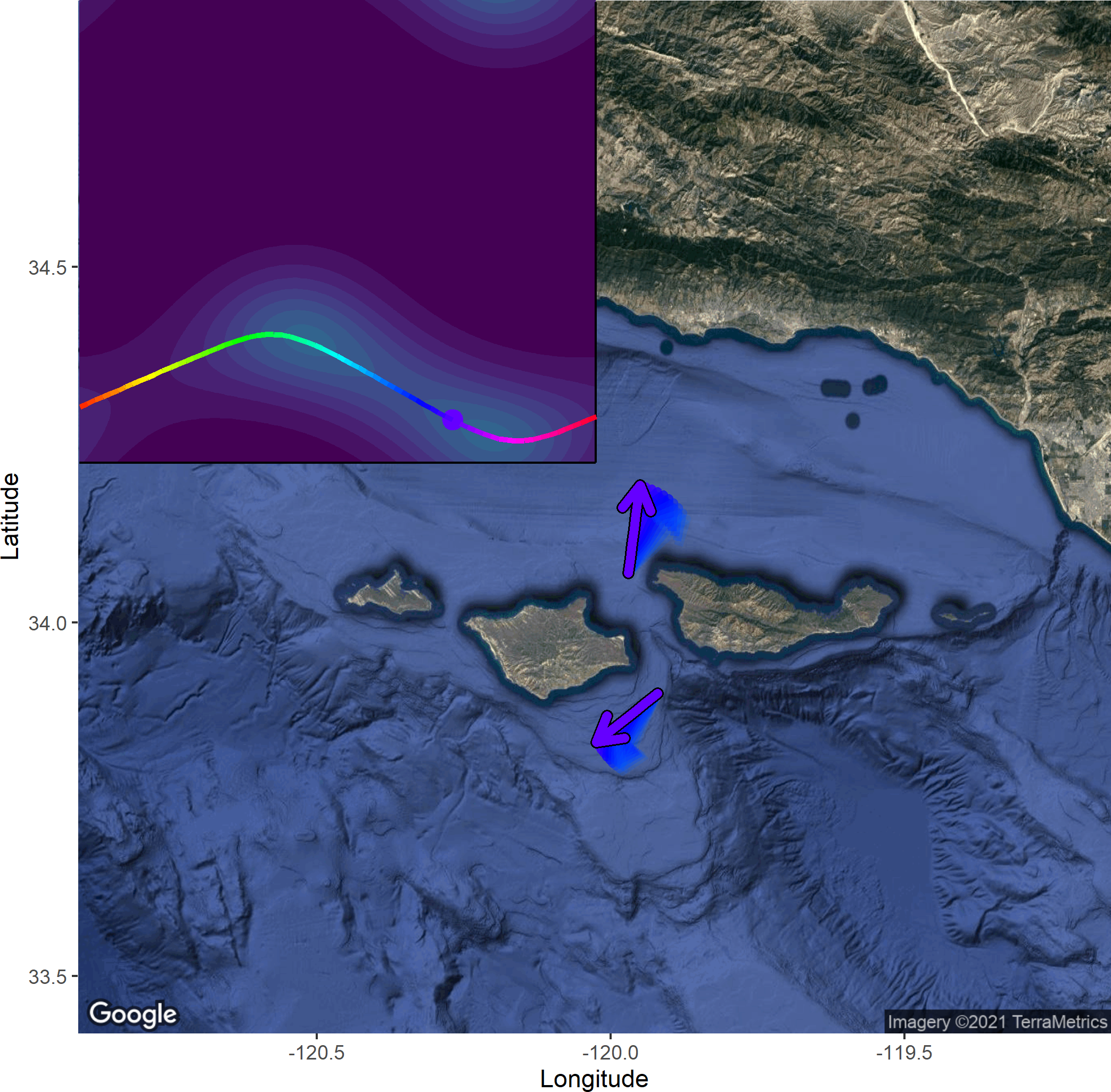}%
	\caption{\small Four snapshots of the march along the ridge curve $\mathcal{R}_{\hat{\boldsymbol{\mu}}}(\hat{f})$ for C--D. The arrows at C (top) and D (bottom) are colored according to their position on the ridge curve. Past directions are also shown with transparencies in order to visualize the movement's direction. The main variability of the currents on C--D is manifested on a pendular-like variation of the current direction at D that is aligned with a counterclockwise variation of the current direction at C. Video available at \url{https://github.com/egarpor/ridgetorus/tree/main/application}.}
	\label{fig:santabarbaraindexing}
\end{figure}

TR-PCA scores were computed in the rightmost column of Figure \ref{fig:santabarbarafits}, resulting in narrower, more concentrated distributions that capture a large part of the total variance, with the PVEs collected in Table \ref{tab:fits} being greater than $74\%$. These scores could be used for clustering purposes to find particular meteorological events or outliers.

\section{Discussion and conclusions}
\label{sec:discussion}

In this work we have advocated the use of density ridges for constructing a well-grounded bivariate toroidal PCA. The construction is based on using the implicit equation approach to determine ridges, which has proven to be more efficient and robust than the Euler algorithm. By tailoring this procedure to bivariate data, we have corroborated empirically that two reference toroidal distributions, the BWC and the BSvM, present stable connected components that go through the distributions' modes. We have proposed a practical way to parametrize $\mathcal{R}_{\boldsymbol{\mu}}(f)$ to yield a tractable computation of PCA-like scores that allows for a full dimension-reduction analysis. The real data application has shown that TR-PCA explains $75\%-80\%$ of the variability of the ocean currents and gives interpretations that are consistent with previous explanations of large-scale water movements in the~area.

An important takeaway of our investigations is that the BWC seems to be, in general, a more robust parametric distribution than the BSvM for TR-PCA. Although the BSvM is recognized as a somewhat canonical choice for ``the normal density on $\mathbb{T}^2$'', for this application, the squarish form of its density contours may introduce ``elbows'' on its associated density ridges, these in turn introducing artifacts on the resulting scores. In comparison, the BWC does not present this problem and tends to yield more flexible and descriptive principal ridges. Nevertheless, the choice of the reference parametric distribution is subject to improvement, as any sufficiently well-behaved density in $\mathcal{C}^2(\mathbb{T}^2)$ could be considered within our methodology. In this regard, we highlight that the ridge parametrization introduced in Section \ref{sec:ridgepar} could be used in other densities apart from BSvM/BWC. Deriving theoretical results giving the conditions under which the existence and well-definedness of a connected ridge for parametric families of densities would be a useful endeavor to be addressed in the future. In this respect, we experienced with another normal analog on the torus, the wrapped normal, for the analyses in Sections 3.1, 3.2, and 5.5. We found numerically that some ridges could be multivalued functions in both variables $\theta_1$ and $\theta_2$, giving convoluted ridge curves. When this does not happen, somehow unpleasant Z-shaped ridges with rough corners arise, thus making this distribution less appropriate for TR-PCA than BSvM and BWC. Although not pursued in this paper, we note that asymptotic inference for $\mathcal{R}_{\boldsymbol{\mu}}(f)$ is readily addressable when using a density model $f$ with tractable maximum likelihood.

TR-PCA has been constructed as a toroidal analog of PCA on $\mathbb{R}^2$ under the following optic. PCA can be seen as a \emph{parametric} dimension-reduction method driven by Gaussianity: in an arbitrary sample, it fits a normal distribution $\mathcal{N}(\boldsymbol{\mu},\boldsymbol{\Sigma})$ by maximum likelihood and extracts the eigendecomposition of $\hat{\boldsymbol{\Sigma}}$; the subspace spanned by the first eigenvector coincides with the $\hat{\boldsymbol{\mu}}$-connected ridge. TR-PCA follows this view of PCA, replacing the normal distribution with the BSvM/BWC model. It might be argued that PCA is a \emph{model-free} technique since $(\hat{\boldsymbol{\mu}},\hat{\boldsymbol{\Sigma}})$ estimate the mean and covariance matrix of a general population. In the torus, these two descriptive summaries are less canonical; e.g., there exist two definitions of circular means (extrinsic or intrinsic). The maximum likelihood estimators of the BSvM and BWC distributions do not necessarily coincide with extrinsic/intrinsic means, and so TR-PCA does not mimic this Gaussian-specific aspect of PCA. However, the maximum likelihood estimators of the BSvM/BWC model $f_{\boldsymbol{\xi}}$ estimate $\boldsymbol{\xi}_0=\arg\min_{\boldsymbol{\xi}}\mathrm{D}_{\mathrm{KL}}(f_0\| f_{\boldsymbol{\xi}})$, the parameter that minimizes the Kullback--Leibler divergence of a general population $f_0$ from $f_{\boldsymbol{\xi}}$. In the Gaussian model $f_{\boldsymbol{\mu},\boldsymbol{\Sigma}}$, $(\boldsymbol{\mu}_0,\boldsymbol{\Sigma}_0)=\arg\min_{\boldsymbol{\mu},\boldsymbol{\Sigma}}\mathrm{D}_{\mathrm{KL}}(f_0\| f_{\boldsymbol{\mu},\boldsymbol{\Sigma}})$ are the mean and covariance matrix of the population $f_0$, highlighting that both TR-PCA and PCA estimate a general population descriptor under a parametric model.

A clear limitation of the present work is its restriction to $p=2$. The concept of ridges can be extended to higher dimensions by redefining the eigenvector matrix, the projected gradient, and the density ridge of a multivariate toroidal density $f\in \mathcal{C}^2(\mathbb{R}^p)$ as $\mathbf{U}_{(p-q)}(\mathbf{x}):=(\mathbf{u}_{q+1}(\mathbf{x}), \ldots, \mathbf{u}_{p}(\mathbf{x}))$, $\mathrm{D}_{(p-q)} f(\mathbf{x}):=\mathbf{U}_{(p-q)}(\mathbf{x}) (\mathbf{U}_{(p-q)}(\mathbf{x}))^{\prime} \mathrm{D} f(\mathbf{x})$, and $\mathcal{R}_{q}(f):=\{\mathbf{x}\in\mathbb{R}^{p}:\|\mathrm{D}_{(p-q)} f(\mathbf{x})\|=0,\ \allowbreak \lambda_{q+1}(\mathbf{x}),\allowbreak\ldots,\lambda_{p}(\mathbf{x})<0\}$, with $1\leq q\leq p-1$. However, in this multivariate case, obtaining an equivalent of the implicit equation is more challenging, so one may need to rely on the (computationally expensive) Euler algorithm. Still, important concepts such as iterative projections on nested spaces and scores computation would need to be carefully defined for the multivariate case. The multivariate extension of TR-PCA is also currently limited by the scarcity of multivariate toroidal models beyond the multivariate von Mises distribution \citep{Mardia2008} (which extends \eqref{eq:bsvm}). For example, currently there does not exist a multivariate extension of the BWC. Therefore, the larger endeavor of extending TR-PCA to higher dimensions is open for future research.

\section*{Supplementary materials}

The R-package \texttt{ridgetorus} implements TR-PCA on the two-dimensional torus (\texttt{ridge\_pca()}). It provides functions to estimate the parameters of BSvM and BWC models and compute their density ridges. It also contains all the data used for the application in Section 6 (\texttt{santabarbara}), as well as the code for its end-to-end replicability. The package is available at \url{https://CRAN.R-project.org/package=ridgetorus}.

\section*{Acknowledgments}

The authors acknowledge support from grant PID2021-124051NB-I00, funded by MCIN/\-AEI/\-10.13039/\-501100011033 and by ``ERDF A way of making Europe'', and from the Community of Madrid through the framework of the multi-year agreement with Universidad Carlos III de Madrid in its line of action ``Excelencia para el Profesorado Universitario'' (EPUC3M13). Funding for APC: Universidad Carlos III de Madrid (Read \& Publish Agreement CRUE-CSIC 2023).

\appendix

\section{Proofs}
\label{Appendix:Proofs}

\begin{proof}{ of Proposition \ref{prp:1}}
	A standard change of variables gives $f(\mathbf{x};\boldsymbol{\mu},\mathbf{R})
	=f\big(\mathbf{R}^{\prime}(\mathbf{x}-\boldsymbol{\mu})\big)$, for $\mathbf{x}\in\mathbb{R}^p$. Then, $\mathrm{D}f(\mathbf{x};\boldsymbol{\mu},\mathbf{R})=\mathbf{R}\mathrm{D}f(\mathbf{y})$ and $\mathrm{H}f(\mathbf{x};\boldsymbol{\mu},\mathbf{R})=\mathbf{R}\mathrm{H}f(\mathbf{y})\mathbf{R}^{\prime}$, where $\mathbf{y}=\mathbf{R}^{\prime}(\mathbf{x}-\boldsymbol{\mu})$. Now, since $\mathrm{H}f(\mathbf{y})=\mathbf{U}(\mathbf{y}) \boldsymbol{\Lambda}(\mathbf{y}) \mathbf{U}(\mathbf{y})^{\prime}$, we have that
	\begin{align*}
		\mathrm{H}f(\mathbf{x};\boldsymbol{\mu},\mathbf{R})
		&=\left(\mathbf{R} \mathbf{U}(\mathbf{y})\right) \boldsymbol{\Lambda}(\mathbf{y}) \left(\mathbf{R}\mathbf{U}(\mathbf{y})\right)^{\prime}.
	\end{align*}
	Since $\mathbf{R} \mathbf{U}(\mathbf{y})$ is orthogonal, there exists a valid eigendecomposition of $\mathrm{H}f(\mathbf{x};\boldsymbol{\mu},\mathbf{R})$ with eigenvalues given by $\boldsymbol{\Lambda}(\mathbf{y})$. The projected gradient is then 
	\begin{align*}
		\mathrm{D}_{(p-1)} f(\mathbf{x};\boldsymbol{\mu},\mathbf{R})
		=(\mathbf{R}\mathbf{U}_{(p-1)}(\mathbf{y}))(\mathbf{R}\mathbf{U}_{(p-1)}(\mathbf{y}))^{\prime} (\mathbf{R}\mathrm{D}f(\mathbf{y}))=\mathbf{R}\mathrm{D}_{(p-1)} f(\mathbf{y}).
	\end{align*}
	Then, since $\mathbf{R}$ is an orthogonal matrix, $\mathrm{D}_{(p-1)}f(\mathbf{y})=\mathbf{0}$ if and only if $\mathrm{D}_{(p-1)} f(\mathbf{x};\boldsymbol{\mu},\mathbf{R})=\mathbf{0}$.
\end{proof}

\begin{proof}{ of Proposition \ref{prp:2}}
	We prove that all the points spanned by the first principal component satisfy the condition in Definition \ref{eq:ridge}. In view of Proposition \ref{prp:1}, we assume $\boldsymbol{\mu}=\mathbf{0}$.
	
	First, recall that the gradient and Hessian of $f$ are $\mathrm{D}f(\mathbf{x})=g^{\prime}\big(\mathbf{x}^{\prime}\boldsymbol{\Sigma}^{-1} \mathbf{x}\big) \big(2\boldsymbol{\Sigma}^{-1} \mathbf{x}\big)$ and $\mathrm{H}f(\mathbf{x})= g^{\prime \prime}\big(\mathbf{x}^{\prime}\boldsymbol{\Sigma}^{-1} \mathbf{x}\big) \big(2\boldsymbol{\Sigma}^{-1} \mathbf{x}\big) \big(2\boldsymbol{\Sigma}^{-1} \mathbf{x}\big)^\prime+g^{\prime}\big(\mathbf{x}^{\prime}\boldsymbol{\Sigma}^{-1} \mathbf{x}\big)2\boldsymbol{\Sigma}^{-1}$, with $g^{\prime}$ denoting the derivative of $g$ and $\mathbf{x}\in\mathbb{R}^p$. By the spectral decomposition theorem, $\boldsymbol{\Sigma}=\sum_{i=1}^{p} \lambda_{i} \mathbf{v}_{i} \mathbf{v}_{i}^{\prime}$, with $\mathbf{v}_i^{\prime}\mathbf{v}_j=\delta_{ij}$, $i,j=1,\ldots,p$, and $\boldsymbol{\Sigma}^{-1}=\sum_{i=1}^{p} \lambda_{i}^{-1} \mathbf{v}_{i} \mathbf{v}_{i}^{\prime}$. Therefore,
	\begin{align*}
		\mathrm{D}f(c\mathbf{v}_{1})=&\;g^{\prime}\big(c^2\lambda_1^{-1} \big) 2 c \lambda_{1}^{-1} \mathbf{v}_{1},\\
		\mathrm{H} f(c \mathbf{v}_{1})=&\;g^{\prime \prime}\big(c^2\lambda_1^{-1}\big)\big(4c^{2} \lambda_{1}^{-2} \mathbf{v}_{1} \mathbf{v}_{1}^{\prime}\big)+g^{\prime}\big(c^2\lambda_1^{-1}\big)2\left(\sum_{i=1}^{p} \lambda_{i}^{-1} \mathbf{v}_{i} \mathbf{v}_{i}^{\prime}\right).
	\end{align*}
	As a consequence, $\mathrm{H} f(c \mathbf{v}_{1})$ admits the decomposition $\mathrm{H}f(c \mathbf{v}_{1})=\mathbf{V} \boldsymbol{\Lambda}(c \mathbf{v}_{1}) \mathbf{V}^{\prime}$, where $\mathbf{V}=(\mathbf{v}_1,\ldots,\mathbf{v}_p)$ and $\boldsymbol{\Lambda}(c\mathbf{v}_{1})=\big(4c^2\lambda_1^{-2} g^{\prime \prime} (c^2\lambda_1^{-1})+2\lambda_1^{-1}g^{\prime}(c^2\lambda_1^{-1})\big)\mathbf{e}_1\mathbf{e}_1^\prime+2g^{\prime}(c^2\lambda_1^{-1})\,\mathrm{diag}\big(0,\lambda_2^{-1}, \ldots,\lambda_{p}^{-1}\big)$. By assumption, $g^{\prime}(x)<0$ for all $x\geq0$, so all but the first eigenvalue of $\mathrm{H} f(c\mathbf{v}_{1})$ are strictly negative. Finally,
	$\mathrm{D}_{(p-1)}(c \mathbf{v}_{1})
	\propto \mathbf{V}_{(p-1)} \mathbf{V}_{(p-1)}^{\prime} \mathbf{v}_{1} =\mathbf{0}$.
\end{proof}

\begin{proof}{ of Proposition \ref{prp:limit_cases_vm}}
	In \ref{prp:limit_cases_vm_2}, the derivatives with respect to $\theta_1$ and the cross-terms are null. Thus, the implicit equation is $\mathrm{D}_2 (w -  \vert w \vert) = 0 \iff \mathrm{D}_2 = 0 \text{ or } w \leq 0$. The first condition gives the solution $\theta_2 = 0$. The second gives $\theta_2=\vartheta$ such that $\cos \vartheta \geq \big[1 + (1-4\kappa_2^2)^{1/2}\big]/(2\kappa_2)$ or $\cos \vartheta \leq \big[1 - (1-4\kappa_2^2)^{1/2}\big]/(2\kappa_2)$. The eigenvalue condition is $w - \vert w \vert < 0$, which is satisfied by this last solution (except in the boundaries) and by $\theta_2 =0$, where $w = -\kappa_2$. Since these solutions are horizontal lines that do not intersect, $\theta_2=0$ is the unique connected component passing through $\mathbf{0}$, hence giving the claimed form for $\mathcal{R}_\mathbf{0}(f_\mathrm{BSvM})$.
	
	In \ref{prp:limit_cases_vm_1}, $\mathrm{D}_1=\mathrm{D}_2$ and $u=w$, so the implicit equation holds if $\theta_2=\mathrm{sign}(\lambda)\theta_1$. The eigenvalue condition \eqref{eq:eigenvalue} turns to $u-\vert v\vert<0$, which, if $\lambda >0$, leads to $-\kappa_1\cos\theta_1-\lambda(\sin^2\theta_1+\cos^2\theta_1)<0 \Rightarrow \cos\theta_1<-\lambda/\kappa_1$ (the diagonal is not complete when $\kappa_1 > \lambda$). The case $\lambda<0$ follows by symmetry. 
\end{proof}

\begin{proof}{ of Proposition \ref{prop:limit_cases_WC}}
	In \ref{prp:limit_cases_wc_1}, $c_0=1+\xi^2$, $c_1=0$, $c_2=2\xi_2$, $c_3=0$, and $c_4=0$. Analogously to the proof of Proposition \ref{prp:limit_cases_vm}, the implicit equation is $\mathrm{D}_2 (w -  \vert w \vert) = 0$ and the eigenvalue condition reads $w - \vert w \vert < 0$. Since $\mathrm{D}_2 \propto \sin \theta_2$, $\theta_2 = 0$ is the solution of the implicit equation. For this solution, $w = -2\xi_2$ and $\theta_2 = 0 \subset \mathcal{R}_{\mathbf{0}}(f_\mathrm{BWC})$. Similarly to Proposition \ref{prp:limit_cases_vm}, the other solutions are of the form $\theta_2=\vartheta$ such that $\cos \vartheta$ is bounded by a constant value. Again, since the restrictions are independent of $\theta_1$, the other solutions do not intersect in $\theta_2=0$, making it the unique connected component passing through $\mathbf{0}$.
	
	In \ref{prp:limit_cases_wc_2}, we assume that $\rho>0$ ($\rho<0$ is analogous by symmetry). Thus, if $\xi_1=\xi_2$ and $\theta_1 = \theta_2$, then $\mathrm{D}_1 = \mathrm{D}_2$ and $u = w$. Hence, it is evident that the implicit equation is satisfied for the diagonal. To conclude, Equation \eqref{eq:eigenvalue} becomes $u-v<0\Rightarrow \cos \theta_1 < (c_3+c_4)/c_1$. Since the limit of the RHS when $\xi\to0$ is $+\infty$, the entire diagonal is part of the ridge.
\end{proof}

\begin{proof}{ of Proposition \ref{prop:rf}}
	For \ref{prop:rf:1}, it trivially follows that $\tilde{\mathbf{r}}_{f,\boldsymbol{\mu}}(0)=\mathbf{r}_{f,\boldsymbol{\mu}}(0)=\boldsymbol{\mu}$ since $r_{f,\boldsymbol{\mu},j}(\mu_1)=\mu_2$, $j = 1,2$. The first entry of $\tilde{\mathbf{r}}_{f,\boldsymbol{\mu}}(\pm\pi)=\mathbf{r}_{f,\boldsymbol{\mu}}(R/2)$ is $\mathrm{cmod}(\mu_1\pm\pi)$ due to the pointwise symmetry about $\boldsymbol{\mu}$ of $\mathcal{R}_{\boldsymbol{\mu}}(f_\mathrm{BSvM})$ and $\mathcal{R}_{\boldsymbol{\mu}}(f_\mathrm{BWC})$. The second entry is $\mu_2$ or $\mathrm{cmod}(\mu_2\pm\pi)$, since $\rho_m(0),\rho_m(\pm\pi)\in\{0,\pi\}$ and $r_{f,\boldsymbol{\mu},1}(\mu_1+\pi)=\mathrm{cmod}(\mu_2+\rho_m(\pm\pi)-\rho_m(0))$.
	
	The statement \ref{prop:rf:2} follows due to the arc-length parametrization of \eqref{eq:rf2} and the scaling in \eqref{eq:rf3}.
\end{proof}


\begin{thebibliography}{}
	
	\bibitem[Allen et~al., 2012]{allen2012influence}
	Allen, L.~Z., Allen, E.~E., Badger, J.~H., McCrow, J.~P., Paulsen, I.~T.,
	Elbourne, L. D.~H., Thiagarajan, M., Rusch, D.~B., Nealson, K.~H.,
	Williamson, S.~J., Venter, J.~C., and Allen, A.~E. (2012).
	\newblock Influence of nutrients and currents on the genomic composition of
	microbes across an upwelling mosaic.
	\newblock {\em ISME J.}, 6(7):1403--1414.
	
	\bibitem[Auad et~al., 1998]{auad1998wind}
	Auad, G., Hendershott, M.~C., and Winant, C.~D. (1998).
	\newblock Wind-induced currents and bottom-trapped waves in the {Santa Barbara
		Channel}.
	\newblock {\em J. Phys. Oceanogr.}, 28(1):85--102.
	
	\bibitem[Boomsma et~al., 2008]{Boomsma2008}
	Boomsma, W., Mardia, K.~V., Taylor, C.~C., Ferkinghoff-Borg, J., Krogh, A., and
	Hamelryck, T. (2008).
	\newblock A generative, probabilistic model of local protein structure.
	\newblock {\em Proc. Natl. Acad. Sci. U.S.A.}, 105(26):8932--8937.
	
	\bibitem[Chen et~al., 2015]{Chen2015}
	Chen, Y.-C., Genovese, C.~R., and Wasserman, L. (2015).
	\newblock Asymptotic theory for density ridges.
	\newblock {\em Ann. Stat.}, 43(5):1896--1928.
	
	\bibitem[Delicado, 2001]{delicado2001}
	Delicado, P. (2001).
	\newblock Another look at principal curves.
	\newblock {\em J. Multivar. Anal.}, 77:84--116.
	
	\bibitem[Delicado, 2003]{delicado2003}
	Delicado, P. (2003).
	\newblock Principal curves of oriented points: Theoretical and computational
	improvements.
	\newblock {\em Comput. Stat.}, 18(2):293--315.
	
	\bibitem[DiGiacomo et~al., 2004]{DiGiacomo2004}
	DiGiacomo, P.~M., Washburn, L., Holt, B., and Jones, B.~H. (2004).
	\newblock Coastal pollution hazards in southern {California} observed by {SAR}
	imagery: stormwater plumes, wastewater plumes, and natural hydrocarbon seeps.
	\newblock {\em Mar. Pollut. Bull.}, 49(11):1013--1024.
	
	\bibitem[Eltzner et~al., 2018]{Eltzner2018}
	Eltzner, B., Huckemann, S., and Mardia, K.~V. (2018).
	\newblock Torus principal component analysis with applications to {RNA}
	structure.
	\newblock {\em Ann. Appl. Stat.}, 12(2):1332--1359.
	
	\bibitem[Fletcher et~al., 2004]{Fletcher2004}
	Fletcher, P.~T., Lu, C., Pizer, S.~M., and Joshi, S. (2004).
	\newblock Principal geodesic analysis for the study of nonlinear statistics of
	shape.
	\newblock {\em IEEE Trans. Med. Imaging}, 23(8):995--1005.
	
	\bibitem[García-Portugués et~al., 2023]{garciaportugues2021}
	García-Portugués, E., Lafaye~de Micheaux, P., Meintanis, S.~G., and
	Verdebout, T. (2023).
	\newblock Nonparametric tests of independence for circular data based on
	trigonometric moments.
	\newblock {\em Stat. Sin.}, to appear.
	
	\bibitem[Genovese et~al., 2014]{Genovese2014}
	Genovese, C.~R., Perone-Pacifico, M., Verdinelli, I., and Wasserman, L. (2014).
	\newblock Nonparametric ridge estimation.
	\newblock {\em Ann. Stat.}, 42(4):1511--1545.
	
	\bibitem[Hall et~al., 1992]{hall1992ridge}
	Hall, P., Qian, W., and Titterington, D.~M. (1992).
	\newblock Ridge finding from noisy data.
	\newblock {\em J. Comput. Graph. Stat.}, 1(3):197--211.
	
	\bibitem[Hastie and Stuetzle, 1989]{Hastie1989}
	Hastie, T. and Stuetzle, W. (1989).
	\newblock Principal curves.
	\newblock {\em J. Am. Stat. Assoc.}, 84(406):502--516.
	
	\bibitem[Jona-Lasinio et~al., 2012]{Jona-Lasinio2012}
	Jona-Lasinio, G., Gelfand, A., and Jona-Lasinio, M. (2012).
	\newblock Spatial analysis of wave direction data using wrapped {G}aussian
	processes.
	\newblock {\em Ann. Appl. Stat.}, 6(4):1478--1498.
	
	\bibitem[Jung et~al., 2012]{Jung2012}
	Jung, S., Dryden, I.~L., and Marron, J.~S. (2012).
	\newblock Analysis of principal nested spheres.
	\newblock {\em Biometrika}, 99(3):551--568.
	
	\bibitem[Kato and Jones, 2015]{Kato2015}
	Kato, S. and Jones, M.~C. (2015).
	\newblock A tractable and interpretable four-parameter family of unimodal
	distributions on the circle.
	\newblock {\em Biometrika}, 102(1):181--190.
	
	\bibitem[Kato and Pewsey, 2015]{Kato2015a}
	Kato, S. and Pewsey, A. (2015).
	\newblock A {M}\"obius transformation-induced distribution on the torus.
	\newblock {\em Biometrika}, 102(2):359--370.
	
	\bibitem[Kent and Mardia, 2009]{Kent2009}
	Kent, J.~T. and Mardia, K.~V. (2009).
	\newblock Principal component analysis for the wrapped normal torus model.
	\newblock In Gusnanto, A., Mardia, K.~V., and Fallaize, C.~J., editors, {\em
		{LASR} 2009 -- Statistical Tools for Challenges in Bioinformatics}, pages
	39--41, Leeds. Department of Statistics, University of Leeds.
	
	\bibitem[Ley and Verdebout, 2018]{Ley2018}
	Ley, C. and Verdebout, T., editors (2018).
	\newblock {\em Applied Directional Statistics}.
	\newblock Chapman \& Hall/CRC Interdisciplinary Statistics Series. CRC Press,
	Boca Raton.
	
	\bibitem[Mardia et~al., 2008]{Mardia2008}
	Mardia, K.~V., Hughes, G., Taylor, C.~C., and Singh, H. (2008).
	\newblock A multivariate von {M}ises distribution with applications to
	bioinformatics.
	\newblock {\em Can. J. Stat.}, 36(1):99--109.
	
	\bibitem[Mardia and Jupp, 1999]{Mardia1999a}
	Mardia, K.~V. and Jupp, P.~E. (1999).
	\newblock {\em Directional Statistics}.
	\newblock Wiley Series in Probability and Statistics. Wiley, Chichester.
	
	\bibitem[Mardia et~al., 2007]{Mardia2007}
	Mardia, K.~V., Taylor, C.~C., and Subramaniam, G.~K. (2007).
	\newblock Protein bioinformatics and mixtures of bivariate von {M}ises
	distributions for angular data.
	\newblock {\em Biometrics}, 63(2):505--512.
	
	\bibitem[Mu et~al., 2005]{Mu2005}
	Mu, Y., Nguyen, P.~H., and Stock, G. (2005).
	\newblock Energy landscape of a small peptide revealed by dihedral angle
	principal component analysis.
	\newblock {\em Proteins}, 58(1):45--52.
	
	\bibitem[Nodehi et~al., 2015]{Nodehi2015}
	Nodehi, A., Golalizadeh, M., and Heydari, A. (2015).
	\newblock Dihedral angles principal geodesic analysis using nonlinear
	statistics.
	\newblock {\em J. Appl. Stat.}, 42(9):1962--1972.
	
	\bibitem[Ozertem and Erdogmus, 2011]{Ozertem2011LocallyDP}
	Ozertem, U. and Erdogmus, D. (2011).
	\newblock Locally defined principal curves and surfaces.
	\newblock {\em J. Mach. Learn. Res.}, 12(34):1249--1286.
	
	\bibitem[Pewsey and Garc\'ia-Portugu\'es, 2021]{Pewsey2021}
	Pewsey, A. and Garc\'ia-Portugu\'es, E. (2021).
	\newblock Recent advances in directional statistics.
	\newblock {\em Test}, 30(1):1--58.
	
	\bibitem[Qiao and Polonik, 2016]{Qiao2016}
	Qiao, W. and Polonik, W. (2016).
	\newblock Theoretical analysis of nonparametric filament estimation.
	\newblock {\em Ann. Stat.}, 44(3):1269--1297.
	
	\bibitem[Riccardi et~al., 2009]{Riccardi2009}
	Riccardi, L., Nguyen, P.~H., and Stock, G. (2009).
	\newblock Free-energy landscape of {RNA} hairpins constructed via dihedral
	angle principal component analysis.
	\newblock {\em J. Phys. Chem. B}, 113(52):16660--16668.
	
	\bibitem[Singh et~al., 2002]{Singh2002}
	Singh, H., Hnizdo, V., and Demchuk, E. (2002).
	\newblock Probabilistic model for two dependent circular variables.
	\newblock {\em Biometrika}, 89(3):719--723.
	
	\bibitem[Sittel et~al., 2017]{Sittel2017}
	Sittel, F., Filk, T., and Stock, G. (2017).
	\newblock Principal component analysis on a torus: theory and application to
	protein dynamics.
	\newblock {\em J. Chem. Phys.}, 147(24):244101.
	
	\bibitem[Wehrly and Johnson, 1980]{Wehrly1980}
	Wehrly, T.~E. and Johnson, R.~A. (1980).
	\newblock Bivariate models for dependence of angular observations and a related
	{M}arkov process.
	\newblock {\em Biometrika}, 67(1):255--256.
	
	\bibitem[White et~al., 2010]{white2010ocean}
	White, C., Selkoe, K.~A., Watson, J., Siegel, D.~A., Zacherl, D.~C., and
	Toonen, R.~J. (2010).
	\newblock Ocean currents help explain population genetic structure.
	\newblock {\em Proc. R. Soc. B: Biol. Sci.}, 277(1688):1685--1694.
	
	\bibitem[Winant et~al., 2003]{winant2003characteristic}
	Winant, C.~D., Dever, E.~P., and Hendershott, M.~C. (2003).
	\newblock Characteristic patterns of shelf circulation at the boundary between
	central and southern {California}.
	\newblock {\em J. Geophys. Res.}, 108(C2):3021.
	
	\bibitem[Zoubouloglou et~al., 2022]{zoubouloglou2021scaled}
	Zoubouloglou, P., Garc\'ia-Portugu\'es, E., and Marron, J.~S. (2022).
	\newblock Scaled torus principal component analysis.
	\newblock {\em J. Comput. Graph. Stat.}, to appear.
	
\end{thebibliography}

\end{document}